\documentclass[12pt,preprint]{aastex}

\def\arcsec {$^{\prime \prime}$}

\def\etal   {{et~al.\/}}

\def\mo     {{$M_{\odot}$}}

\begin{document}

\shorttitle{Galactic Mid-IR Excesses}
\shortauthors{Uzpen, B. \etal}

\title{The Frequency of Mid-Infrared Excess Sources in Galactic Surveys}

\author{B. Uzpen, \altaffilmark{1} H. A. Kobulnicky, \altaffilmark{1}
A. J. Monson, \altaffilmark{1} M. J. Pierce, \altaffilmark{1}
D. P. Clemens,\altaffilmark{2} D. E. Backman, \altaffilmark{3}
M. R. Meade,\altaffilmark{4} B. L. Babler,\altaffilmark{4}
R. Indebetouw,\altaffilmark{5} B. A. Whitney,\altaffilmark{6}
C. Watson,\altaffilmark{7} M. G. Wolfire,\altaffilmark{8}
R. A. Benjamin,\altaffilmark{9} S. Bracker,\altaffilmark{4}
T. M. Bania,\altaffilmark{2} M. Cohen,\altaffilmark{10}
C. J. Cyganowski, \altaffilmark{4} K. E. Devine,\altaffilmark{4}
F. Heitsch, \altaffilmark{11} J. M. Jackson,\altaffilmark{2}
J. S. Mathis,\altaffilmark{4} E. P. Mercer,\altaffilmark{2}
M. S. Povich, \altaffilmark{4} J. Rho, \altaffilmark{12}
T. P. Robitaille, \altaffilmark{13} M. Sewilo, \altaffilmark{4}
S. R. Stolovy,\altaffilmark{12} D. F. Watson, \altaffilmark{4}
M. J. Wolff,\altaffilmark{6} \and E. Churchwell \altaffilmark{4} }


\altaffiltext{1}{University of Wyoming, Dept. of Physics \&
Astronomy, Dept. 3905, Laramie, WY 82071}

\altaffiltext{2}{Boston University, Institute for Astrophysical
Research, 725 Commonwealth Ave., Boston, MA 02215}

\altaffiltext{3}{SOFIA, MS 211-3, NASA-Ames Research Center, Moffett Field, CA
94035-1000}

\altaffiltext{4}{University of Wisconsin-Madison, Dept. of Astronomy,
475 N. Charter St., Madison, WI 53706}

\altaffiltext{5}{University of Virginia, Dept. of Astronomy,
P.O. Box 3818, Charlottesville, VA, 22903-0818}

\altaffiltext{6}{Space Science Institute, 4750 Walnut St. Suite 205,
Boulder, CO 80301}

\altaffiltext{7}{Manchester College, Dept. of Physics, North Manchester, IN 46962}

\altaffiltext{8}{University of Maryland, Dept. of Astronomy,
College Park, MD 20742-2421}

\altaffiltext{9}{University of Wisconsin-Whitewater, Physics Dept.,
800 W. Main St., Whitewater, WI 53190}

\altaffiltext{10}{University of California-Berkeley, Radio Astronomy
Lab, 601 Campbell Hall, Berkeley, CA 94720}

\altaffiltext{11}{University of Michigan, Department of Astronomy
500 Church St, Ann Arbor, MI 48109-1042}

\altaffiltext{12}{Caltech, Spitzer Science Center, MS 314-6, Pasadena,
CA 91125}

\altaffiltext{13}{School of Physics and Astronomy, University of St Andrews, North Haugh, St Andrews, Fife KY16 9SS}





\begin{abstract}

We have identified 230 Tycho-2 Spectral Catalog stars that exhibit 8
$\mu$m mid-infrared extraphotospheric excesses in the MidCourse Space
Experiment (MSX) and \textit{Spitzer Space Telescope} Galactic Legacy
MidPlane Survey Extraordinaire (GLIMPSE) surveys.  Of these, 183 are
either OB stars earlier than B8 in which the excess plausibly arises
from a thermal bremsstrahlung component or evolved stars in which the
excess may be explained by an atmospheric dust component.  The
remaining 47 stars have spectral classifications B8 or later and
appear to be main sequence or late pre-main-sequence objects harboring
circumstellar disks.  Six of the 47 stars exhibit multiple signatures
characteristic of pre-main-sequence circumstellar disks, including
emission lines, near-infrared K-band excesses, and X-ray emission.
Approximately one-third of the remaining 41 sources have emission
lines suggesting relative youth.  Of the 25 GLIMPSE stars with $SST$
data at $\lambda\geq24$ $\mu$m, 20 also show an excess at 24 $\mu$m.
Three additional objects have 24 $\mu$m upper limits consistent with
possible excesses, and two objects have photospheric measurements at
24 $\mu$m.  Six MSX sources had a measurement at $\lambda > 8$
$\mu$m. We modeled the excesses in 26 stars having two or more
measurements in excess of the expected photospheres as
single-component blackbodies.  We determine probable disk temperatures
and fractional infrared luminosities in the range $191 < T < 787$ and
3.9 $\times$ 10$^{-4}$ $<$ $\frac{L_{IR}}{L_{*}}$ $<$ 2.7 $\times$
10$^{-1}$.  Six stars exhibit fractional infrared luminosities less
than or equal to 10$^{-3}$ and are consistent with having debris disks
similar to that around $\beta$ Pictoris.  Six stars have
$\frac{L_{IR}}{L_{*}}$ $>$ 5 $\times$ 10$^{-2}$ and are almost
certainly Class II protostars.  The majority of our sample (14 stars)
have $10^{-3} < \frac{L_{IR}}{L_{*}} < 10^{-2}$ and are consistent
either with transition disks in late-stage protostars or massive
debris disks.  These objects have fractional infrared luminosities and
temperatures between those of $\beta$-Pic type debris-disk systems
($\frac{L_{IR}}{L_{*}}$ $\leq$ 10$^{-3}$) and Class II
pre-main-sequence systems ($\frac{L_{IR}}{L_{*}}$ $\simeq$ 10$^{-1}$).
The modeled temperatures for excesses in emission-line stars are all
$>430$~K, while the best-fit temperatures for excesses in non
emission-line stars are almost exclusively $<430$~K.  We estimate a
lower limit on the fraction of Tycho-2 Spectral Catalog main-sequence stars
having mid-IR, but not near-IR, excesses to be 1.0$\pm0.3$$\%$.

\end{abstract}

\keywords{circumstellar matter --- planetary systems: }

\section{Introduction}

The formation of planetary systems is one of the major unsolved
problems in modern-day astronomy. A necessary step in understanding
planet formation is understanding the circumstellar disks in which the
planets form (Lada \& Wilking 1984). During evolution of a Class
II\footnote{Using the stellar formation classes of Wilking \etal\
(1989), Class I objects are accreting protostars, Class II are
pre-main-sequence stars with circumstellar disks, Class III are
pre-main-sequence stars with their circumstellar disk nearly
dissipated.}  pre-main-sequence star (T-Tauri or Herbig AeBe star),
the original optically thick circumstellar disk dissipates as a star
approaches the main-sequence. These disks re-radiate stellar energy at
longer wavelengths (e.g., Zuckerman 2001). Because of this
re-radiation, pre-main-sequence stars are easily detected in the
far-IR. When the primordial disk is nearly depleted, the star becomes
a class III pre-main-sequence star. After the initial primordial disk
clears nearly all of its gas and dust a new disk can form
(e.g., Lagrange \etal\ 2000). This disk is not due to the initial
collapse of the interstellar cloud, but rather it is a
second-generation structure produced by fragmentation of planetesimals
that formed in the original protoplanetary disk.  These
second-generation disks are composed primarily of dust and are known
as debris disks (e.g., Backman \& Paresce 1993).

The detection of an excess in the infrared does not necessarily imply
the presence of a disk, but investigations of infrared excesses often
reveal the existence and nature of disks. The fractional infrared
luminosity, $\frac{L_{IR}}{L_{*}}$, can be used to differentiate
between protoplanetary and debris disks.  Protoplanetary disk systems
have $\frac{L_{IR}}{L_{*}}$$\leq$10$^{-1}$, while debris disk systems
have $\frac{L_{IR}}{L_{*}}$$\simeq$10$^{-3}$ (Lagrange \etal\ 2000).
Debris disk prototypes are the 200$\pm$100~Myr old Vega system
(Barrado y Navascu\'{e}s 1998), which has a far-IR excess at
$\lambda\geq24$ $\mu$m (Su \etal \ 2005), and the 20$\pm$10~Myr
(Barrado y Navascu\'{e}s \etal \ 1999) old $\beta$ Pictoris system,
which has an excess at mid- and far-IR wavelengths ($\lambda>$4.5-10)
$\mu$m (Backman \etal \ 1992).  Sylvester \etal\ (1996) investigated
23 main-sequence stars exhibiting large fractional infrared excesses
at far-IR wavelengths and found that 9 of these sources exhibited
near-IR excesses ($\lambda<3$) $\mu$m, yielding similar near-IR colors
to pre-main-sequence stars. The stars exhibiting near-IR excesses had
greater fractional infrared luminosities than typical Vega-like
systems, consistent with the presence of remnants from the original
protoplanetary disk. The wavelength where the excess first appears
provides an indication of the nature and evolutionary status of the
disk. Sylvester \etal\ (1996) found that the stars that exhibited an
excess only at $\lambda>3.5$ $\mu$m were more similar to Vega-like
systems.  The largest fractional infrared luminosities appear at the
shortest wavelengths and indicate the most massive and optically thick
disks, while the smallest excesses appear only at long wavelengths and
are signatures of debris disks.  A complete evolutionary progression
from protoplanetary disk systems to Vega-like disk systems has yet to
be determined.  Dusty debris may cause the IR excess at any point
after planetesimals have formed and the disks have cleared their
gaseous component, a process that takes 3--30 Myr (Meyer \etal \ 2006
and references therein).

The detection of dust around main-sequence stars provides insight into
the planetary formation process. Aumann \etal \ (1984) reported the
first detection of an IR excess owing to dust in the Vega system, and
this excess was later explained as a debris disk.  Infrared excesses
are the signatures that have allowed the \textit{Infrared Astronomical
Satellite} (\textit{IRAS}), \textit{Infrared Space Observatory}
(\textit{ISO}), and \textit{Spitzer Space Telescope} (\textit{SST};
Werner \etal \ 2004) to identify nearby stars with disk systems,
including protoplanetary disks and debris disk (Werner \etal \ 2006;
Lagrange \etal \ 2000; Mannings \& Barlow 1998; Backman \etal \ 1993
and references therein).  Using the \textit{IRAS} Point Source Catalog
matched with the SAO catalog, Oudmaijer \etal \ (1992) identified 462
stars that exhibited infrared excesses based on photometric
colors. This same technique was utilized by Clarke \etal \ (2005) in
which the Mid-Course Space Experiment (MSX; Egan \etal\ 2003) catalog
was matched with the Tycho-2 Spectral Catalog to identify 1938 sources that
exhibited an infrared excess based on photometric colors.  Both the
Oudmaijer \etal \ (1992) and Clarke \etal \ (2005) samples provide a
wealth of sources to further investigate the nature of the stellar
excesses. Understanding the nature of the excess allows statistical
surveys into the frequency of occurrence of circumstellar material
located around stars. At least 15\% of nearby A-K main-sequence stars
have dusty debris disks that were detectable to \textit{IRAS} and
\textit{ISO} sensitivities in the far-infrared (Meyer \etal \ 2006;
Lagrange \etal \ 2000; Backman \& Paresce 1993). Plets \& Vynckier
(1999) found the excess fraction for both main-sequence stars and
their descendants also to be $\sim$13$\pm$10\% at 60 $\mu$m, and
Bryden \etal\ (2006) found an excess fraction of 2$\pm$2$\%$ at 24
$\mu$m, for main-sequence field FGK stars with
$\frac{L_{IR}}{L_{*}}$$\geq$10$^{-4}$.

Excesses at mid-IR wavelengths are much less common for main-sequence
stars (Meyer \etal\ 2006).  Aumann \& Probst (1991) investigated 548
nearby main-sequence stars using 12 $\mu$m measurements from
\textit{IRAS}, but out of the 60 stars with possible V-[12] excesses
targeted for ground based follow-up, only $\beta$ Pictoris and $\zeta$
Leporis exhibited genuine excesses once confusing sources within the
large $IRAS$ beam were excised.  Fajardo-Acosta \etal\ (2000)
investigated 2834 \textit{IRAS} sources, of which 296 were
main-sequence B9 through M stars, and found 8 possible 12 $\mu$m
excesses.  Stars that exhibit a mid-infrared excess have
characteristic disk temperatures of 200--1000 K and dust within a few
AU of the star, similar to our asteroid belt.  By comparison,
long-wavelength excesses have cooler $\sim$100 K dust located farther
from the central star and are more comparable to the Kuiper Belt.  The
warmer dust identified in mid-IR surveys may also be a signature of a
disk in a transitional state from protoplanetary to debris disk (Meyer
\etal\ 2006).

``Transition disks'' are especially interesting as possible
evolutionary links between the optically thick (Class II) disk systems
and the older optically thin debris disk systems.  Transition disks
may be evolved protoplanetary disks with inner holes caused by
clearing of the initial gas-dominated disks (Strom \etal \ 1989).
Recent photometric work of Silverstone \etal \ (2006) found 5 young
($<10~Myr$) solar type stars with optically thick disks out of 74
searched at mid-IR wavelengths. They note that they have not found any
optically thin disks that emit in the mid-IR, indicating that their
sample of excesses consists entirely of pre-main-sequence classical
T-Tauri objects.  Megeath \etal\ (2005) identified 4 candidate solar-
and lower-mass transition disks in $\eta$ Cha, of the 17 stars
investigated. Three of the 4 sources indicate accretion based on their
H$\alpha$ profiles.  Sicilia-Aguilar \etal\ (2006) have also
identified a number of candidate transition disks, including
intermediate-mass stars that exhibit mid-IR excesses. Hern\'{a}ndez
\etal \ (2006) identify a B9 star which exhibits an excess at both
[8.0] and [24] and has a spectral energy distribution (SED) similar to
$\beta$ Pictoris. Identifying and studying such a sample of mid-IR
excess sources is an important step in understanding disk evolution.

In this paper we present results of a search for mid-IR excesses in
the \textit{SST} Galactic Legacy Infrared MidPlane Survey
Extraordinaire (GLIMPSE; Benjamin \etal \ 2003) and MSX catalogs to
determine the fraction of mid-IR excess sources among field stellar
populations.  We begin with stellar sources from the Tycho-2 Spectral
Catalog (Wright \etal\ 2003), as this is the largest available
compilation of stellar classifications needed to model the SED and
identify extraphotospheric emission. In Section 2 we describe the
sample selection and source matching between the optical and IR
catalogs.  We also describe new optical spectroscopy used to determine
and/or confirm temperature and luminosity classifications for a small
subsample of mid-IR excess sources. In Section 3 we identify stars
with mid-IR excesses by comparing photometric colors to model SEDs.
In Section 4 we confirm the mid-IR excess for some sources through 24
$\mu$m Multiband Imaging Spectrometer for Spitzer (MIPS; Rieke \etal \
2004) measurements from the MIPSGAL (PID 20597) survey. For stars with
mid-IR excesses and at least one measurement at $\lambda>8$ $\mu$m, we
model the excess as a single temperature blackbody to estimate the
probable disk temperatures and fractional infrared luminosities in
Section 5.

\section{Data}

\subsection{GLIMPSE}
\subsubsection{IR Photometry}

\begin{figure}
    \plotone{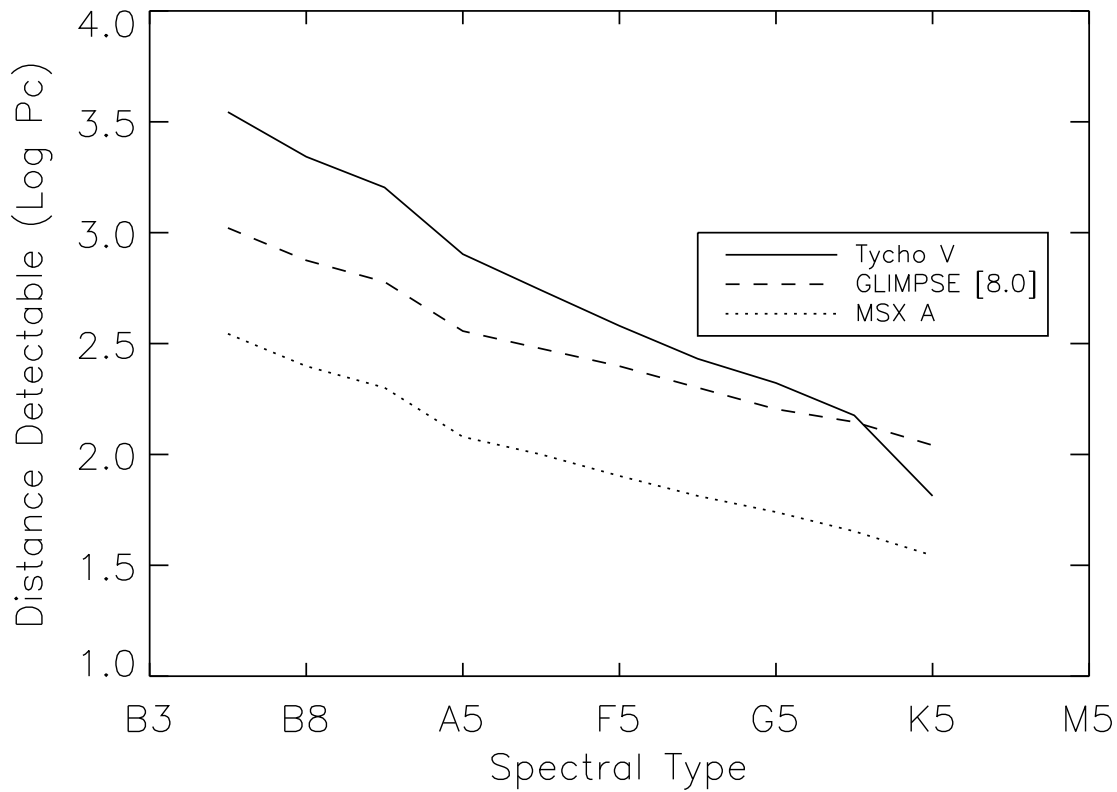}
    \caption{Limiting distance for main-sequence stars, as a function of spectral type, for
    inclusion in the Tycho-2, GLIMPSE, and MSX catalogs. The solid line
    shows the distance at which Tycho magnitudes reach V=11.5, 95
    \% completeness.  The dashed line is the distance at which IRAC
    [8.0] reaches a flux of 10 mJy. The dotted line is the distance at
    which MSX A band reaches a flux of 100 mJy.}
    \label{sens}
\end{figure}

The GLIMPSE project is one of the \textit{SST} Legacy
Programs. GLIMPSE mapped the Galactic Plane in four infrared array
camera (IRAC; Fazio \etal \ 2004) bandpasses, $[3.6]$, $[4.5]$,
$[5.8]$, and $[8.0]$ $\mu$m from $|l|=10^\circ-65^\circ$ and
$|b|\leq1^\circ$ degrees (Benjamin \etal \ 2003). This survey has
generated a Point Source Catalog of 3$\times$$10^{7}$ objects.
To be included in the GLIMPSE Point Source Catalog, a source must have
a signal-to-noise ratio greater than 5:1 with at least two detections
in one band, at least one detection in an adjacent band, and fluxes
densities greater than 0.6, 0.6, 2, and 10 mJy (in bands [3.6], [4.5],
[5.8] and [8.0] respectively).  See the GLIMPSE Data Products
Description\footnote{GLIMPSE Team Webpage
http://www.astro.wisc.edu/sirtf/docs.html} for further documentation.
Within the GLIMPSE data reduction pipeline, JHK photometry was
performed using 2MASS images (Cutri 2003), and sources were
cross-correlated with GLIMPSE sources. The GLIMPSE program is the
largest, most sensitive, mid-infrared survey of the Galactic Plane to
date. Given the sensitivity of GLIMPSE, the survey is able to detect
at 8 $\mu$m the photospheres of unreddened main-sequence A stars to
$\sim$$600$ pc, F stars to $\sim$$300$ pc, G stars to $\sim$$200$ pc,
and K stars to $\sim$$140$ pc.  Figure~\ref{sens} shows the distance
at which unreddened main-sequence stars are detectable in all three
catalogs used in this investigation.  The solid line shows the maximum
detection distance as a function of spectral type for the V=11.5 (95\%
completeness) magnitude limit of the Tycho-2 Catalog (H\/{o}g \etal \
2000) from which the Tycho-2 Spectral Catalog sources and magnitudes
are drawn.  The dashed line shows the maximum detection distance for
GLIMPSE at [8.0].  The dotted line shows the detection limit for the
MSX Catalog A band which is centered at $\sim$ 8.28 $\mu$m.  This
figure shows that the vast majority of the unreddened mid-IR sources
detected in GLIMPSE or MSX should have optical counterparts in the
Tycho-2 Catalog.  Given these distances, the combined catalogs allow a
survey for stars exhibiting mid-IR excesses over a much larger volume
than any previous study.

Of the 4043 Tycho-2 Spectral Catalog stars within the GLIMPSE region,
we matched 3361\footnote{The Tycho sources not matched with a GLIMPSE
source were, in all but the instances to be discussed below, the
brightest Tycho stars which are saturated in GLIMPSE and not included
in the catalog.  } with sources in the GLIMPSE Point Source Catalog
(v1.0) to produce a working database with B\footnote{Johnson B and V
were derived from the linear transformation of Tycho B$_{T}$ and
V$_{T}$, V$_{J}$=V$_{T}$-0.090(B-V)$_{T}$, and
(B-V)$_{J}$=0.850(B-V)$_{T}$; see Tycho Catalog for descriptions.}, V,
2MASS J, H, K, and GLIMPSE [3.6], [4.5], [5.8], [8.0] photometry.  The
matching criteria required that the positions of GLIMPSE and Tycho-2
Spectral sources were within 3\arcsec\ of each other with no
additional sources inside this radius.  Only 37 Tycho-2 sources had
multiple GLIMPSE sources within a 3\arcsec \ radius, and these were
rejected.  The median separation for Tycho-2 Spectral and GLIMPSE
sources is 0.4\arcsec, and 99.6\% of the sources were within 1\arcsec.
The probability of a false match can be estimated by determining the
ratio of the average source density to the area searched for an
infrared counterpart to the Tycho star. However, the limiting
magnitude at [3.6] is $\sim$14 magnitudes while a Tycho star will be
brighter than 10.5 magnitudes at [3.6] using very conservative
estimates. With 3.0$\times$10$^{7}$ sources in the GLIMPSE Point
Source Catalog over an area of 220 sq. degrees, there are 0.011
sources per square arcsecond or an average area of 91 square
arcseconds per source. If we limit the GLIMPSE catalog to sources that
would be similar in brightness to Tycho sources ($[3.6]>10.5$) we are
left with 1.5$\times$10$^{6}$ sources over a 220 sq degree area.  The
source density of possible confusing sources is 5.3$\times$10$^{-4}$
per square arcsec. Allowing a maximum separation of 1.0 arcseconds
between the Tycho source and the matched GLIMPSE source, there is a
0.2$\%$ chance of positional coincidence (i.e.,
$\pi$$\times$1.0$^{2}$$\times$0.00053 = 0.002) for
GLIMPSE sources distributed randomly. Another way of stating this is
that the average source separation for possible confusing GLIMPSE
sources is $\sim$24 arcseconds, whereas the median separation between
Tycho and GLIMPSE counterparts is 0.4 arcseconds.  This probability,
multiplied by the number of Tycho sources (3361), predicts that,
at most, 7 objects could be chance matches between a Tycho
star and an unrelated GLIMPSE object.  One common signature of a false
match or confusion with an unrelated background object would be a
discontinuity in the spectral energy distribution between the Tycho
bands and 2MASS bands, or between the 2MASS bands and the GLIMPSE
bands.  We found some $\sim$30 objects where the SEDS showed such
discontinuities and these objects were discarded as unreliable matches
or possibly objects with photometric variability.

Some objects lack photometry at one or more of the nine bandpasses,
primarily because of saturation in GLIMPSE at magnitudes brighter than
7 at [3.6] or because of the 5$\sigma$ sensitivity limit of 10 mJy at
[8.0].  Objects with no [8.0] measurement, 218 stars, were removed
leaving 3143 stars in the GLIMPSE sample.  The GLIMPSE sample includes
1203 main-sequence or near-main-sequence stars with luminosity
classification V, IV, or IV/V.  The remainder are either have no
luminosity classification or are evolved stars with luminosity class
I, II, or III.

\subsubsection{Optical Spectra of Selected Sources}

We obtained optical spectroscopy of a small subset of 38 stars from
the GLIMPSE sample at declinations higher than -20$^\circ$ with the
WIRO-Spec instrument on the Wyoming InfaRed Observatory (WIRO) on the
nights of 2005 September 17-19 using the 2400 $l/mm$ volume-phase
holographic grating. The instrument is an integral field unit
spectrograph consisting of 293 densely packed fibers in a 19 $\times$
20 configuration with each fiber projecting to 1\arcsec \ on the sky.
This grating provides a dispersion of 0.38 \AA/pixel, a spectral
resolution of 0.83 \AA,\ and wavelength coverage of $\sim$$780$ \AA \
from $\sim$$4150-4900$ \AA. This wavelength regime covers hydrogen
Balmer $\beta$ and $\gamma$, several helium lines, and a few metal
lines which are useful for luminosity and temperature
classification. Standard reduction techniques were utilized, including
flat fielding using a continuum lamp and wavelength calibration using
a CuAr arc lamp, but no flux calibration was performed.  A typical
signal-to-noise of 100:1 per pixel was achieved.

We also obtained optical spectroscopy of an additional subsample of 27
stars from the GLIMPSE sample using the WIRO-Longslit spectrograph on
the night of 2005 November 20. Of the 27 stars, 19 had been observed
with WIRO-Spec. The WIRO-Longslit is a low resolution spectrograph
with a dispersion of $1.1$ \AA/pixel, a resolution of $2.6$ \AA \
using a 2\arcsec \ slit width, and covers $\sim$$3800-6100$ \AA. The
longslit data provided more metal lines for classification and was
useful for comparison with the WIRO-Spec data and spectral
classifications from the literature. The same reduction procedures
were performed for the longslit data as WIRO-Spec data.  A typical
signal-to-noise in excess of 50:1 per pixel was achieved.

Our goal was to obtain classification spectra and confirm the
cataloged spectral types for a small subset of objects.  This
subsample covered a range of spectral and luminosity classes but
preferentially consisted of B stars.  Combining the higher resolution
of WIRO-Spec with the greater wavelength coverage of WIRO-Longslit, we
were able to modify or improve upon both the temperature and
luminosity classification for 12 of the 46 stars.  Those 12 stars are
listed in Table 1 along with our improved spectral 
classifications. The stars were classified by eye using comparison
spectra (Yamashita \etal \ 1976).  Two of the 12 stars in Table 1 show
a significant deviation from the original classification. The first of
the two sources is G010.4468+00.0629, which was classified as a B3/5Ib
in the Tycho-2 Spectral Catalog, but our spectrum lacks the \ion{C}{2}
$\lambda$4267 line which is present in supergiants. The spectrum for
this source more closely resembles a B3/5V star. The other star is
G015.0981+00.3409, which was classified as a B7/8 (III), but the
\ion{He}{1} lines are similar to those of a main-sequence star. This
star also has Balmer emission, and we reclassified it as a B5Ve
star. Most stellar classifications were consistent within 2
temperature subclasses. Since less than 5\% of the sample, 2 of 46,
have a significant change in spectral type, we will assume that the
majority of the literature classifications from the Tycho-2 Spectral
Catalog are approximately correct and use them for analysis.

\subsection{MSX}

MSX surveyed the entire Galactic Plane within b=$\pm$ 5 degrees, and
it covered additional regions including the large and small Magellanic
Clouds in six bands: 4.29 (B1), 4.35 (B2), 8.28 (A), 12.13 (C), 14.65
(D) and 21.34 $\mu$m (E) (Price \etal \ 2001). The MSX survey was less
sensitive than GLIMPSE, having greatest sensitivity at 8.28 $\mu$m
(0.1 Jy; Egan \etal \ 2003) corresponding to 200 pc for unreddened
main-sequence A stars, 100 pc for F stars, 65 pc for G stars, and 45
pc for K stars.  We matched 8688 Tycho-2 Spectral Catalog sources to
MSX catalog positions. The matching criteria required that the
positions of the MSX and Tycho-2 Spectral sources were within 6\arcsec
\ with no additional sources inside this radius, the beamsize at 8
$\mu$m is 5\arcsec. The median separation for Tycho-2 Spectral and MSX
sources is 0.8\arcsec\ with 68.8\% of matches within 1\arcsec. The one
sigma positional accuracy of MSX is 2\arcsec\ (Egan 2003). We then
matched the 8688 MSX and Tycho-2 Spectral sources to the nearest 2MASS
source requiring that the 2MASS source is within 1\arcsec \ of the
Tycho-2 Spectral source position. The probability of a false match can
be estimated by determining the ratio of the average source density to
the area searched for an infrared counterpart to the Tycho star. The
MSX Catalog has $\frac{1}{7}$ the number of sources as the GLIMPSE
Catalog. As a (very) conservative estimate, we use the GLIMPSE source
density of 0.00054 sources per square arcsecond when determining
probability of a chance coincidence.  The actual MSX source density is
approximately a factor of 10 lower. Using twice the median separation
of sources, 1.6\arcsec\, as a realistic average separation, yields a
probability of a random positional coincidence of 0.4$\%$, (e.g.,
$\pi$$\times$1.6$^{2}$$\times$0.00054 = 0.004).

\section{Identifying the Excess}

The goal of our analysis is to identify stars with mid-IR excesses in
the GLIMPSE and MSX samples for the purpose of finding stars with
circumstellar disks.  We expect circumstellar disks to exist primarily
around main-sequence stars or protostars.  Evolved stars may exhibit
IR excesses owing to dust in cool extended atmospheres or thermal
bremsstrahlung emission (Barlow \& Cohen 1977).  Therefore, we
separate the GLIMPSE sample into main-sequence or near-main-sequence
objects (luminosity class IV or V) and evolved stars (luminosity
classes I, II, III).  Main-sequence O and early-B stars
often produce localized \ion{H}{2} regions with thermal bremsstrahlung
components that may contribute to an extraphotospheric excess at IR
wavelengths (e.g., Gehrz \etal\ 1974).  In Section 3.3 we
discuss in more detail the signatures thermal bremsstrahlung emission
in the SEDs.  We further separate main-sequence stars into those
earlier than B8 and those B8 or later. In the GLIMPSE main-sequence
sample there are 253 B stars (B8 or later), 391 A stars, 247 F stars,
109 G stars, and 24 K stars of luminosity class V, IV, or IV/V.  For
the MSX sample we removed all luminosity class I, II, III and unknown
objects, leaving 872 main-sequence and near-main-sequence stars
(luminosity class V, IV, or IV/V). We do not tabulate evolved stars in
the MSX sample because of the large number such objects and the
limited interest in mid-IR excesses among this population. 
See Clarke \etal\ (2005) for a tabulation of such objects. We further
separate the MSX main-sequence stars into those earlier than B8 and
those B8 or later.  The MSX main-sequence sample consists of 45 B
stars (B8 or later), 118 A stars, 238 F stars, 233 G stars, and 151 K
stars.

\subsection{GLIMPSE}

The photometric data for each object in the entire GLIMPSE subsample
of 3143 stars were fit with Kurucz ATLAS9 models using temperatures
and effective gravities corresponding to the nearest spectral type
(Kurucz 1993). This method is similar to that used by Uzpen \etal \
(2005) to identify excess sources in the field of RCW49.  We
calculated $K-[8.0]$ color excesses, $E(K-[8.0])$, by measuring the
differences between the Kurucz photospheric model and the photometric
$K-[8.0]$. The Kurucz photospheric model colors were determined by
passing the stellar model through digitized transmission curves of the
selected bandpasses. We chose the $K-[8.0]$ color because
main-sequence stars should exhibit minimal color for these wavelengths
since both bandpasses are on the Rayleigh-Jeans tail of the stellar
blackbody spectrum. This method is similar to that employed by Aumann
\& Probst (1991) who used $K-[12]$ to identify warm disk candidates.

We required that the ``excess'' stars have an
$\frac{E(K-[8.0])}{\sigma_{(K-[8.0])}}$ $\geq$ 3 where
$\sigma_{(K-[8.0])} = \sqrt{\sigma^{2}_{K} + \sigma^{2}_{[8.0]} +
\sigma^{2}_{cal}}$. Included in this uncertainty is a 5\% absolute
calibration uncertainty, $\sigma_{cal}$, for IRAC (Reach \etal \
2005).  Among the 3143 GLIMPSE stars we find 167 objects with mid-IR
excesses.  Table~2 gives the spectral types, magnitudes, and
uncertainties for the 20 luminosity class IV or V stars and 8 unknown
luminosity class objects B8 or later that met the above
criteria. Table~3 contains spectral types, magnitudes, and
uncertainties for 139 stars earlier than B8 and post main-sequence
stars.

\begin{figure}
    \plotone{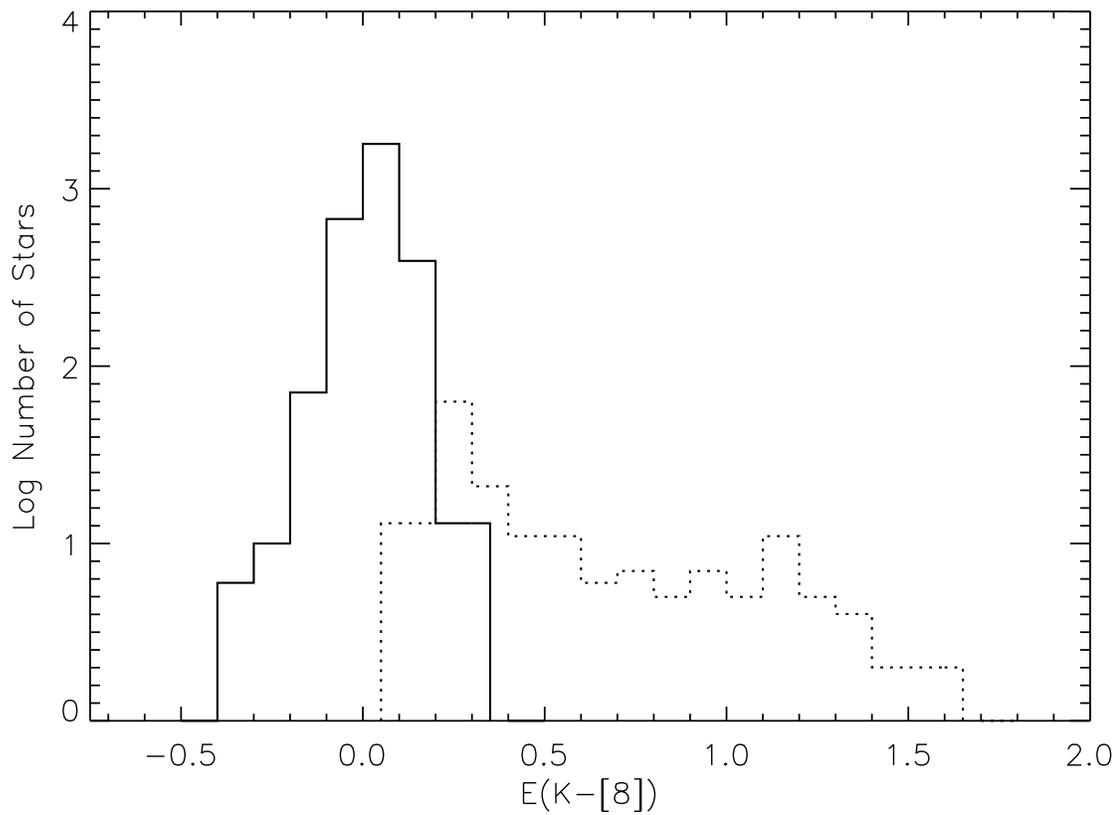}
    \caption{Histograms of $E(K-[8.0])$ for both mid-IR excess (dotted)
      and non-excess (solid) stars from the GLIMPSE sample.  
	 The excess sources form a positive tail to the
      distribution of non-excess sources. The overlap between the two
      distributions is a result of photometric uncertainties.}
    \label{glimp}
\end{figure}

\begin{figure}
    \plotone{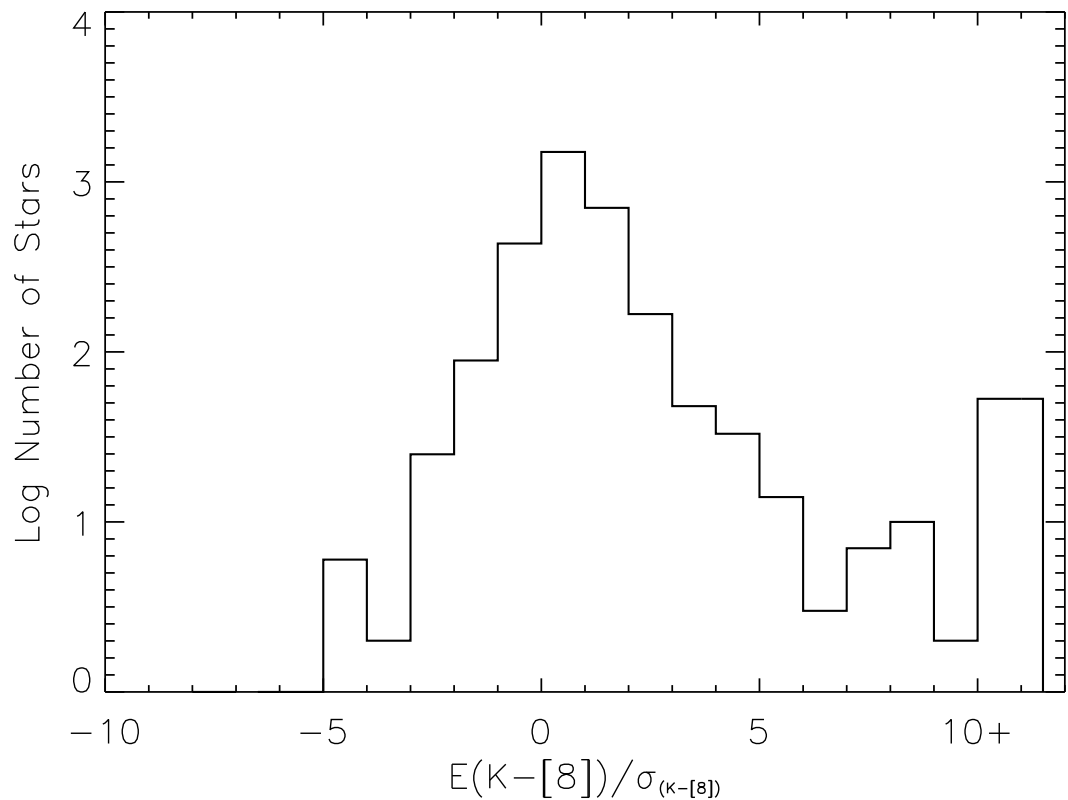} 
    \caption{Histograms of the uncertainty
    distribution for 3086 GLIMPSE objects.  The distribution is
    approximately Gaussian with a substantial positive tail. Only 10
    stars have $\frac{E(K-[8.0])}{\sigma_{(K-[8.0])}}$ $\leq$ -3
    consistent with the expectations of a normal error
    distribution. The 167 stars with
    $\frac{E(K-[8.0])}{\sigma_{(K-[8.0])}}$ $\geq$ 3 are listed in
    Tables 2 and 3.}  \label{sig}
\end{figure}

\begin{figure}
    \plotone{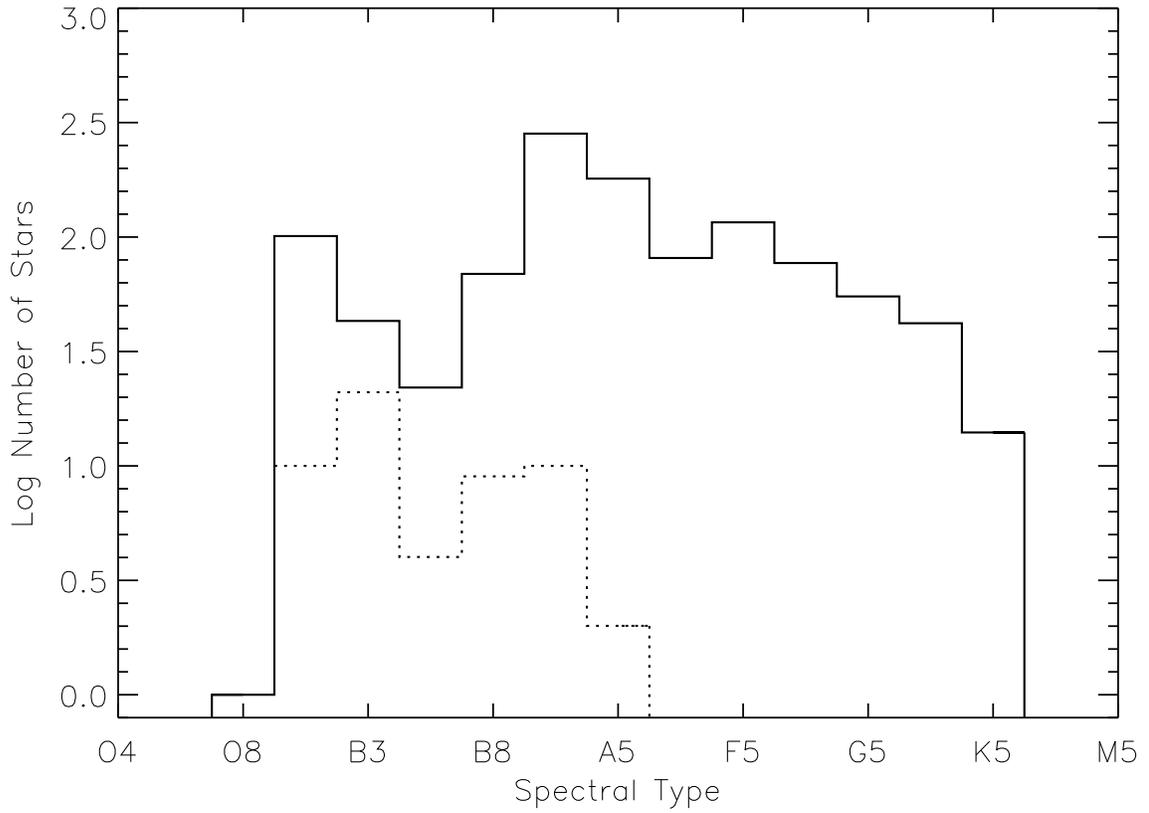}
    \caption{Histograms of main-sequence excess (dotted) and 
	non-excess (solid)
      sources by spectral type from the GLIMPSE sample. 
	The distribution is broad peaks near
      A0 for the non-excess sources. 
	The majority of the excess sources are earlier than A5.}
    \label{glimp2}
\end{figure}

Figure~\ref{glimp} shows a histogram of $E(K-[8.0])$ for both the 167
excess sources (dotted line) and 2919 non-excess sources (solid line),
demonstrating that the excess sources form a positive tail to the
distribution. Photometric uncertainties cause some overlap between the
excess and non-excess sources.  Stars with genuine excesses can have
small $E(K-[8.0])$ as long as the photometric uncertainties are small.
This population of objects will overlap in Figure~\ref{glimp} with
non-excess objects that have large $E(K-[8.0])$ but large
uncertainties.  For example, the non-excess star G035.1185--00.7465
has $E(K-[8.0])$=0.48 and $\sigma_{(K-[8.0])}$=0.19 while the excess
star G014.4239-00.7657 also has $E(K-[8.0])$=0.48 and
$\sigma_{(K-[8.0])}$=0.12. Figure~\ref{sig} shows a histogram of
$\frac{E(K-[8.0])}{\sigma_{(K-[8.0])}}$, revealing a large
population of stars that have $\frac{E(K-[8.0])}{\sigma_{(K-[8.0])}}$
$\geq$ 3. Some sources also appear to have $E(K-[8.0])$
deficits. There are only 10 sources with a deficit
$\frac{E(K-[8.0])}{\sigma_{(K-[8.0])}}$ $\leq$ -3. In these sources
the [8.0] measurement lies below the best-fitting photospheric model.
The occurrence of $E(K-[8.0])$ deficits less than 3$\sigma$ is
$\sim$0.3$\%$, consistent with the expected fraction exceeding
3$\sigma$ from a normal distribution.   Conversely we would also expect
10 stars, 6$\%$ of our 167-star excess sample, to exhibit a false [8.0]
excess on statistical grounds. We will discuss the removal of
these false excess sources in Sec. 4. M stars (27 objects) that are
poorly fit by Kurucz models and stars which show zero point
photometric offsets between IRAC and 2MASS ($\sim$30 objects),
possibly due to variability, were omitted from Figure~\ref{glimp} and
Figure~\ref{sig}.  Figure~\ref{glimp2} shows the distribution among
spectral types for the 50 main-sequence excess objects (dotted line)
and 1153 non-excess objects (solid line).  The distribution of
non-excess objects is broad and peaks near A0, while virtually all of
the excess sources are earlier than A5.  A two-sided K-S test on the
distributions of excess and non-excess spectral types
shows that there is less than a 10$^{-8}$ probability of obtaining
the excess distribution from the non-excess population.
Therefore, the population exhibitting mid-IR excesses preferentially
consists of late-B and early A stars, as would be expected
for systems capable of producing hot dust excesses.
The spectral type distribution of excess stars inconsistent with
being drawn at random from among the Tycho-2 Spectral Catalog 
parent population,
as should be expected if the excesses were
caused by confusion with unrelated background objects.  

\subsection{MSX}

We fit the MSX sample stars with a Kurucz model in the same manner as
the GLIMPSE sample. We used the same reliability constraints on the
MSX sources, $\frac{E(K-[8.0])}{\sigma_(K-[8.0])}$ $\geq$ 3, and
include an absolute photometric uncertainty in $\sigma_{K-[8.0]}$ of
5\% (Price \etal \ 2001).  There are 63 stars that exhibit a
photometric excess of $\frac{E(K-[8.0])}{\sigma_(K-[8.0])}$ $\geq$ 3
in the MSX sample. The 19 stars of spectral type B8 or later are
listed in Table 4, while the 44 stars earlier than B8 are listed in
Table 5.

\begin{figure}
    \plotone{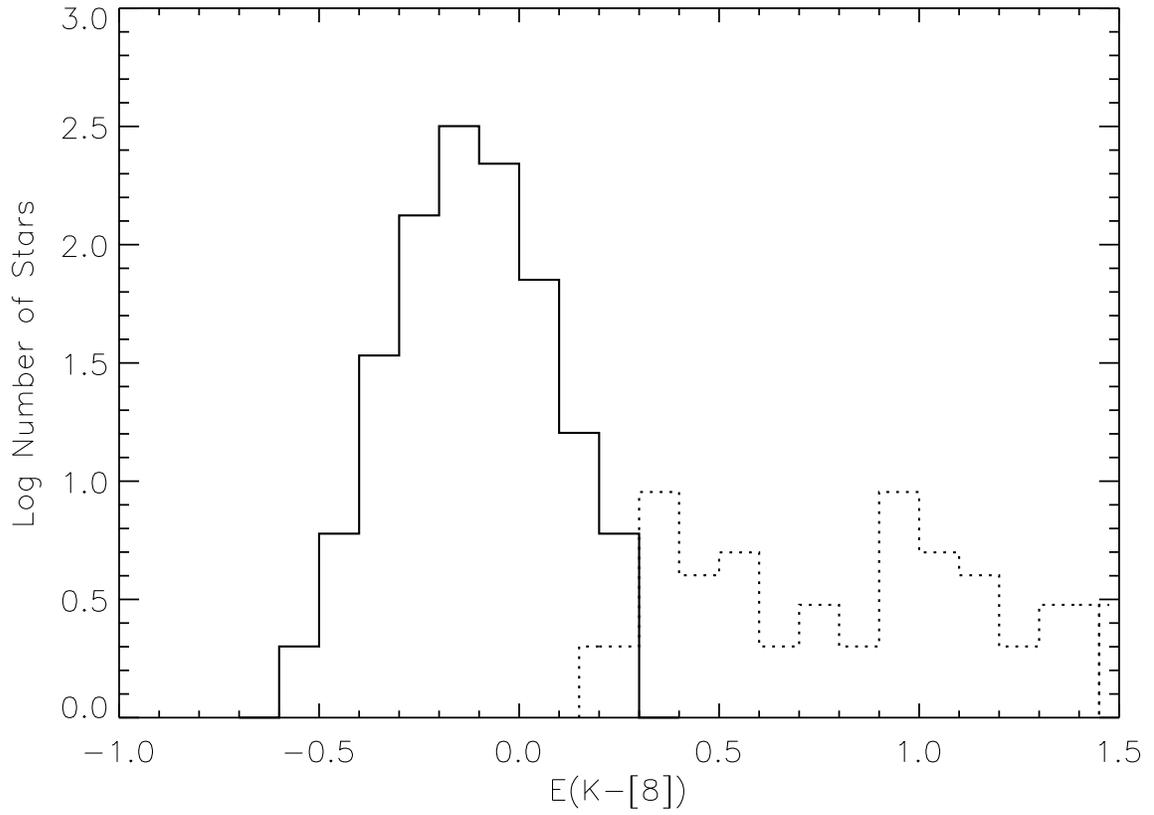}
    \caption{Histograms for main-sequence excess
      (dotted) and non-excess (solid) sources from the MSX sample. 
	The excess sources form positive a tail to the non-excess distribution.}
    \label{msx}
\end{figure}

\begin{figure}
    \plotone{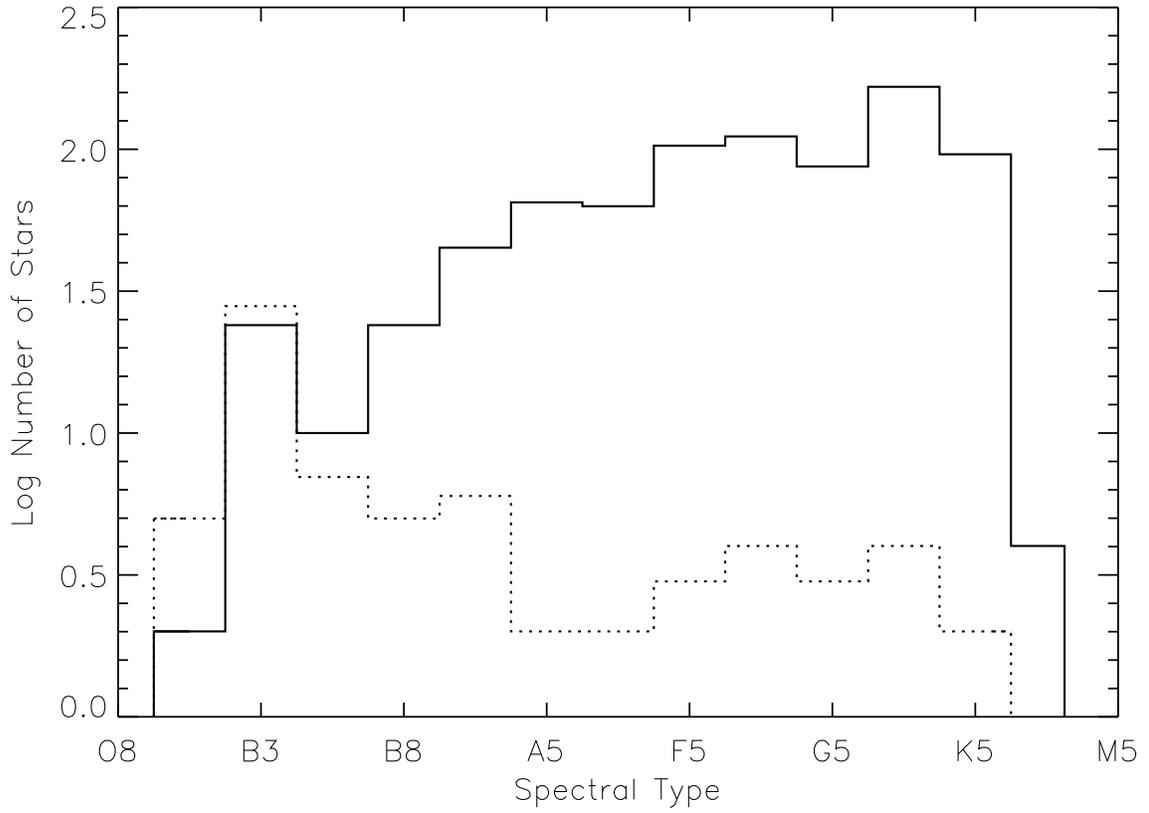}
    \caption{Histograms for main-sequence excess (dotted) and
      non-excess (solid) sources by spectral type for the MSX sample.
      A larger fraction of the MSX sources are of later spectral type
      relative to the distribution of GLIMPSE sources in
      Figure~\ref{glimp2}.  For the earliest stars excesses are more
      common than not.}
    \label{msx2}
\end{figure}

Figure~\ref{msx} is a histogram of $E(K-[8.0])$ for the main-sequence
MSX sources.  Objects identified as having an excess (dotted line)
form a positive tail to the distribution of non-excess sources (solid
line). The two histograms overlap slightly for the same reason as the
overlap in Figure~2.  The MSX sample peaks at $E(K-[8.0])$=$-$0.12,
implying a small systematic calibration uncertainty may be present
given that normal stars should have zero color.\footnote{For the MSX
band A we use a photometric zero point of 58.49 Jy from Cohen \etal\
(2000). } Since the median $E(K-[8.0])$ is negative, the stars with
the largest deficits, 8 sources, have $E(K-[8.0])<$ $-$0.4. Two of the
8 sources have an $E(K-[8.0])$ deficit but a C band excess, implying a
possible contribution from a structured background. One of the 8
sources is an M star and is poorly fit by the Kurucz models.  The
remaining deficit sources have only an A band measurement and deficit.
Given that the fractional occurrence for the deficit sources is less
than 1$\%$ and excess sources over 4$\%$, we include all excess
sources but note that some excess sources may be
spurious. Figure~\ref{msx2} shows the distribution of spectral types
in the MSX sample. The distribution peaks among late-type stars, as is
expected given the nature of the stellar initial mass function and the
biases of a magnitude-limited survey.  The MSX sample shows excesses
among all spectral types and has proportionally more late-type excess
sources than the GLIMPSE sample. We speculate that GLIMPSE contains
proportionally more early-type stars as a result of saturation in the
IRAC bands.  The IRAC [8.0] saturates at $\simeq$ 0.7 Jy in GLIMPSE
exposure times.  For a 5,500 K blackbody, stars brighter than
V$\simeq$ 7 will saturate, and for a 10,000 K blackbody, stars
brighter than V$\simeq$ 5 will saturate.  Thus, late-type stars are
preferentially missing from GLIMPSE compared to early-type stars.  The
distribution of spectral types in the MSX sample (Figure~\ref{msx2})
essentially confirms this hypothesis, given that virtually all of the
MSX stars would be saturated in GLIMPSE.

\subsection{The Mid-IR Excesses}

\begin{figure}
    \plotone{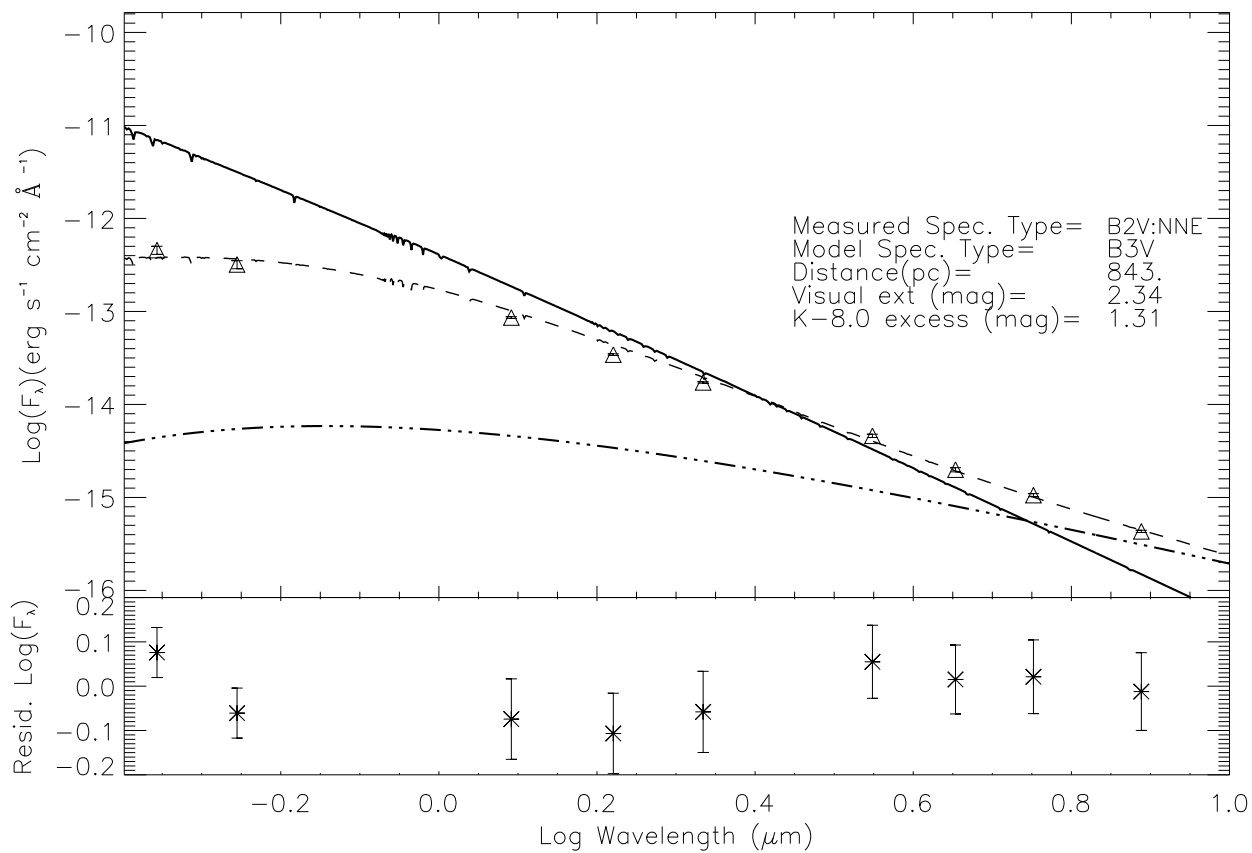}
    \caption{A spectral energy distribution of an early B
    main-sequence star, G058.7277-00.4505, in which the IR excess is best explained by
    thermal bremsstrahlung emission.  The Kurucz model for a B3V star
    is shown by the thick solid curve. The dash-dot curve is the
    thermal bremsstrahlung model component. The dashed curve is the
    B3V stellar model with applied extinction and the thermal
    bremsstrahlung component. The lower panel shows the residuals from
    the Kurucz model after the inclusion of the bremsstrahlung
    component, extinction and distance normalization.}
    \label{brem}
\end{figure}

The origin of the mid-IR excesses may be something other than
absorption and re-emission of photospheric energy by small dust
particles in a circumstellar disk.  One alternative process could be
thermal bremsstrahlung emission, most likely to arise among the early
O and B stars stars as well as supergiants listed in Table 3. To test
whether free-free emission could be viable explanation for the excess,
an optically thin, thermal bremsstrahlung component was included
without a blackbody disk component in the modeling of the stars'
SEDs. The normalization of the bremsstrahlung component is a free
parameter along with extinction and distance.  Models with
bremsstrahlung components fit the SEDs of O and early B stars, which
typically have all four IRAC bands in excess of the model photosphere,
better than the models with a single blackbody component.  An example
of an SED with the thermal bremsstrahlung component is shown in
Figure~\ref{brem}.  A blackbody component with a single-temperature
provides a poor fit to the four IRAC measurements in excess of the
photosphere. Since early-type stars are ionizing stars and they can be
fit better by photospheric plus thermal bremsstrahlung models,
free-free emission becomes an increasingly likely explanation for the
mid-IR excess of stars earlier than B8. We use B8 or later as our
cut-off for systems with excesses plausibly explained by circumstellar
disks.

IR excesses may instead be caused by diffuse interstellar emission or
multiple confusing sources in the telescope beam rather than any form
of circumstellar disk. A few examples of previously identified excess
sources from \textit{IRAS} with interstellar rather than circumstellar
dust origins are found in Kalas \etal\ (2002). However, given that the
angular resolution of IRAC at [8.0] is approximately 10 times better
than $IRAS$ at 12 $\mu$m, the probability of spurious excesses from
unrelated sources in the beam is greatly reduced. Without additional
measurements, either spectroscopic or coronographic, this possibility
cannot entirely be ruled out. However, there is an argument against
some unrelated background source that appears only at 8 microns or
longer wavelengths (e.g., highly reddened stars or dusty post-AGB
stars or protostars).  If such sources were the case for the majority
of objects, we would expect the histogram of excess objects in
Figure~4 to look just like the histogram of all Tycho-2 Spectral
Catalog sources---that
is, the excesses should be randomly distributed across all spectral
types.  The fact that the excess stars are preferentially A and late B
stars, the very objects where hot dust is most likely to be seen,
argues against a large fraction of excesses due to source confusion in
the beam.

\begin{figure}
    \plotone{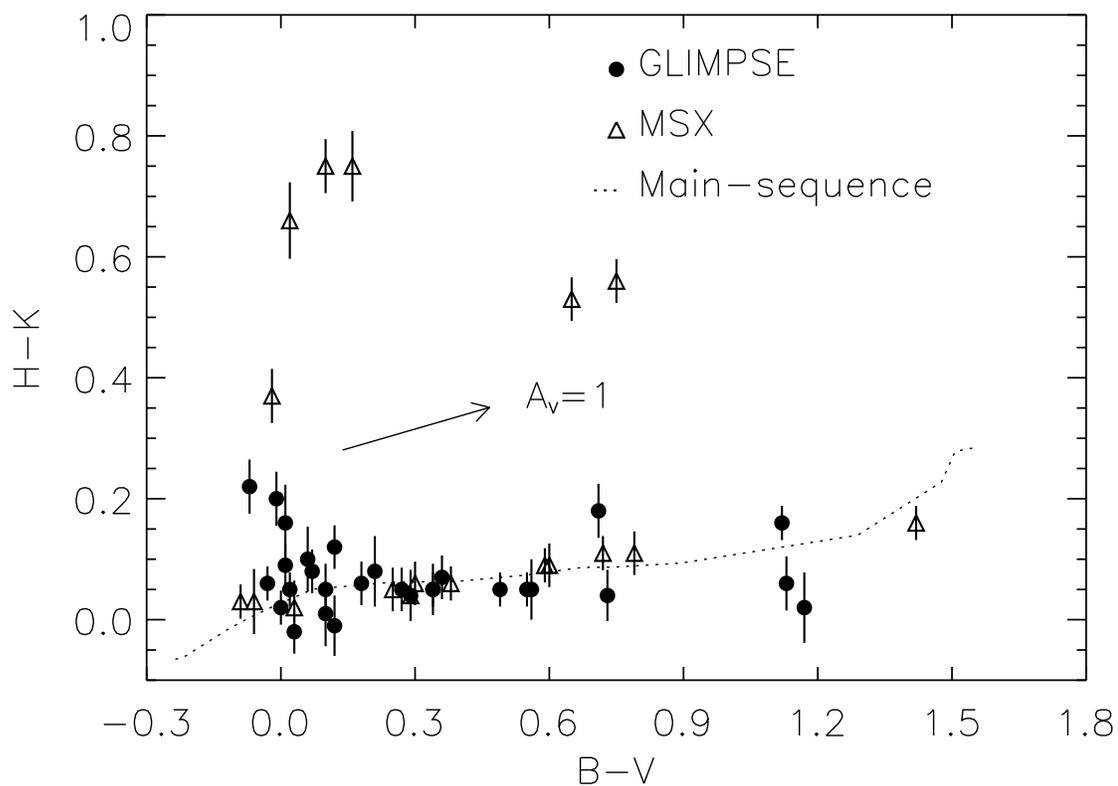}
    \caption{A color-color plot of $B-V$ versus $H-K$ for excess sources
    B8 or later from GLIMPSE (circles) and MSX (triangles).  The
    dotted line denotes the main-sequence and the arrow shows the
    reddening vector for A$_{V}$=1.  Six stars show large
    near-IR excesses, $H-K>$ 0.3, however the majority (87$\%$) of the
    stars from both samples are consistent with having minimal near-IR
    excess.}
    \label{bvhk}
\end{figure}

Another possible origin for an excess is a pre-main-sequence
circumstellar disk. During pre-main-sequence stellar evolution, stars
accrete circumstellar material via equatorial disks that produce
near-IR excesses. Figure~\ref{bvhk} shows a $B-V$ versus $H-K$
color-color plot of the 47 sources from Tables 2 and 4 (i.e., B8 and
later). GLIMPSE stars are the filled circles and MSX stars are triangles.
The main-sequence is shown by the dotted curve while the reddening
vector for $A_V=1$ is shown by the arrow. With the exception of six
objects, discussed in Sec 5.2, most stars are consistent with having
no near-IR excess. Since the remaining 41 stars do not have a near-IR
excess, the mid-IR excess is likely to arise within either a
dissipating pre-main-sequence disk or a debris disk.  

\section{Confirming the Excesses at 24 $\mu$m and Identifying False Excesses}

We performed PSF-fitting photometry on post-basic-calibrated
data (pbcd) from the MIPSGAL survey (S14.4) on images containing
mid-IR excess sources. We utilized IRAF/DAOPHOT point source fitting
with an aperture size of 15 pixels (36.8\arcsec \ ) and an aperture
correction factor of 1.08 that we obtained by interpolating Table 3.13
of the MIPS Data Handbook version 3.2.1. We matched the [24] sources
to GLIMPSE/MSX sources using an 8\arcsec \ search radius.  We modeled
the sources with the additional [24] measurement to determine
$E(K-[24]$). This was determined in the same manner as
$E(K-[8.0]$). We adopt 10\% as an absolute calibration uncertainty on
the [24] data (MIPS Data Handbook).

We detected 22 of the 28 GLIMPSE sources and were able to put upper
limits on 3 of the GLIMPSE sources.  The upper limits were calculated
using 3$\times$(RMS per pixel)$\times$$\sqrt{N_{pix}}$, where
N$_{pix}$ is the aperture area of 707 pixels.  Three sources were in
regions not yet available in the $SST$ archives. Twenty sources
detected at [24] had a measurement above the expected photosphere but
not all exhibited an excess at the 3$\sigma$ or greater level. Two
sources, G042.9876+00.4201 and G322.6930+00.0572, were found to have
photospheric [24] measurements.  Given that we expect 6$\%$ of our
167-star excess sample (0.3\% of the total GLIMPSE sample) to exhibit
a false excess due to statistical deviations, 1--2 of the 28 excess
GLIMPSE stars later than B8 should exhibit a false excess. Also, given
that we would expect at most 4$\%$ (7 of 167 stars) to have excesses
due to source correlation mismatches, we expect 1 of the 28 excess
GLIMPSE stars may be spurious.  Therefore, we expect that 2--3 of the
28 GLIMPSE excess sources may be false excesses.  Given that we have
identified the two aforementioned sources as false [8.0] excesses, we
feel that we have identified all the false [8.0] excesses among the 28
GLIMPSE sources in Tables~2 and 6.  Two MSX sources with mid-IR
excesses were within the MIPSGAL survey region and shown to be false
excess sources as well.  The false sources will be discussed
individually in Sec. 5.3. We modeled the disk temperatures and
fractional infrared luminosities for the 20 GLIMPSE sources having
[24] excesses using all available photometry.

\section{Examining the Excesses}

\subsection{Modeling}

We modeled the SEDs of stars exhibiting mid-IR excesses (Tables 2 and
4) to estimate the stellar distances, extinctions, putative
circumstellar disk temperatures, and fractional infrared luminosities.
The photometric data points were fit with Kurucz ATLAS9 models using
temperatures and effective gravities corresponding to the nearest
spectral type (Kurucz 1993).  Since all of the stars lie within the
solar circle, we adopt the solar metallicity models.  The Kurucz model
surface fluxes were scaled by a factor, $X$, so that the model K-band
fluxes matched the observed K-band 2MASS photometry.  This scale
factor represents $X=R^2/D^2$, the ratio of the stellar radius squared
over the distance to the star squared. Adopting stellar radii
appropriate for the observed spectral types (Drilling \& Landolt
2002), we calculated a spectrophotometric distance to each star.
These distances and their uncertainties, estimated by taking into
account the K-band photometric errors added in quadrature with an
additional 15\% uncertainty in the stellar radii, are listed in the
Table 7.  The K-band measurements were used as the basis for the
distance computation because the effects of extinction are nearly
negligible at 2.2 $\mu$m for most of our stars. Our derived $A_V$
values are all $<2.5$ mag implying $A_K<0.25$ mag, with most having
$A_V$ $<1.0$. The model atmospheres were reddened with variable
amounts of extinction using the Li \& Draine (2001) extinction curve,
except in the mid-infrared ($\lambda>3$ $\mu$m) where the GLIMPSE
extinction results were used (Indebetouw \etal \ 2005).  Distance and
extinction were fit iteratively as free parameters covering a grid of
all reasonable values.  The best fit parameters for each star are
determined by the minimum of the $\chi^2$ statistic.  Modeling the
star with the next earlier or later spectral type model available in
the Kurucz library (i.e., an A3 with an A0 or an A5) produced similar
reduced $\chi^2$ values. This allows small errors in classification to
yield similar results. This also provides a check on
classification. If a star is grossly misclassified, the fit is poor,
resulting in photometric data becoming grossly inconsistent with the
SED, and the classification is re-visited.

To estimate the fractional infrared luminosities and temperatures of
disks surrounding the mid-IR excess stars, we modeled the excess
in each star as a single blackbody with the temperature and
fractional infrared luminosity, $\frac{L_{IR}}{L_{*}}$, as free
parameters. This assumption, although simplistic, yields an
approximation to the temperature and $\frac{L_{IR}}{L_{*}}$ of
the dust disks, given the limited number of photometric points
available. We searched a grid of blackbody temperatures and fractional
infrared luminosities to find best fits to the data. The temperature
grid ranged from 50 K to 1050 K in steps of 1 K. Because the IRAC data
are sensitive only to the hottest dust in the inner portions of disks,
fitting the mid-IR points in the absence of longer-wavelength
measurements essentially provides upper limits on the disk
temperatures and fractional infrared luminosities.  Chen \& Jura
(2001) found the color temperature of the disk around $\zeta$ Leporis
to be 370~K using mid-IR data, but they note that fits including data
at wavelengths longer than 10 $\mu$m indicate a disk temperature more
consistent with a 230~K blackbody. Owing to this effect we only
model sources which were measured at [24], for GLIMPSE sources, or
bands C, D or E in MSX, in addition to exhibiting
an $E(K-[8.0]$).

We investigated the likelihood of other disk temperature/fractional
infrared luminosity combinations using a Monte Carlo simulation to
determine the uncertainties on our derived disk and star
parameters. The Monte Carlo code adds noise to each of the photometric
data points based on idealizations drawn from Gaussian models of their
1$\sigma$ photometric errors and then re-computes the most probable
distance, visual extinction, blackbody temperature, and fractional
infrared luminosity. We then used the linear distributions of key
parameters from the simulations to determine their uncertainties.

Table 7 lists derived parameters and their uncertainties based on the
median values of the Monte Carlo simulations for both the GLIMPSE and
MSX excess sources B8 or later. Unfortunately, most MSX sources are
detected only at A band and fall outside the MIPSGAL survey region and
were not modeled. Also in Table 7 is the measured radial separation
between the optical and IR source.

\begin{figure}
\hbox{
    \includegraphics[width=3.0in]{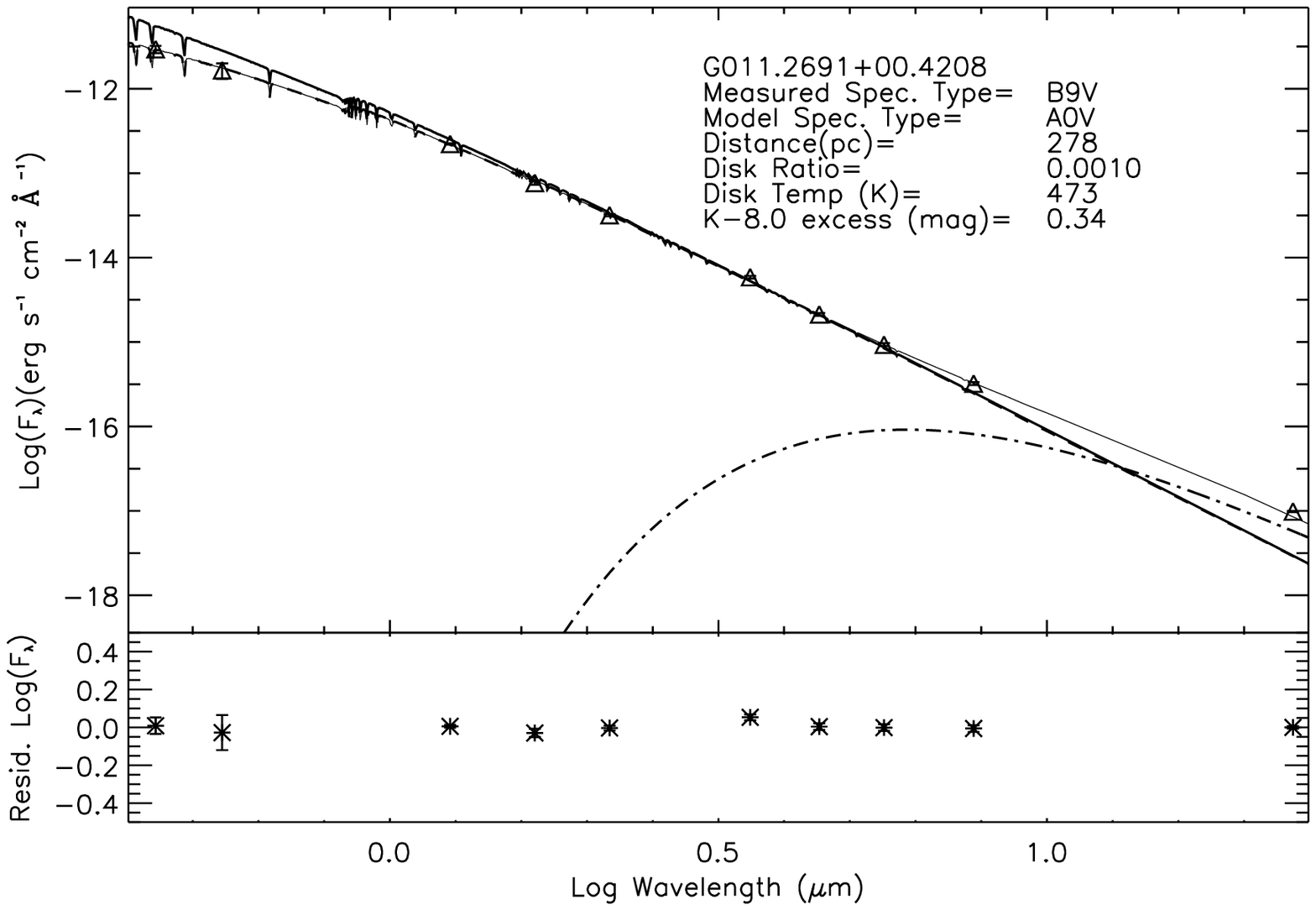}
    \includegraphics[width=3.0in]{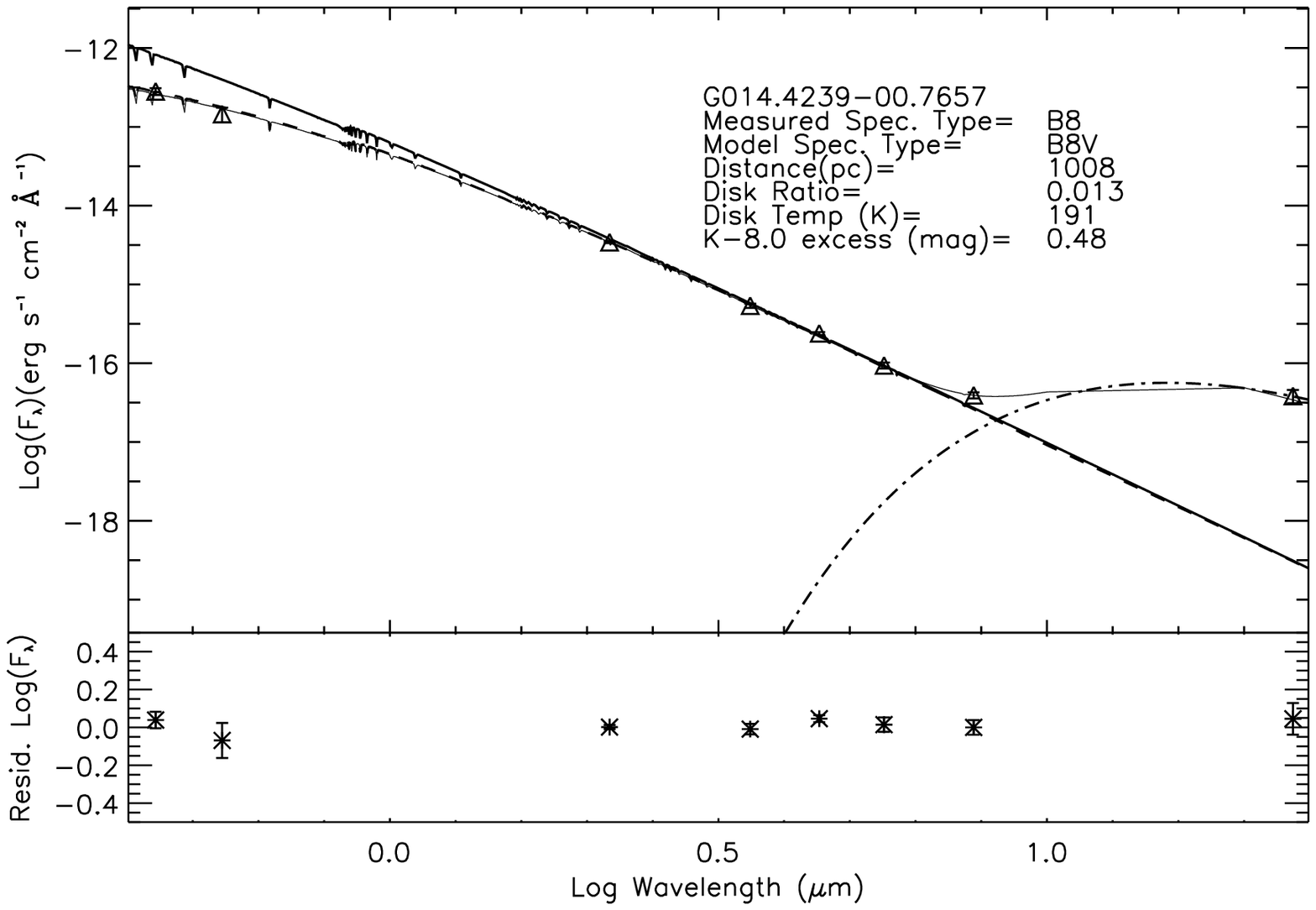}
}
\hbox{
    \includegraphics[width=3.0in]{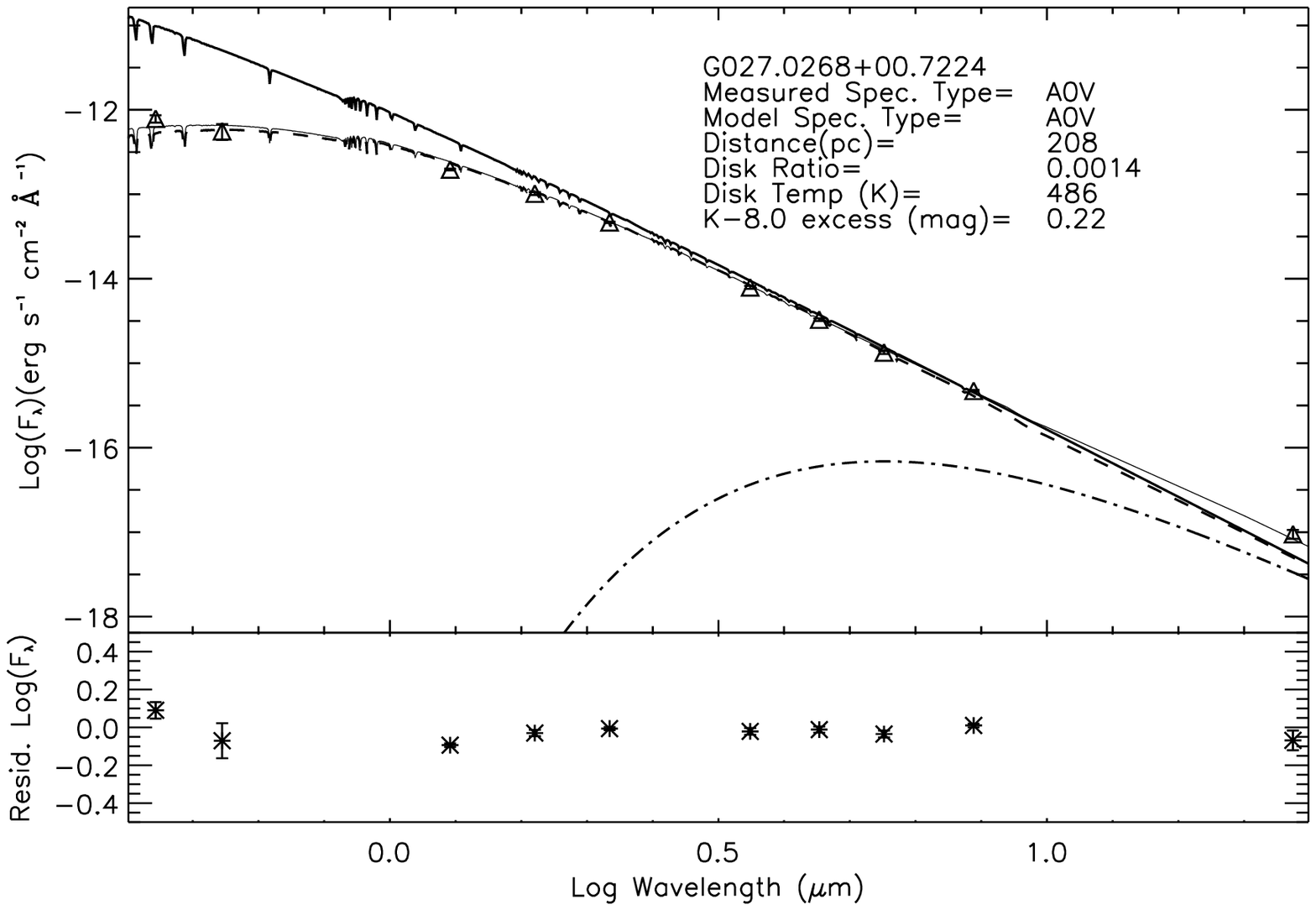}
    \includegraphics[width=3.0in]{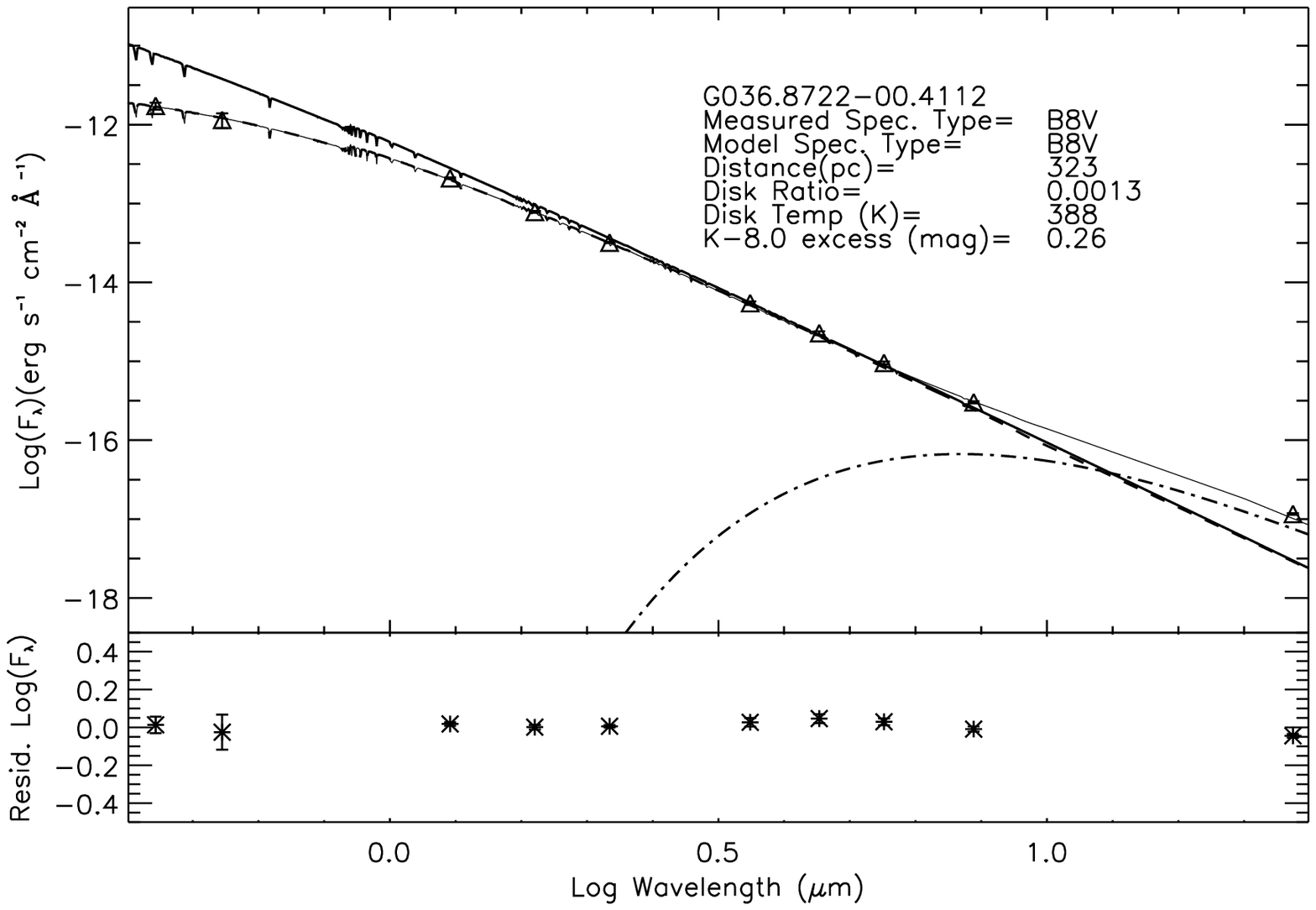}
}
    \caption{(upper left) An SED of the mid-IR excess B9V star
G011.2691+00.4208. The Kurucz model atmosphere is shown by the solid
curve, the model atmosphere with applied extinction is the dashed
curve, the dash-dot curve is the blackbody representing the disk, and
the thin curve is the extinguished model atmosphere before the
addition of the blackbody component. This source is well modeled by a
single blackbody component. (upper right) An SED of the mid-IR excess
B8 star G014.4239--00.7657. This star is well modeled by a single
temperature blackbody. (lower left) An SED of the A0V star
G027.0268+00.7224. The star is adequately modeled by a single
temperature blackbody. (lower right) An SED of the B8V star
G036.8722--00.4112.  This source is well modeled by a single
temperature blackbody.}
    \label{disks}
\end{figure}

\begin{figure}
\hbox{
    \includegraphics[width=3.0in]{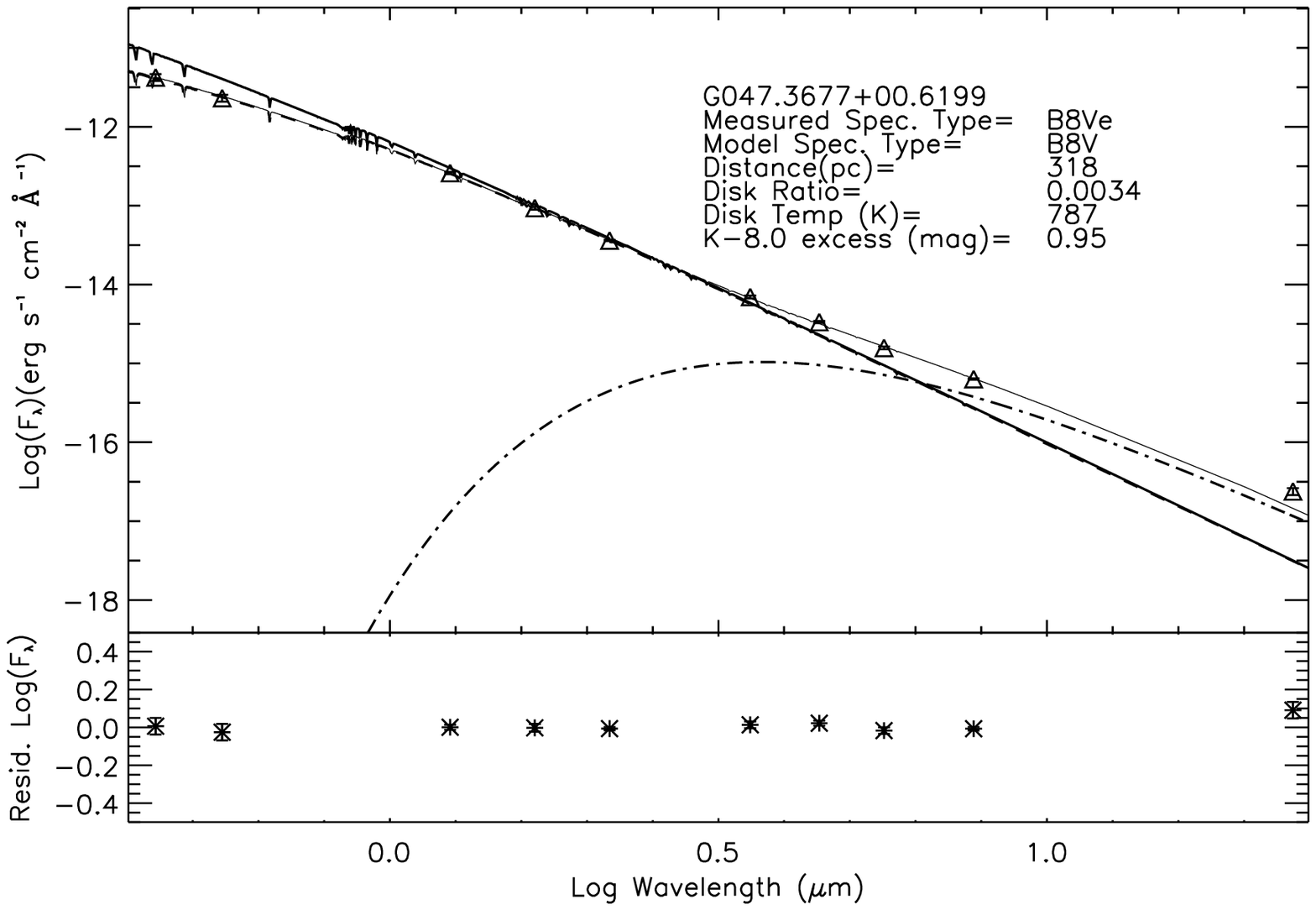}
    \includegraphics[width=3.0in]{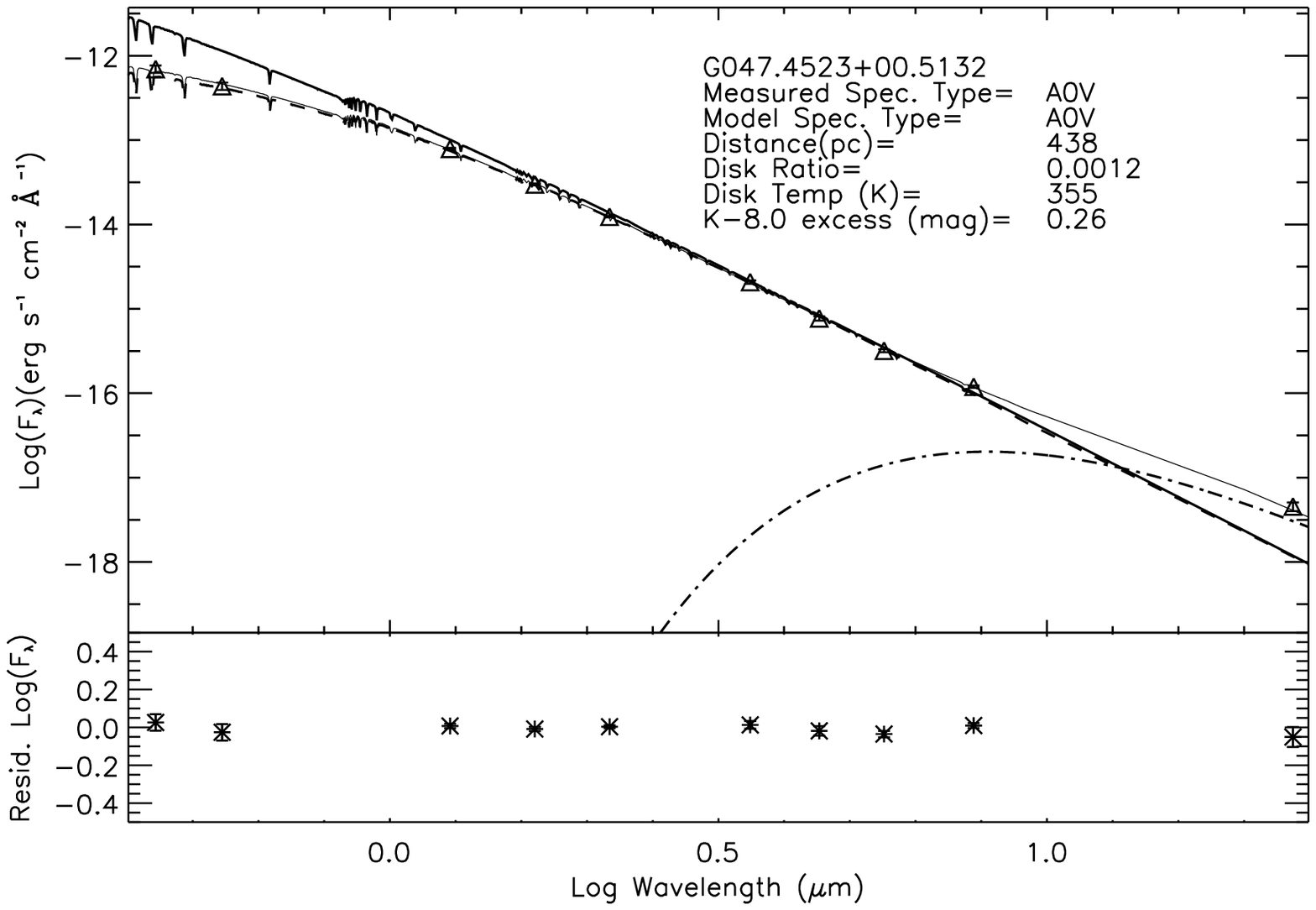}
}
\hbox{
    \includegraphics[width=3.0in]{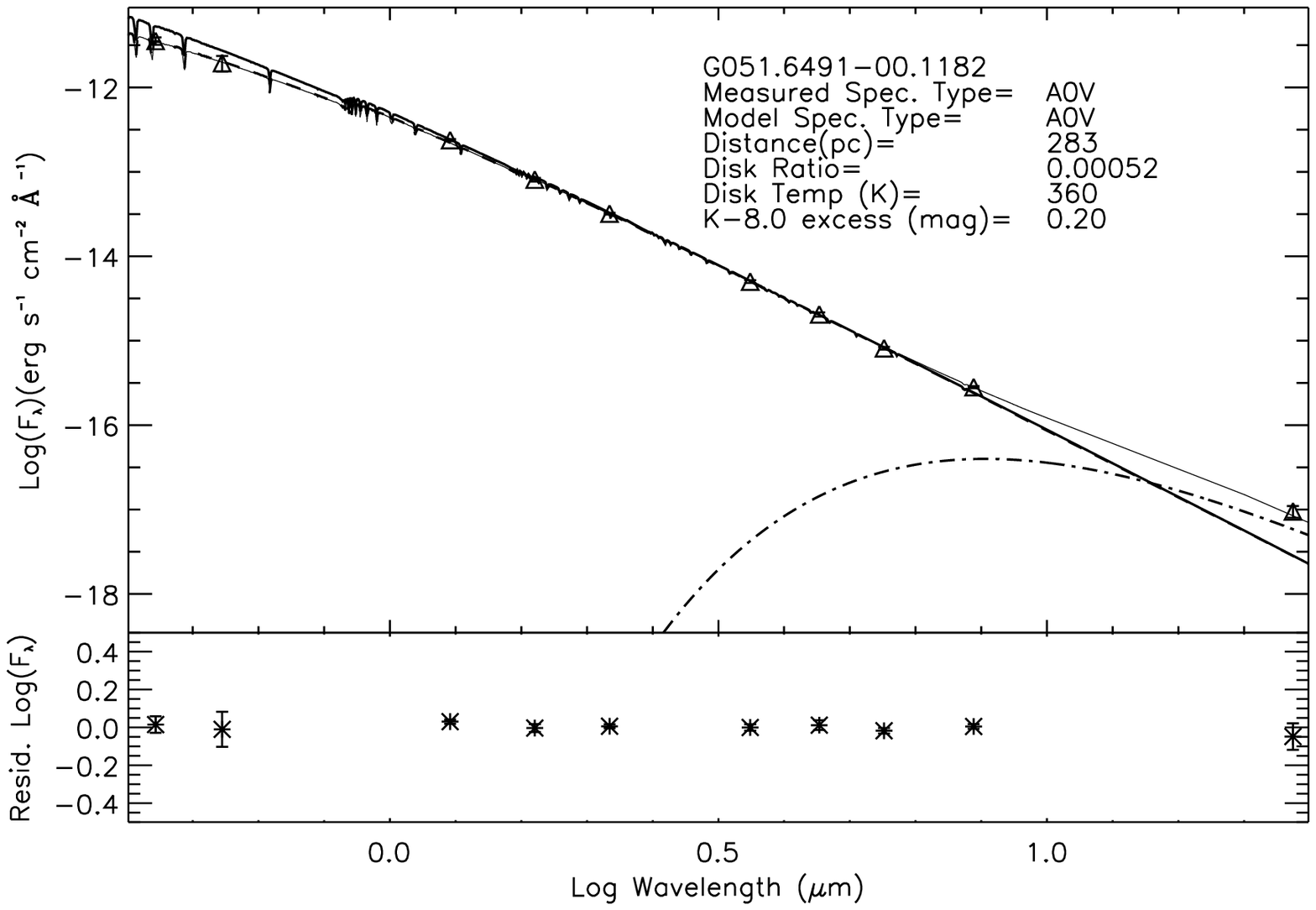}
    \includegraphics[width=3.0in]{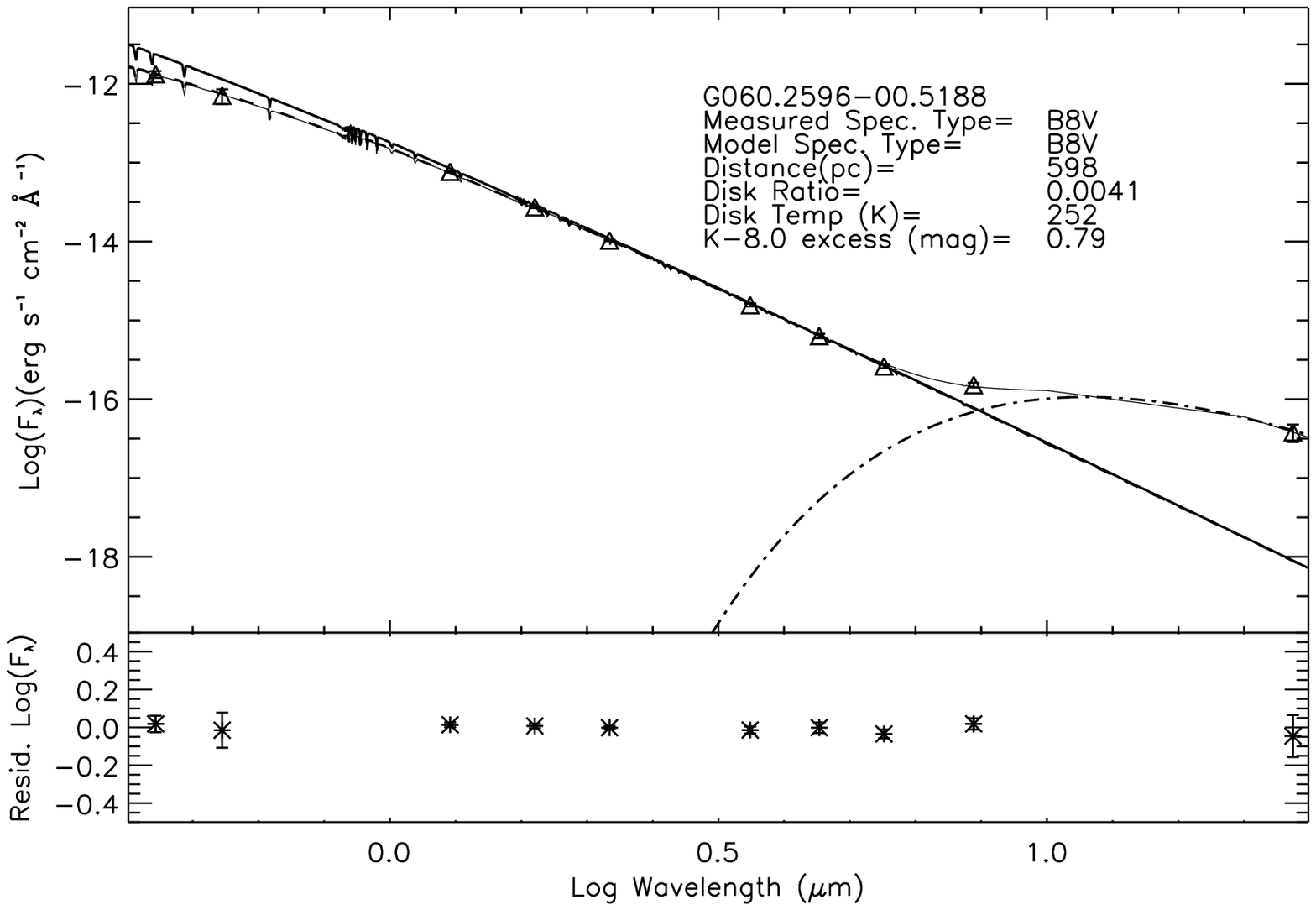}
}
 \caption{(upper left) An SED of the B8Ve star G047.3677+00.6199.  The
 source is poorly fit at [24]. (upper right) An SED of the A0V star
 G047.4523+00.5132. This star is well fit by a single-temperature
 blackbody. (lower left) An SED of the A0V star G051.6491--00.1182.
 This star is well fit by a single-temperature blackbody. (lower
 right) An SED of G060.2596--00.5188 a B8V star. The SED of this star
 is adequately modeled with an additional single-temperature blackbody.}
    \label{disks2}
\end{figure}

\begin{figure}
\hbox{
    \includegraphics[width=3.0in]{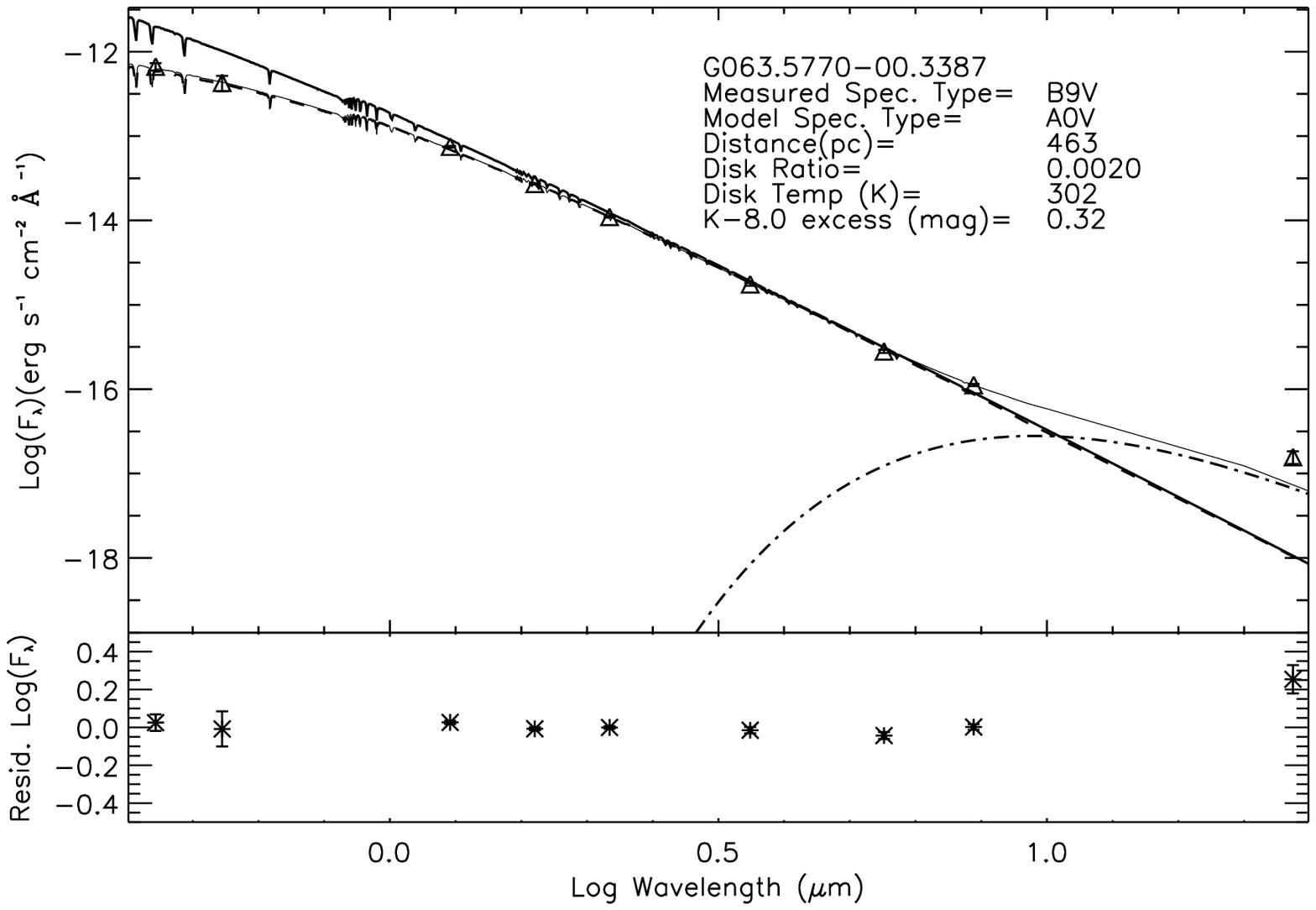}
    \includegraphics[width=3.0in]{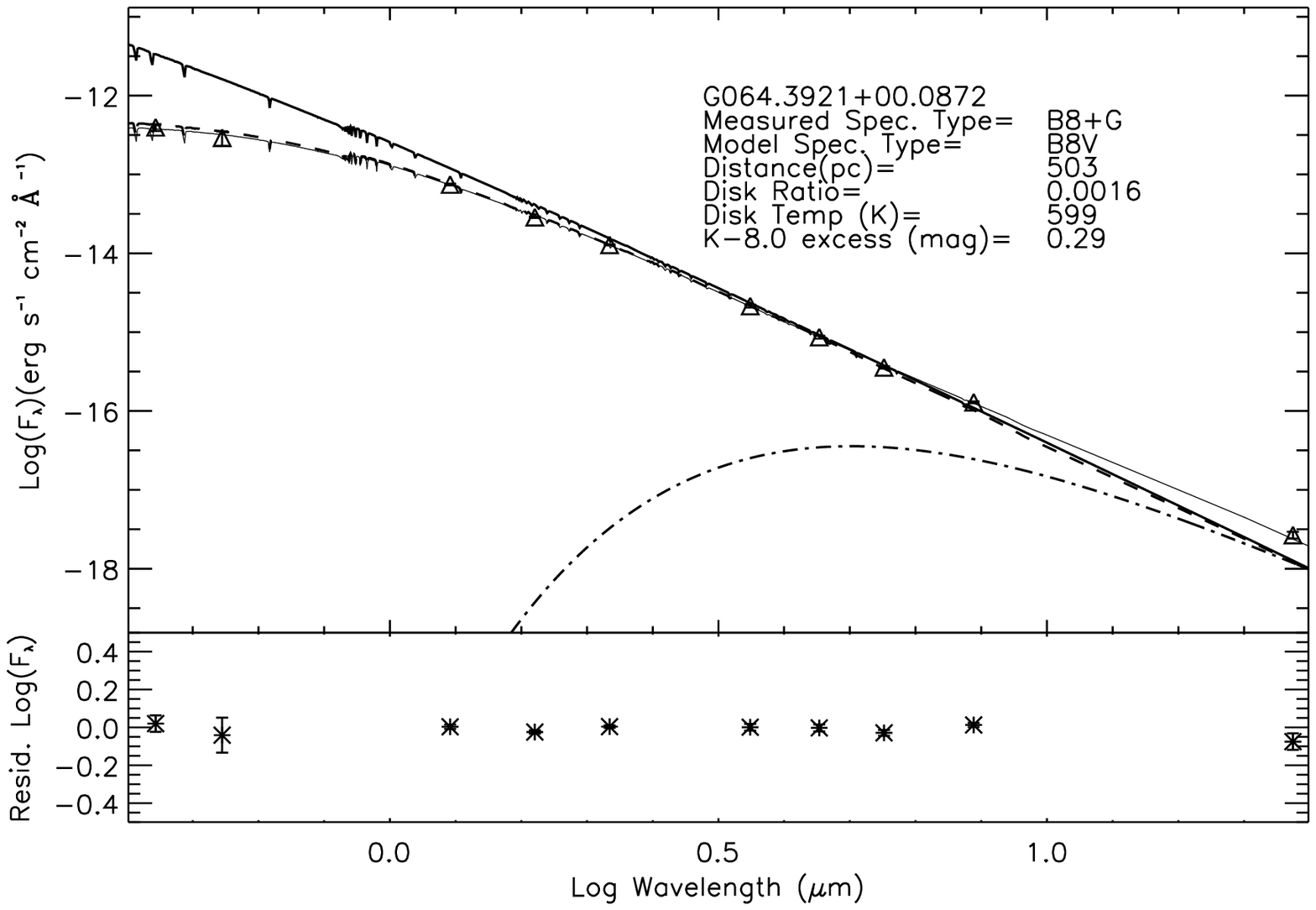}
}
\hbox{
    \includegraphics[width=3.0in]{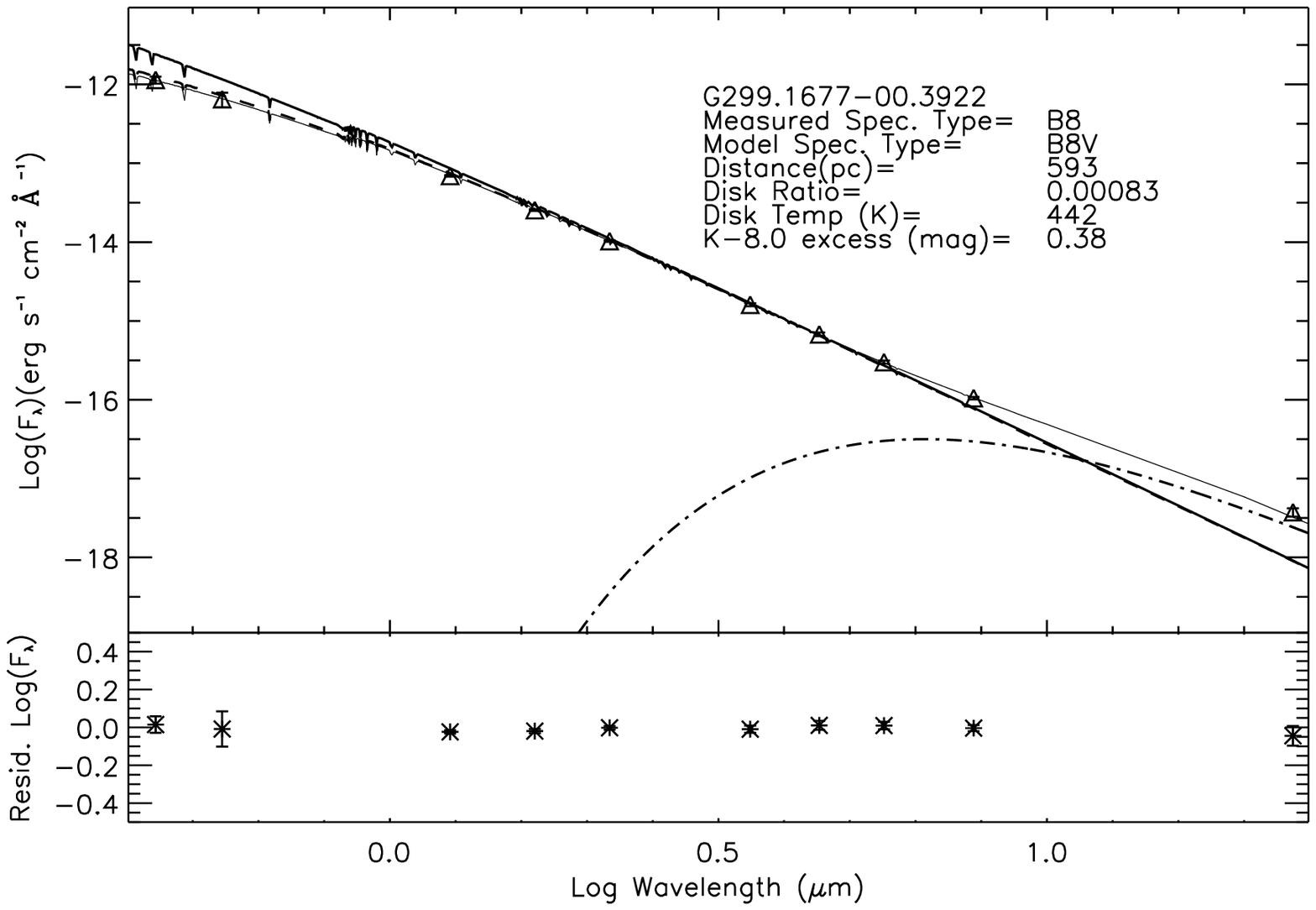}
    \includegraphics[width=3.0in]{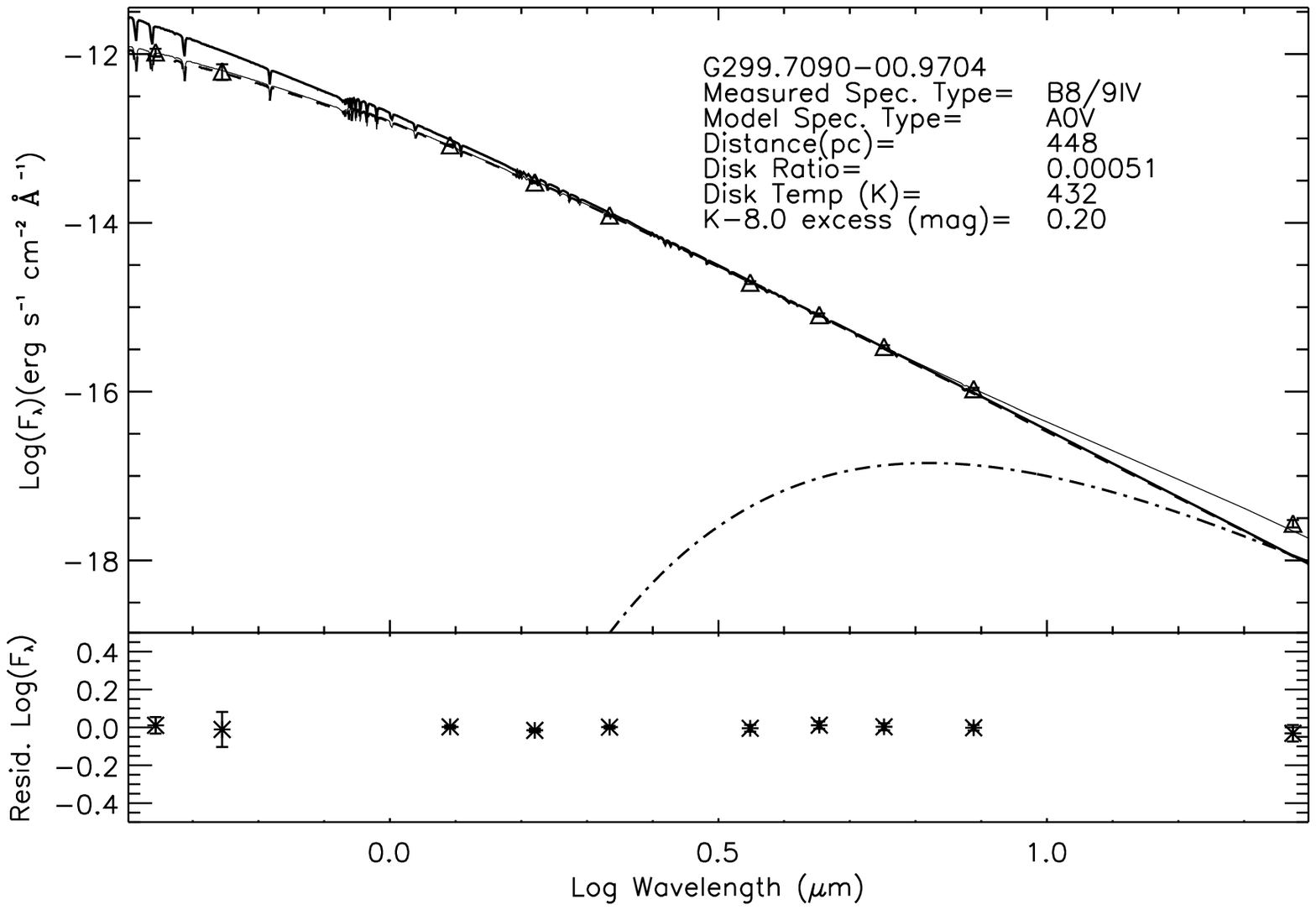}
}
 \caption{(upper left) An SED of the B9V star G063.5770--00.3387. The
model is defecient at [24].  (upper right) An SED of the B8+G star
G064.3921+00.0872. The measurements are well-fit by a single
temperature blackbody.  (lower left) An SED of G299.1677--00.3922 a
B8. This source is well modeled by a single temperature blackbody.
(lower right) An SED of G299.7090--00.9704 a B8/9 iV star. The model
adequately fits the observed fluxes.}
    \label{disks3}
\end{figure}

\begin{figure}
\hbox{
    \includegraphics[width=3.0in]{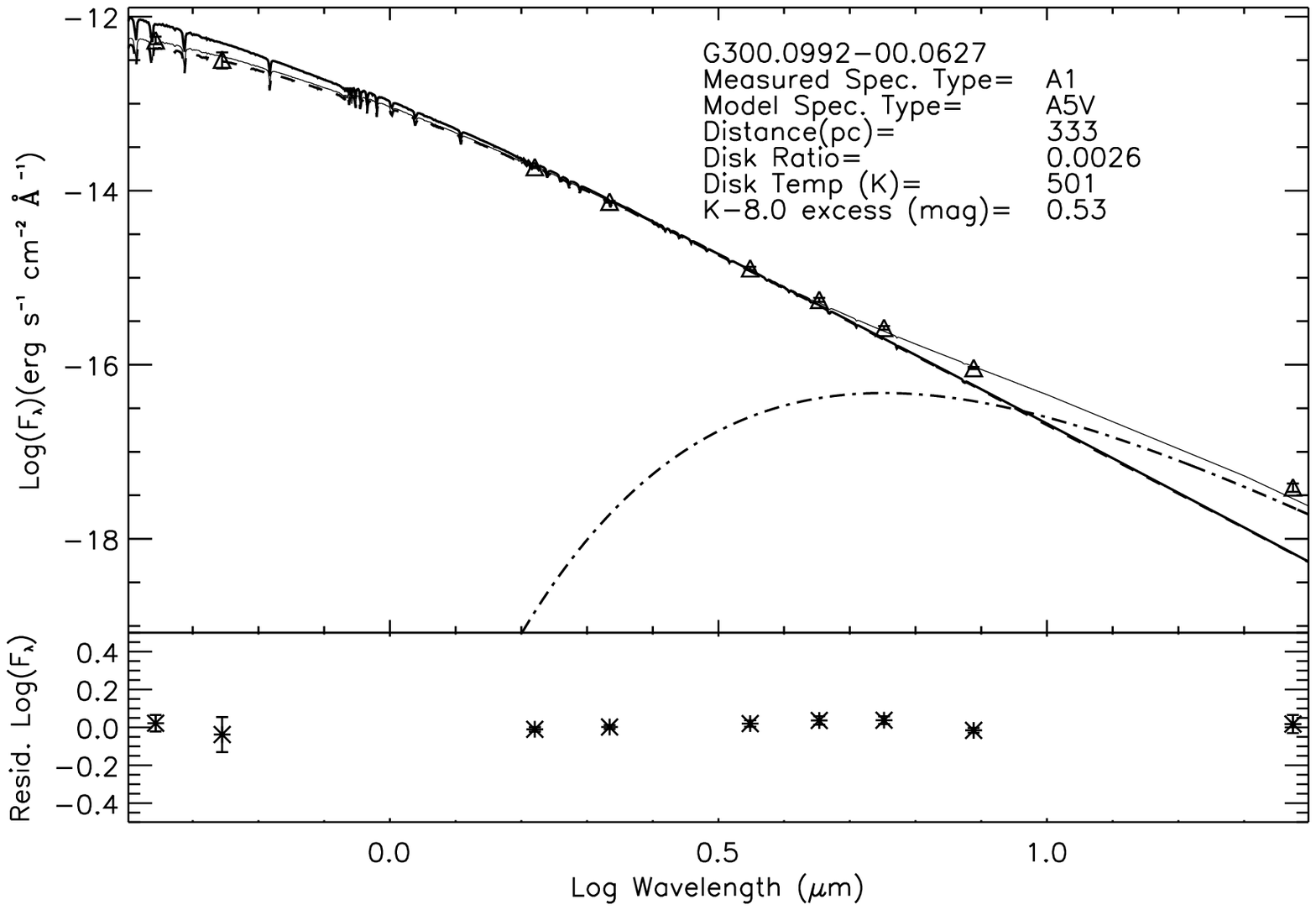}
    \includegraphics[width=3.0in]{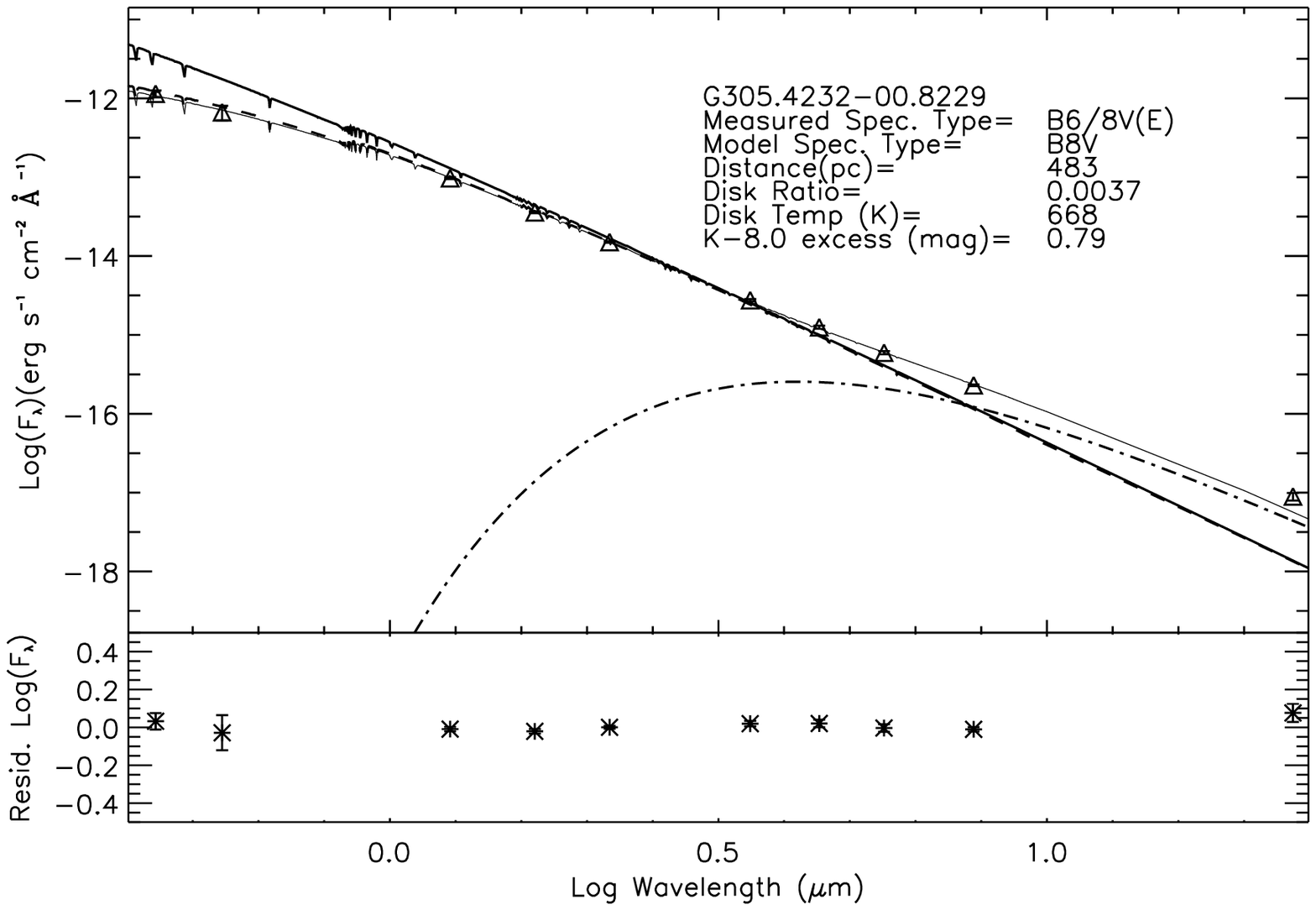}
}
\hbox{
    \includegraphics[width=3.0in]{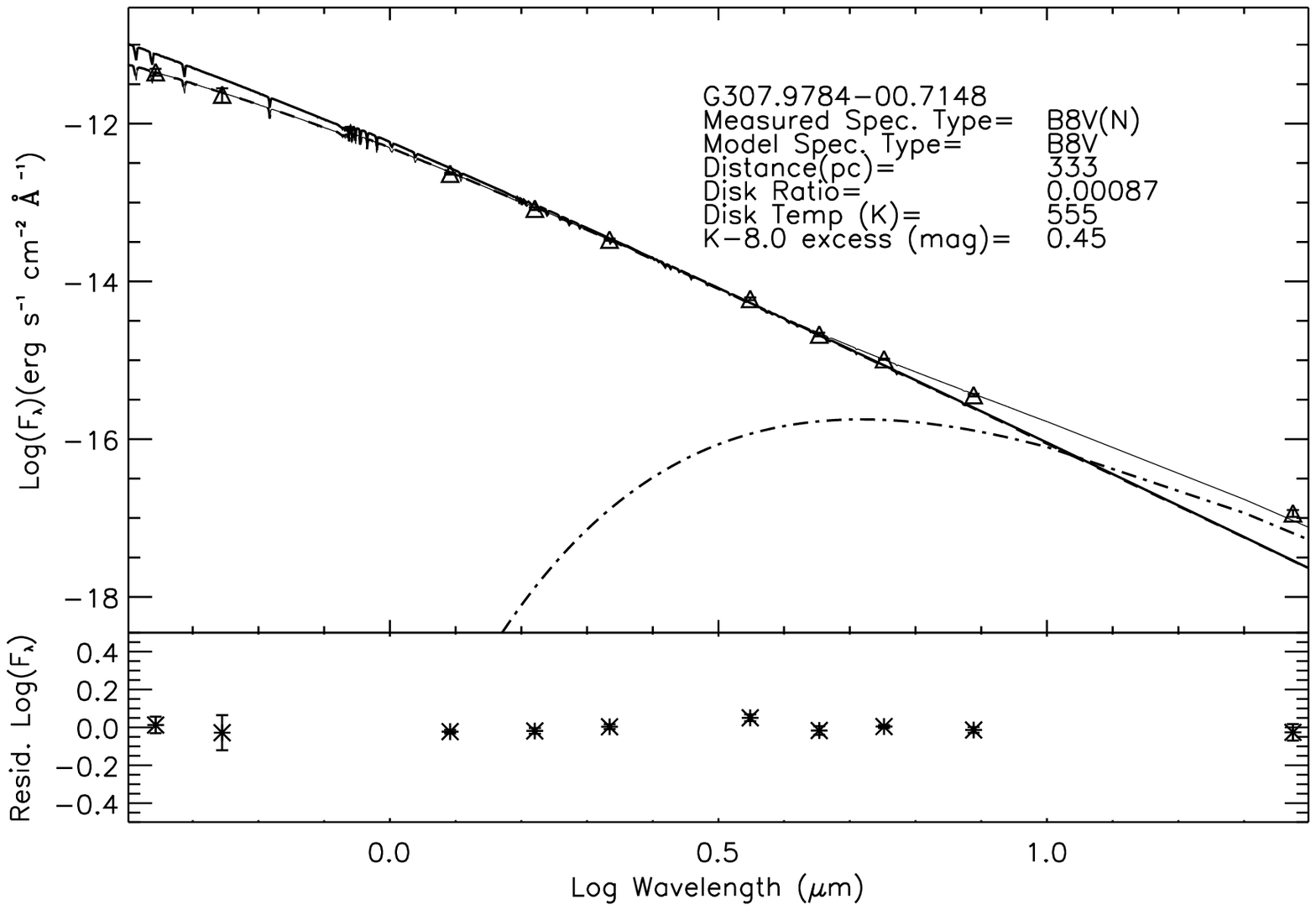}
    \includegraphics[width=3.0in]{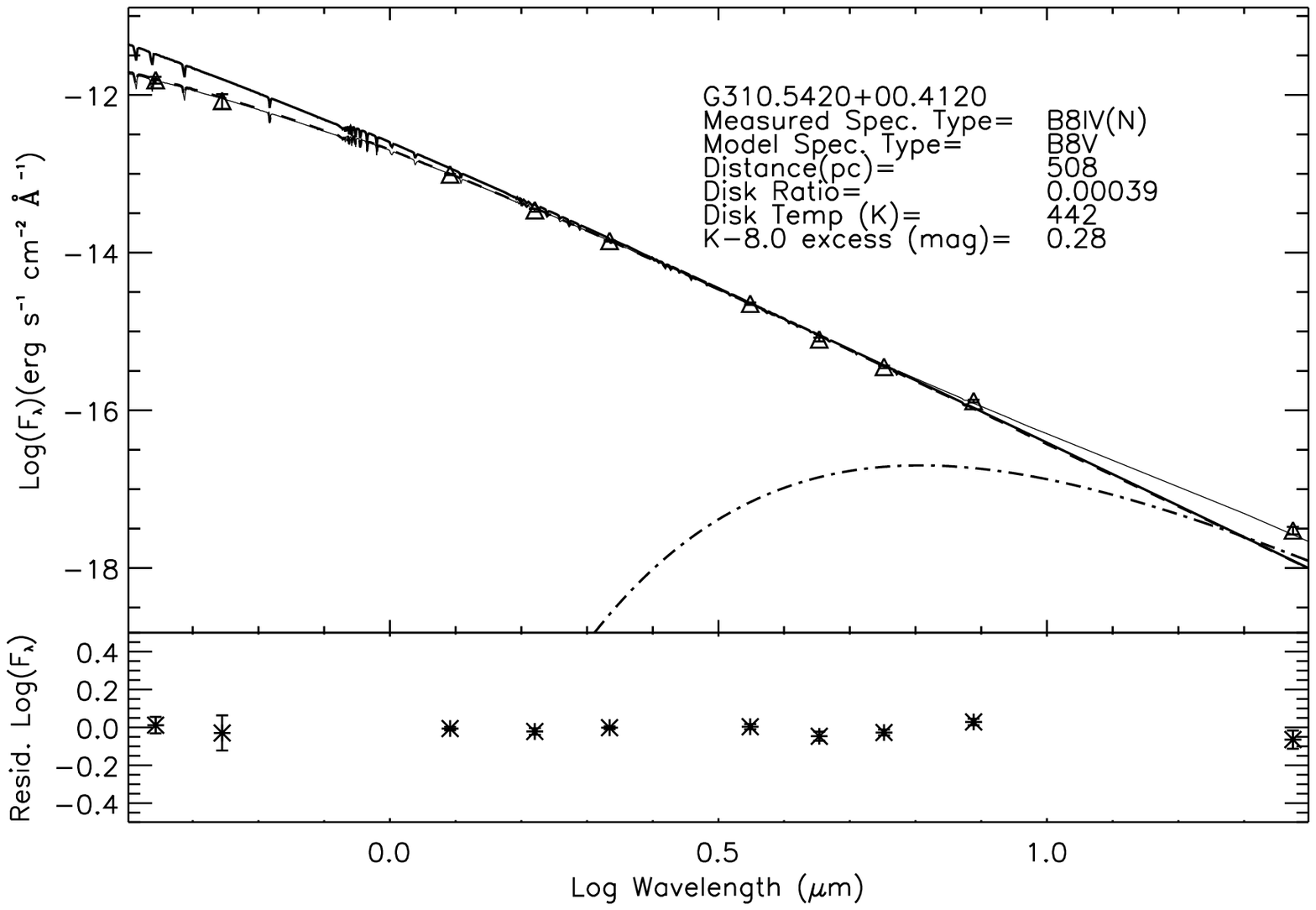}
}
 \caption{(upper left) An SED of the A1 star G300.0992--00.0627. The
model fits the observed measurements. (upper right) An SED of the B6/8
V(E) star G305.4232--00.8229. The model is defecient for the [24]
measurement.  (lower left) An SED of G307.9784--00.7148 an
B8V(N). This star is well modeled with an additional
single-temperature blackbody. (lower right) The SED of
G310.5420+00.4120 a B8IV(N) star. The model for this star fits the
observered measurements.}
    \label{disks4}
\end{figure}

\begin{figure}
\hbox{
    \includegraphics[width=3.0in]{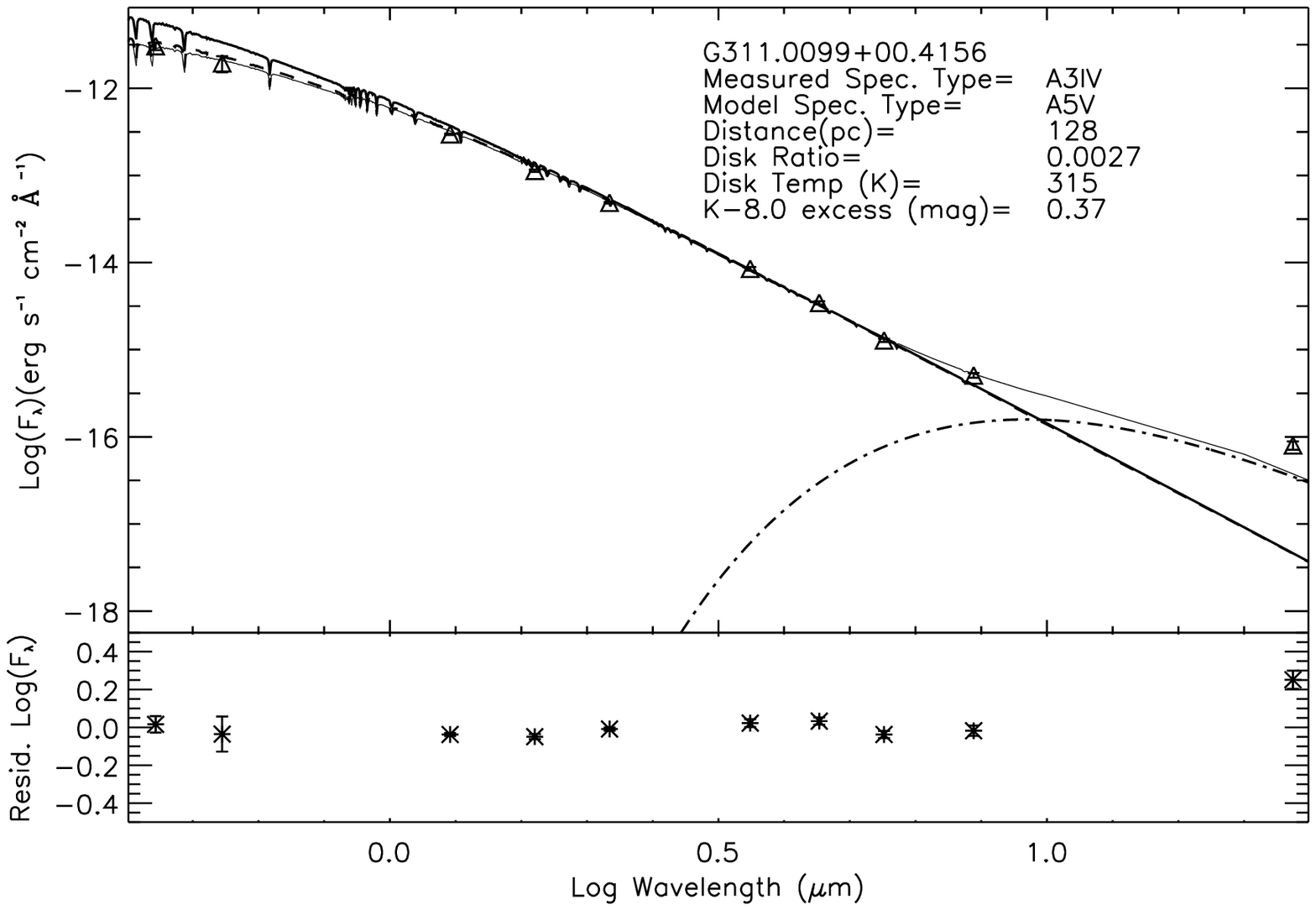}
    \includegraphics[width=3.0in]{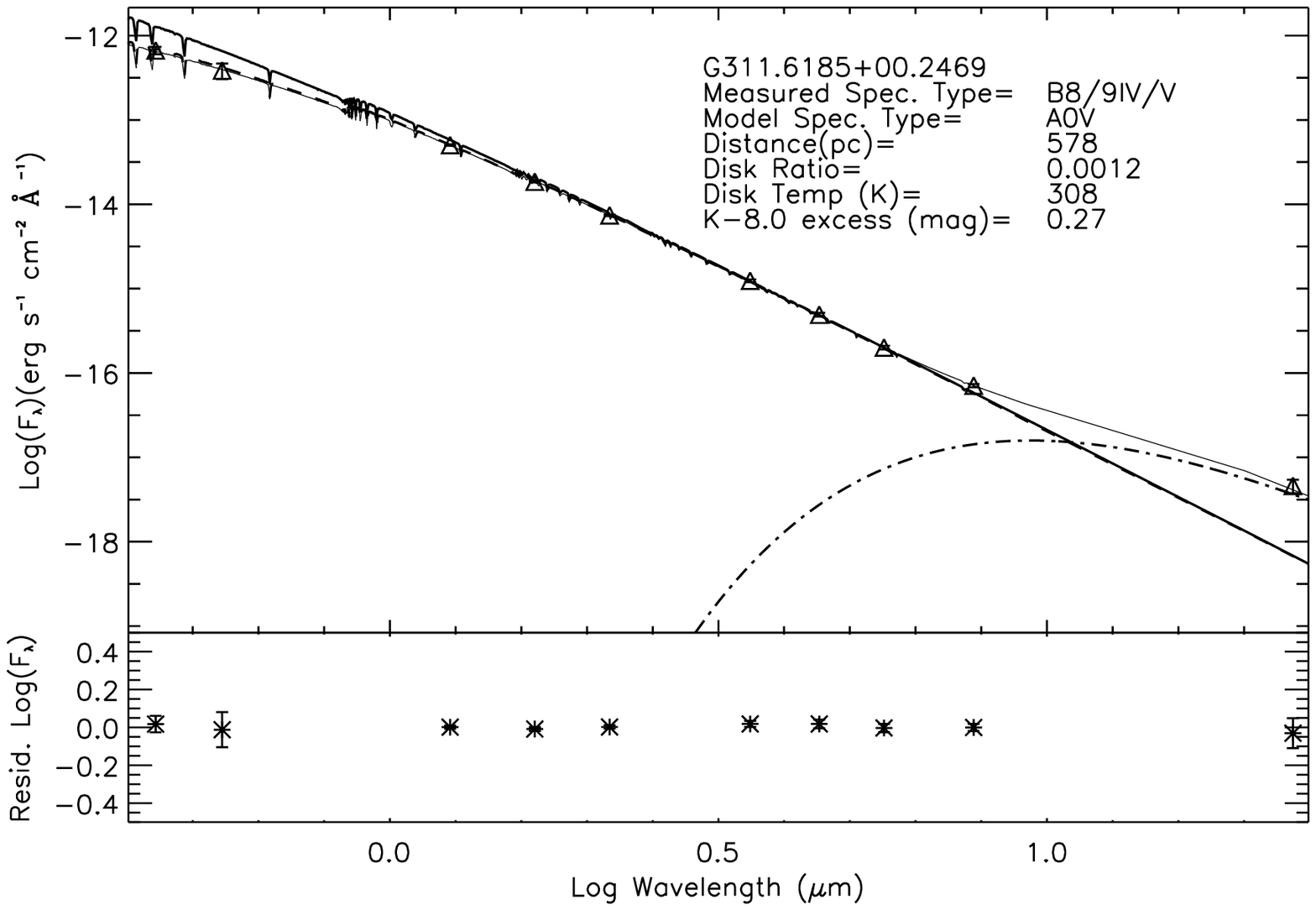}
}
\hbox{
    \includegraphics[width=3.0in]{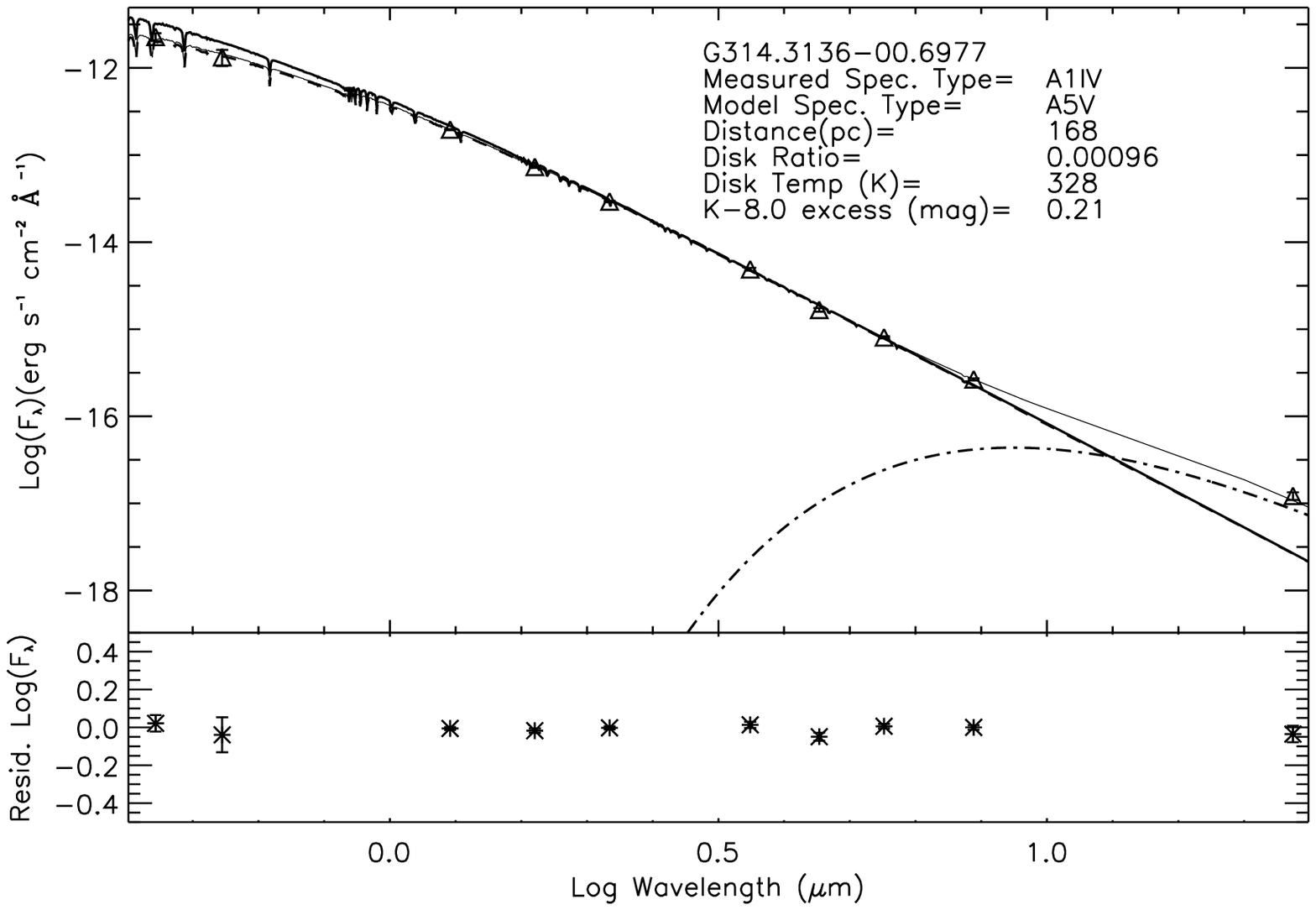}
    \includegraphics[width=3.0in]{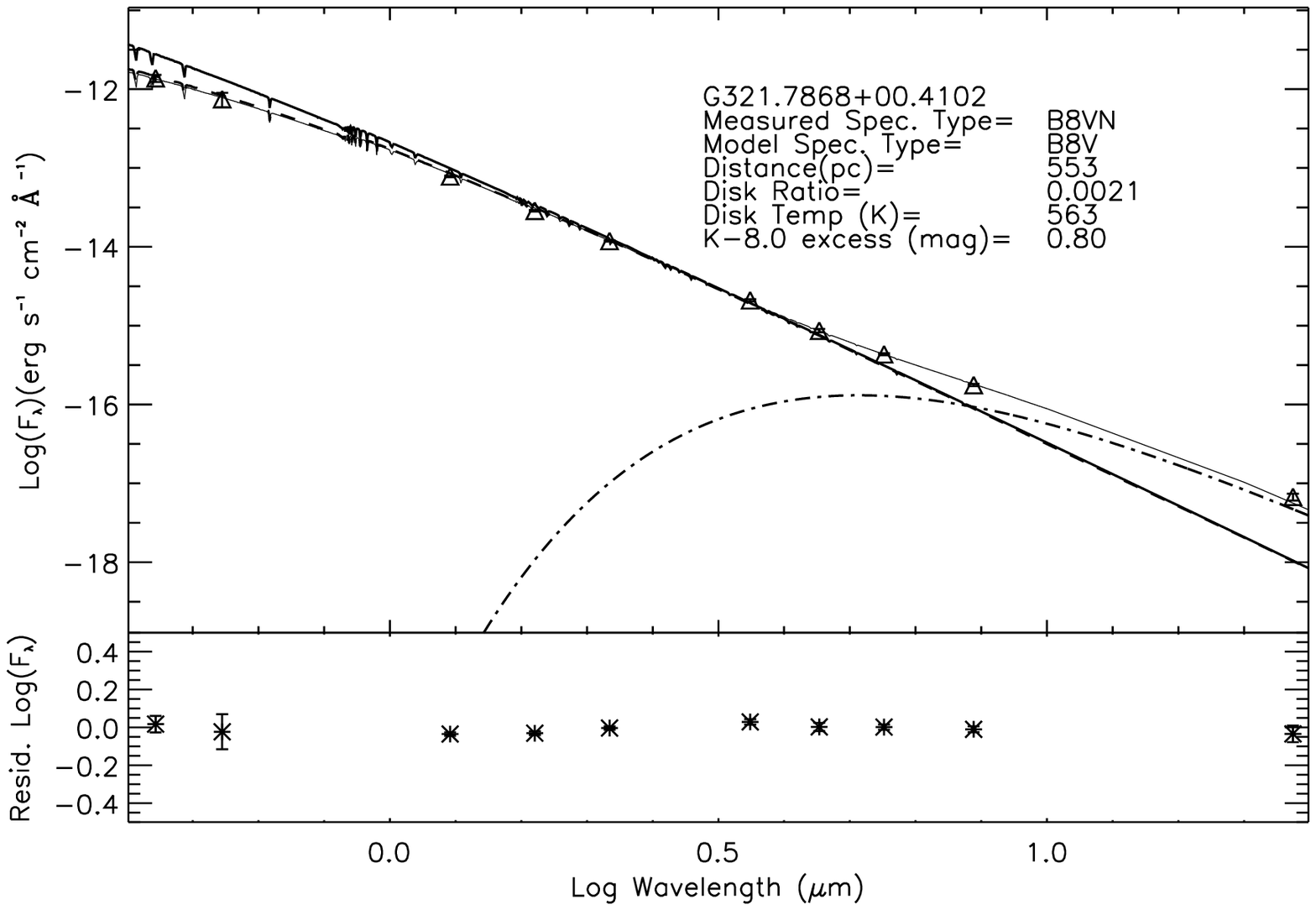}
}
 \caption{(upper left) An SED of the A3IV star G311.0099+00.4156. The
model is deficient at [24]. (upper right) An SED of the B8/9IV/V star
G311.6185+00.2469. This star is well modeled with an additional single
temperature blackbody.  (lower left) An SED of G314.3136--00.6977 an
A1IV. This star is well modeled with an additional single-temperature
blackbody. (lower right) The SED of G321.7868+00.4102 a B8VN
star. The star is well modeled at [24].}
    \label{disks5}
\end{figure}

Figures~\ref{disks}--\ref{disks5} show the SEDs of mid-IR excess
sources in GLIMPSE that are confirmed in MIPSGAL.  The Kurucz model
atmosphere is shown by the solid curve, the model atmosphere with
applied extinction is the dashed curve, the dash-dot curve is the
blackbody representing the disk, and the thin curve is the sum of the
extincted model atmosphere and the applied blackbody. The plot below
each SED shows residuals from the best-fit model.  In most cases, the
single-temperature blackbody component provides a good fit to the
data.  The $\sim$4 sources that are not well modeled by a
single-temperature blackbody will be discussed in Sec. 5.3. In most
cases, a plausible explanation for the poor fits is the existence of
multiple temperature (i.e., cooler) components within the disks that
contribute to the excesses at wavelengths $\lambda \geq 8$ $\mu$m.

\begin{figure}
    \plotone{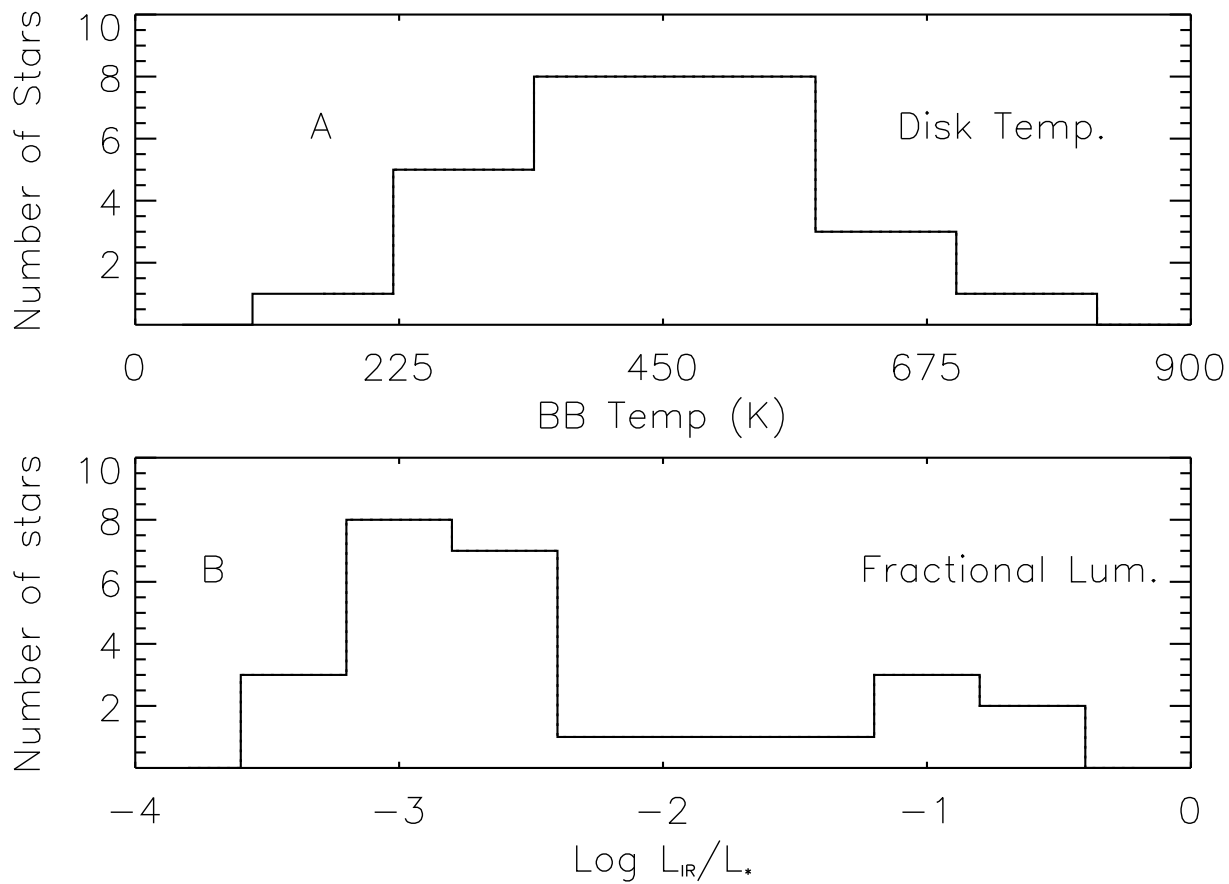}
    \caption{Histograms of disk temperature (A) and fractional
    infrared luminosity (B).  The distribution of disk temperatures is
    broadly peaked at $\sim450$K but temperatures are found between
    $\sim190-800$ K. The majority of stars have fractional infrared
    luminosities between 10$^{- 3}$--10$^{- 2}$.  However there is
    still a sizable population with fractional infrared luminosities
    greater than 10$^{- 2}$ including the the probable Class II
    protostars.}
	 \label{tempratiohist}
\end{figure}

Figure~\ref{tempratiohist} shows a summary histogram of all model disk
temperatures and fractional infrared luminosities. Disk temperatures
are broadly peaked at $\sim450$ K but span a range from
$\sim190-800$K. Assuming blackbody radiation an estimation of the
inner radius for circumstellar material can be made, and is found to
range from 0.4-27 AU for these sources. The majority of the sources
have inner radii of 1-6 AU for the late B and early A stars of our
sample. The distribution of fractional infrared luminosities is
bimodal, having peaks near $10^{-3}$ and $10^{-1}$.  The majority of
stars have fractional infrared luminosities between 10$^{-
3}$--10$^{-2}$. Figure~\ref{tempratiohist} clearly shows we have two
different populations of mid-IR sources, with 1 or 2 sources that
bridge this gap.

\subsection{Protostellar Disks or Debris Disks?}

Lagrange \etal\ (2000) suggested a definition of debris disks using
four criteria. These criteria require that the fractional infrared
luminosity is small, i.e., $\frac{L_{IR}}{L_{*}}$ $\leq$ 10$^{-3}$, the
mass of the gas and dust must be below 10$^{-2}$ \mo , the dust mass
must be significantly greater than the gas mass, and the grain
destruction time must be much less than the stellar age. Our limited
data only allow comparisons based on the fractional infrared
luminosities.  Artymowicz (1996) gives a more liberal criterion,
allowing maximum fractional infrared luminosities of
$\frac{L_{IR}}{L_{*}}$ $<$ 10$^{-2}$ for debris disks.  Protostars, on
the other hand, have $\frac{L_{IR}}{L_{*}}$ $\geq$ 0.1 (e.g., Baud
\etal\ 1984).

Six stars exhibit fractional infrared luminosities less than or equal
to 10$^{-3}$ and have fractional infrared luminosities consistent with
being debris disks.  Six stars have $\frac{L_{IR}}{L_{*}}$ $>$
5$\times$10$^{-2}$ and are almost certainly Class II protostars.  The
majority of our sample (14 stars) have $10^{-3} < \frac{L_{IR}}{L_{*}}
< 10^{-2}$ and are consistent either with transition disks in
late-stage protostars or massive debris disks.

Optical spectra can be used to further discriminate between the
pre-main-sequence and debris disk nature of the IR excess.
Pre-main-sequence stars exhibit Balmer emission produced by accretion from
their circumstellar disks, whereas debris disks should not exhibit
accretion characteristics.  Three of the 6 stars with
$\frac{L_{IR}}{L_{*}}$ $\leq$ 10$^{-3}$ exhibit Balmer emission. If the
emission line classification in the SIMBAD database for these
sources is reliable, then this signature of ongoing accretion would
imply a pre-main-sequence circumstellar disk despite the small
fractional infrared luminosity. Four of the 6 stars that have
$\frac{L_{IR}}{L_{*}}$ $\geq$ 0.05 also exhibit Balmer emission,
supporting the existence of pre-main-sequence circumstellar disks as
the origins of their excesses. Of the remaining 14 stars with $10^{-3}
< \frac{L_{IR}}{L_{*}} < 10^{-2}$, 6 have Balmer emission suggesting
ongoing accretion from a circumstellar disk.  Given the heterogeneous
nature of the spectral information compiled in SIMBAD, stars with very
weak Balmer emission (e.g., weak-line T-Tauri stars or more massive
equivalents) may not have been classified as such in the database.
Therefore, we caution that the fraction of emission line sources in
our sample is best regarded as a lower limit.  Only 4 of the 21
sources without 24 $\mu$m detections exhibit Balmer emission.
Approximately half of the sources that were modeled have
Balmer emission, regardless of fractional infrared luminosity. In
summary, fully one-third of sources identified as having mid-IR
excesses also exhibit Balmer emission.

\begin{figure}
    \plotone{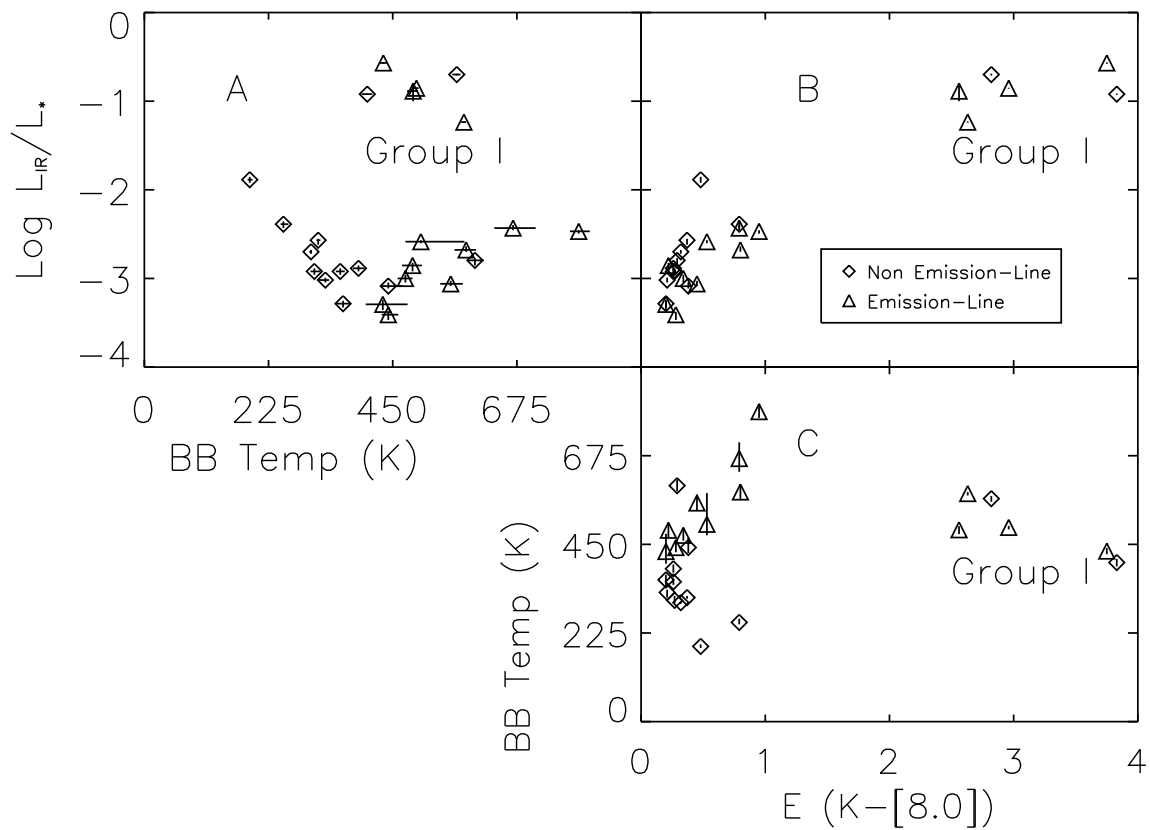}
    \caption{Plots of $E(K-[8.0])$, fractional infrared luminosity,
    and disk temperature for mid-IR excess objects later than
    B8. Diamonds are GLIMPSE and MSX stars without emission
    lines. Triangles are emission-line GLIMPSE and MSX stars.
    Emission-line and non emission-line sources show a strong
    correlation between $\frac{L_{IR}}{L_{*}}$ and $E(K-[8.0]$). The
    correlation arises because both measurements are an indicator of
    extra-photospheric excess.}
    \label{3in1}
\end{figure}

Figure~\ref{3in1} is an inter-comparison of $E(K-[8.0])$, fractional
infrared luminosity, and disk temperature for the 26 modeled sources.
Diamonds are GLIMPSE and MSX stars lacking evidence of emission
lines. Triangles are emission-line GLIMPSE and MSX stars.  The upper
left panel (A) of Figure~\ref{3in1} plots the modeled
$\frac{L_{IR}}{L_{*}}$ versus the modeled blackbody temperatures, the
upper right panel (B) shows $\frac{L_{IR}}{L_{*}}$ versus
$E(K-[8.0])$, and the lower right panel (C) shows temperature versus
$E(K-[8.0])$.  In all three panels there is a distinct grouping of six
objects which have high fractional infrared luminosities $>$0.05,
large $E(K-[8.0])>2$, and temperatures $>$400~K. This grouping is
noted in Table 7 in column 15 by ``g'' and in Figure~\ref{3in1} as
``Group I''.  All 6 stars in this group are MSX sources, have near-IR
excesses ($H-K>$0.3), and 4 of them exhibit emission lines.  There are
four known T-Tauri and HAeBe stars in this group.

In panel A, emission-line stars are found exclusively at temperatures
$>$430~K, and, with three exceptions, non emission-line stars have
temperatures $<$430~K.  Emission-line sources also show a correlation
between $\frac{L_{IR}}{L_{*}}$ and $T$ when Group I sources are
excluded.  Such a correlation might arise physically if the hottest
disks are also youngest and most luminous with the disks closest to
the parent stars. The observed correlation between $T$ and
$\frac{L_{IR}}{L_{*}}$ is consistent with an evolutionary progression
as the inner disk is cleared of materials (e.g., Strom \etal\
1989). 

Panel B shows a correlation between $\frac{L_{IR}}{L_{*}}$ and
$E(K-[8.0])$.  This correlation is a consequence of the fact that both
quantities are measures of the extraphotospheric excess, the former
encompassing all wavelengths and the latter only the [8.0]
measurement. That is, objects with large [8.0] excesses must also have
large fractional excesses overall.  The distribution of $E(K-[8.0])$
is bimodal with Group I objects, by definition, having $E(K-[8.0])>2$
and the other 20 sources having $E(K-[8.0])<1$.  The absence of
objects with intermediate $E(K-[8.0])$ is likely to be a consequence
of selection criteria, stemming from the relative sensitivities and
spatial coverage of the MSX and GLIMPSE surveys.  The limited
sensitivity of MSX necessarily means that only nearby objects with
large 8 $\mu$m excesses and detections at wavelengths $\lambda>8$
$\mu$m are included. In fact, Group I sources lie at an average
distance of $\sim100$ pc, while the other sources are more distant.
GLIMPSE, although more sensitive than the MSX Galactic Plane Survey,
did not include many of the nearby star-forming regions where young
stars with large $E(K-[8.0])$ would preferentially reside.

\begin{figure}
    \plotone{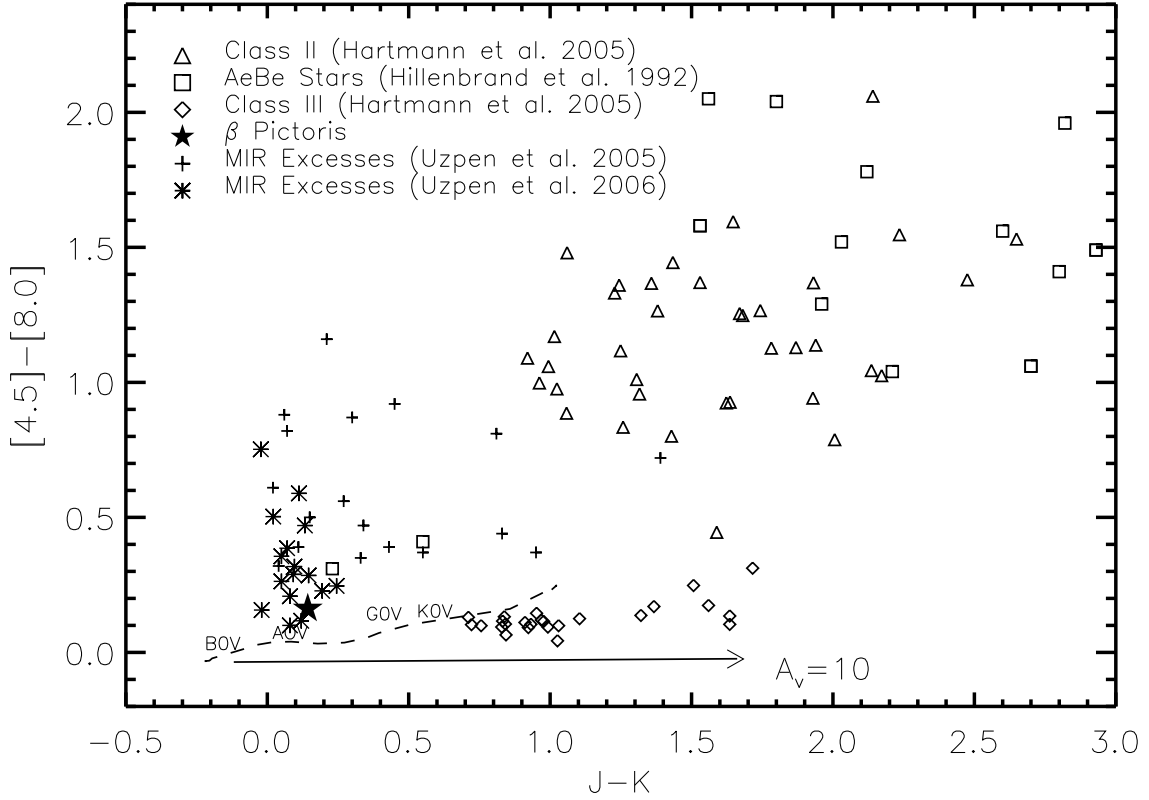}
    \caption{Infrared color-color diagram, \textit{SST} 4.5 $\mu$m
  - 8.0 $\mu$m vs. 2MASS J-K showing the loci of Class II protostars
  (triangles), Class III protostars (diamonds), AeBe stars (squares),
  our sample of mid-IR excess sources from Uzpen \etal \ (2005)
  (pluses) and stars with IRAC measurements and [24] detections from this sample
  (asterisks). There is a clear gap between Class II protostars, Class
  III protostars and the main-sequence stars. Our samples from Uzpen
  \etal \ (2005) and this paper partially bridge this gap. A filled
  star marks the position of $\beta$ Pictoris, a prototype debris-disk
  object. The solid line shows the reddening vector for 10
  magnitudes of visual extinction. The dashed-line marks the
  main-sequence.}
    \label{iraccolor}
\end{figure}

Figure~\ref{iraccolor} is a color-color diagram, \textit{SST} [4.5] -
[8.0] vs. 2MASS J-K showing the loci of Class II protostars, Class III
protostars, the sample of mid-IR excess sources from Uzpen \etal \
(2005) and stars with IRAC measurements and [24] detections from this
sample.  A filled star marks the position of $\beta$ Pictoris, a
prototype debris-disk object. The reddening vector is shown for 10
magnitudes of visual extinction. The main-sequence is labeled by the
dashed line.  We adopt the MSX A band measurement in lieu of an [8.0]
measurement for the Class II sources from Hillenbrand \etal \
(1992). For a selected sample of stars we compared GLIMPSE [8.0] and
MSX A measurements and found that they agreed within 0.1
magnitudes. This error is comparable to the uncertainty in estimating
the [8.0] flux by interpolation between K and MSX band A.  Two
Hillenbrand \etal \ (1992) stars fall in a similar region of color
space as our mid-IR excess sources.  These are B0--B3 stars from their
Group III objects which are classified as young disk-less early B
stars with excesses attributable to free-free emission.  The Class II
and III sources from Hartmann \etal \ (2005) are only late-type
stars. There is a clear gap between Class II protostars, Class III
protostars and main-sequence stars. Our samples from this paper and
Uzpen \etal \ (2005) appear to partially bridge this gap. The location
of our sample in color-space is consistent with the identification of
these sources as either late-stage protostars or massive debris disks.

The stars known to have emission are more likely to be
pre-main-sequence rather than main-sequence sources and may be in a
short-lived transitional state between primordial and debris
disks. Further analysis of both emission-line and non emission-line
sources is necessary to understand their evolutionary status and
relationship to one another.  Mid-IR spectral analysis of the stars in
Table 7 would reveal whether the 10 $\mu$m silicate feature is present
and would be a useful tool in the characterization of mid-IR excesses
around main-sequence stars. The absorption (embedded protostar),
emission (pre-main-sequence), or lack of the 10 $\mu$m silicate
feature (debris disk) is loosely related to the evolutionary
progression of a circumstellar disk (Kessler-Silacci \etal \
2005). However, the debris disks studied by Kessler-Silacci \etal \
(2005) contained debris disks that only exhibited far-IR
excesses. $\beta$ Pictoris, a mid-IR debris disk system, does have a
10 $\mu$m silicate emission feature (Weinberger \etal \ 2003). A
useful comparison would be to determine whether this sample of mid-IR
excess systems also exhibits a 10 $\mu$m silicate emission
feature. Detection of gas signatures would also reveal whether these
sources are late-stage transitional disks or debris
disks. Identification of PAH features in spectra would reveal insight
into the nature of the excess, indicating possible circumstellar or
interstellar origins. Scattered light observations and resolution of
the putative disks would rule out the possibility that the excesses
originate in interstellar rather than circumstellar grains (Kalas
\etal \ 2002).  Mo\'{o}r \etal \ (2006) studied a sample of the
highest fractional infrared luminosity debris disks and found that
they are young, having ages less than 100 Myr.  Age measurements of
our sample would allow these objects to be placed within an
evolutionary sequence.

\subsection{Individual Sources}

G007.2369+01.4894 (HD 163296) is a well studied pre-main-sequence star
present in over 200 papers and we place it in Group I.  It is
classified as an A1V star with a Hipparcos distance of 122$^{+ 16}_{-
13}$.  This star has been studied in mid-IR spectroscopy papers such
as van Boekel \etal \ (2005) and Kessler-Silacci \etal \ (2005). van
Boekel \etal \ (2005) derive a continuum temperature of 461 K which is
near our temperature of 493$^{+ 4}_{- 5}$ K.  In comparison, Oudmaijer
\etal \ (1992) derived a color temperature of 220 K for this source.
The presence of crystalline enstatite indicates dust processing in the
circumstellar disk. Enstatite is a processed mineral and is made in a
circumstellar disk via silicate reactions. We determined
$\frac{L_{IR}}{L_{*}}$ at 0.14$^{+ 0.002}_{- 0.002}$ which is
consistent with its pre-main-sequence classification.

G008.3752-03.6697 (HD 167905) is an MSX source with Group I
designation which is also listed in Oudmaijer \etal \ (1992).  This
F3V star has an \textit{IRAS} excess with ``good'' quality
measurements at 12 and 25 $\mu$m and ``low'' quality at 60 and 100
$\mu$m. Oudmaijer \etal \ found a temperature of 480 K for the color
excess. We find a temperature for the excess to be 566$^{+ 7}_{- 8}$ K
when utilizing the MSX data. We measured $\frac{L_{IR}}{L_{*}}$ for
this star at 0.20$^{+ 0.004}_{- 0.005}$. Given the warmer temperature
and larger fractional infrared luminosity, this star is likely
pre-main-sequence.

G047.3677+00.6199 (HD 180398--Figure~\ref{disks2} upper left) is
classified as a B8Ve star and found in the GLIMPSE catalog. The model
fit for this source at [24] is below the measured source.  In this
source, the magnitude of the excess increases with wavelength.  An
additional cooler disk component may be necessary to fully fit the
observed fluxes at all wavelengths.

G063.5770--00.3387 (HD339086--Figure~\ref{disks3} upper left) is
a B9V star in the GLIMPSE sample. This source is located in a
small region of extended emission. The model for this source at
[24] is below the measurement. A possible explanation is the
addition of flux from the extended source to the measurement. 
An alternate explanation is that an additional cooler disk component
may be necessary to fit the observed flux at [24].

G229.4514+01.0145 (HD 58647) is listed as an emission-line star in the
SIMBAD database and we place it in Group I. This star is classified as
B9V and its Hipparcos distance is 277$^{+ 78}_{- 60}$ pc. This MSX
source is also an \textit{IRAS} source and has ``good'' quality 12,
25, and 60 $\mu$m measurements and ``low'' quality 100 $\mu$m
measurements. Oudmaijer \etal \ measured a color temperature of 650 K
for this star, while we measured a blackbody temperature of 579$^{+
4}_{- 5}$ K. This source has also been investigated for CO emission in
Dent \etal \ (2005), who were only able to place upper limits on the
CO intensity. The fractional infrared luminosity is 0.058$^{+
0.001}_{- 0.001}$, smaller than typical for a pre-main-sequence star
but this star has a near-IR excess.  This star may be at the end of
its pre-main-sequence phase.

G257.6236+00.2459 (HD 72106) is listed as an A0IV star but is also listed
as a multiple or double star in the SIMBAD database.  Hipparcos
parallax places this star at a distance of 288$^{+ 202}_{- 84}$
pc. This MSX source and Group I member has appeared in numerous
\textit{IRAS} far-IR excess searches (e.g., Oudmaijer \etal \
1992). This source has good quality \textit{IRAS} 12, 25, and 60
$\mu$m measurements with a low quality 100 $\mu$m measurement and
Oudmaijer \etal \ (1992) measured the color temperature to be 250 K
while we measured the temperature at 404$^{+ 9}_{- 10}$ K. Given the
additional longer wavelength data used by Oudmaijer \etal \ (1992), it
is not unexpected that they derived a lower temperature. We measured
$\frac{L_{IR}}{L_{*}}$=0.12$^{+ 0.002}_{- 0.002}$ for this source.
The southern component of the binary system is identified as an HAeBe
source (Vieira \etal \ 2003). Sch\"{u}tz \etal \ (2005) note that the
mid-IR spectrum lacks small amorphous silicates, has large amorphous
silicates, and contains crystalline silicates forsterite and
enstatite.  The later species are evidence of grain processing in the
circumstellar disk.  They also note that the mid-IR spectrum is
similar to solar system comets and this object is interesting in the
context of planet formation.

G294.1242+01.4656 (HD 101412), a B9/A0 V star, is listed as an
emission-line star in the SIMBAD database. This MSX source and Group I
member has appeared in numerous \textit{IRAS} papers searching for
far-IR excesses (e.g., Oudmaijer \etal \ 1992). This source is also
listed as a possible HAeBe star. The star has ``good'' quality 12 and
25 $\mu$m data and ``intermediate'' quality 60 and 100 $\mu$m
data. The color temperature derived for this source in Oudmaijer \etal
\ (1992) is only 170 K. Given the longer wavelength data used by
Oudmaijer \etal \ (1992), it is not unexpected that we derived a
higher temperature of 433$^{+ 8}_{- 8}$ K using only the mid-IR data
available from MSX. This source has also been investigated using
mid-IR spectroscopy in van Boekel \etal \ (2003, 2005) who measure the
temperature of the mid-IR continuum to be 425 K, consistent with our
temperature. This source also exhibits a large mass fraction of large
grains, and the majority of the crystalline grain component is
enstatite. In comparing the mid-IR spectrum to HD 163296, HD 101412
has a larger fraction of grains in large grains, and more crystalline
grains in enstatite form. We determined $\frac{L_{IR}}{L_{*}}$ at
0.27$^{+ 0.004}_{- 0.006}$ which is consistent with the
pre-main-sequence classification.

G305.4232--00.8229 (HD 114757--Figure~\ref{disks4} upper right) is a
B6/8V(E) star. This source is inadequately fit at [24].  This
deficiency at longer wavelength may imply the need for an additional
cool disk component.

G311.0099+00.4156 (HD 121808--Figure~\ref{disks5} upper left) is a
GLIMPSE star classified as A3IV. The model is deficient at [24]. This
deficiency may imply the need for an additional cool disk component in
the SED.  The star is also in a region of bright extended background
emission.  The model deficiency at [24] could also be explained by
excess background emission at [24].

G347.3777+04.2010 (HD 152404) is a T Tauri-Type star in the Upper
Centaurus Lupus (Chen \etal \ 2005) and in Group I.  It is classified
as an F5V star with a Hipparcos distance of 145$^{+ 38}_{- 25}$ pc.
In addition to having ``good'' quality 12, 25, and 60 $\mu$m
\textit{IRAS} data, this source also has \textit{SST} 24 and 70 $\mu$m
measurements. Acke \etal \ (2004) measured this MSX source to have the
9.7 $\mu$m silicate feature to be in emission using \textit{ISO}.  We
determined $\frac{L_{IR}}{L_{*}}$ at 0.13$^{+ 0.02}_{- 0.03}$ which
is consistent with its pre-main-sequence classification.

G040.9995--00.0423 (HD 177904) is an MSX source that exhibits an
excess at band A. This source was also detected in GLIMPSE,
G040.9998--00.0429, and MIPSGAL.  This source does not exhibit an
excess at [8.0] or [24]. This source is, therefore, a probable false
excess in the A band.

G042.9876+00.4201 (HD 178479) is a GLIMPSE source classified as a B8V
star. This star has photospheric measurements at [24] and the cause of
the [8.0] excess may be due to PAH contamination.

G054.5163+02.4470 (HD 182293) is an MSX source exhibiting
an excess at both A and C bands. The 2MASS K measurement
is flagged ''U'' and provides only an upper limit. In modeling
the excess for this star, we find that, if all 3 2MASS measurements
are included, the fractional infrared luminosity exceeded
unity in all models. However, if only the K measurement
was included in the disk models, the fractional infrared
excess was $\sim10^{-1}$. Given the poor reliability
of the IR measurements for this source we do not model it.

G056.7848+00.8671 (HD 334593) is an MSX source also found in the
GLIMPSE catalog as G056.7850+00.8668. This source is detected strongly
at [24] and does not exhibit an excess based on the GLIMPSE
measurement SED . There is a nearby object seen in the [24] image, 6
\arcsec \ separation which is probably the source of the additional
flux in the larger MSX beam.

G201.9029--01.9739 (HD 44783) is a B8VN source listed as a Be
star (e.g, Neiner \etal\ 2005). The excess for this source
is attributed to a circumstellar shell rather than
circumstellar disk.

G322.6930+00.0572 (HD 136591) is the only late-type star in GLIMPSE
that exhibits an $E(K-[8.0]$). This star does not exhibit a [24]
excess. This star is listed as a variable in SIMBAD.  The [8.0] excess
may be due to interstellar PAH emission or offset between the 2MASS
and IRAC measurements due to variability.

G348.5129-00.1121 (HD 155826) is a G0V MSX source at 31$^{+ 1}_{- 2}$
pc. It appears in numerous papers as a Vega-like system including, 
Dent \etal \ (2005), Jayawardhana \etal \ (2001),
Sylvester \etal\ (1996), and Aumann (1985). Lisse \etal
\ (2002) observed this source with the Long Wavelength Spectrometer
camera on Keck and found an extremely red object in the field of this
star. They note that the red source is either a highly reddened carbon
star or a Class II YSO and that HD 155826 should no longer be
considered a Vega-like source. 

\subsection{Mid-IR excess fraction among MSX and GLIMPSE Stars}

Based on stars detectable at \textit{IRAS} and \textit{ISO}
sensitivities, at least 15\% of nearby A--K main-sequence stars have
dusty debris disks. (Lagrange \etal \ 2000; Backman \& Paresce
1993). Plets \& Vynckier (1999) found the excess fraction for both
main-sequence stars and their descendants to be $\sim$13$\pm$10$\%$,
and Bryden \etal\ (2006) found the excess fraction for main-sequence
FGK stars with $\frac{L_{IR}}{L_{*}} > 10^{-4}$ to be 2$\pm$2$\%$. The
statistics for disk fraction from both of these studies rely
exclusively on longer wavelength data, such as 24 and 60
$\mu$m. Aumann \& Probst (1991) investigated nearby main-sequence
stars using 12 $\mu$m measurements from \textit{IRAS} and found that
less than 0.5 \% of stars exhibit a mid-infrared
excess. Fajardo-Acosta \etal \ (2000) investigated 296 main-sequence
B9 through M stars and found that nearly 3$\%$ of their sample had a
mid-infrared excess at \textit{IRAS} 12 $\mu$m.

Given the large number of sources investigated in this study, we can
place constraints on the fraction of stars that exhibit mid-IR
excesses. Of the 1203 main-sequence Tycho-2 stars in GLIMPSE, 1024 are
later than B8. We found that 19 GLIMPSE sources with V or IV
luminosity classes exhibited mid-IR excesses (omitting HD 178479, a
false excess source).  There are 7 additional GLIMPSE objects with
unknown luminosity class that exhibit mid-IR excesses (omitting HD
136591, a false excess source). These two omitted sources represent
the expected number of false sources due to source correlation errors
and statistical deviations. For the MSX sample, we found that of the
872 main-sequence stars, 785 are of spectral type later than B8. We
found that 16 of the sources of luminosity class V or IV exhibited
mid-IR excesses (omitting HD 155826, HD 177904, and HD 334593 which
are false excess sources).

Omitting the six probable pre-main-sequence objects (Group I), the
stars found to be false excess sources, and requiring luminosity
classification, we determine the fraction of stars showing mid-IR
excesses to be 1.9$^{+0.4}_{-0.5}$$\%$ for the GLIMPSE sample and
1.3$\pm0.4$$\%$ for the MSX sample.  If we remove stars with
main-sequence classifications listed as emission line stars or X-ray
sources, there are 10 remaining GLIMPSE sources, 8 of which have
confirmed excesses at [24]. If we perform the same procedure on the
MSX sample and remove the emission-line and pre-main-sequence sources,
8 stars remain. Given these numbers, we estimate the fraction of
main-sequence stars that exhibit a mid- but not a near-IR excess, at
1.0$\pm0.3$$\%$ for the GLIMPSE sample and 1.0$\pm0.4$$\%$ for the MSX
sample. These statistics might be considered an upper limit if some of
the sample stars turn out to have infrared excesses owing to free-free
emission or contamination from a field object.  On the other hand,
this statistic might be considered a lower limit if luminosity class is
determined for the 7 additional GLIMPSE sources and they are shown to
be main-sequence.  This might also be considered a lower limit if we
include emission-line sources.

\begin{figure}
    \plotone{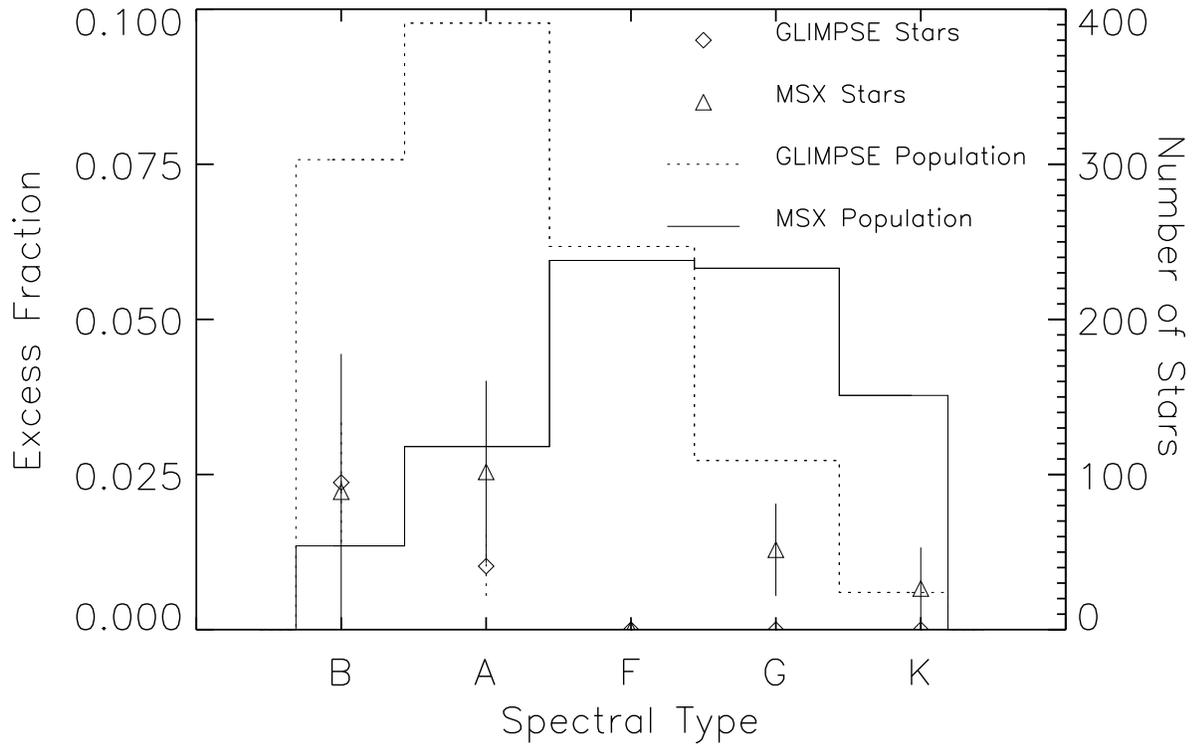}
    \caption{Plot of excess fraction (on the left abscissa) versus
    spectral type for 1024 sources from GLIMPSE (diamonds) and 785
    from MSX (diamonds). The histograms show the number of
    main-sequence GLIMPSE objects (dashed) and MSX objects (solid)
    (numbered on the right abscissa) contained in the Tycho-2 Spectral
    Catalog. For spectral types, B8-K the average fraction of
    main-sequence stars with mid-IR excesses is 1.0$\pm$0.3$\%$,
    varying slightly by type.}
    \label{fracex}
\end{figure}

Figure~\ref{fracex} shows the fraction of stars exhibiting mid-IR
excesses on the left abscissa versus spectral type (points).  Only non
emission-line, non X-ray emitting stars having known luminosity class
are included.  For both the GLIMPSE and MSX samples, the fraction of
excess stars changes with spectral type, with GLIMPSE having the
largest fraction of excess stars among B stars and MSX among A
stars. For the MSX sample, the excess fraction for B, A, and G stars
is similar at $\sim2$$\%$. K stars show a smaller excess fraction, and
there are no F stars with excesses.  The GLIMPSE sample also shows a
decrease in excess fraction with later spectral type. However, there
were no F, G or K stars with an excess in the GLIMPSE sample. The
histogram in Figure~\ref{fracex} shows, on the right abscissa, the
number of stars in the parent population as a function of spectral
type. For example there were 391 stars in the GLIMPSE sample of
spectral type A, and of this subsample, 4 exhibited a mid-IR excess,
yielding an excess fraction of 1.0$\pm0.5$$\%$.

\section{Conclusions}

We investigated stars exhibiting 8 $\mu$m mid-IR excesses in the
GLIMPSE and MSX catalogs. We found 36 probable main-sequence sources
that have a mid-infrared excess but no near-infrared excess (omitting
HD 136591, HD 155826, HD 177904, HD 178479, and HD 334593). All of
these sources lie in a region of mid-IR color space distinct from both
main-sequence and Class II \& III pre-main-sequence stars. We modeled
the properties of the putative circumstellar disks in 20 of the 36
sources. We found putative circumstellar disks temperatures to range
from $191 < T < 787$.  Six stars exhibit fractional infrared
luminosities less than or equal to 10$^{-3}$ and are consistent with
having debris disks similar to $\beta$ Pictoris, although three of
them exhibits Balmer emission lines.  The majority of our sample (14
stars) have $10^{-3} < \frac{L_{IR}}{L_{*}} < 10^{-2}$ and are
consistent either with transition disks in late-stage protostars or
massive debris disks.  These objects have fractional infrared
luminosities and temperatures between those of $\beta$-Pic type
debris-disk systems ($\frac{L_{IR}}{L_{*}}$ $\leq$ 10$^{-3}$) and
Class II pre-main-sequence systems ($\frac{L_{IR}}{L_{*}}$ $\simeq$
10$^{-1}$).  We find that the putative circumstellar disk sources that
exhibit emission-lines have temperatures $>$430~K, while those that do
not exhibit emission lines are almost exclusively $<$430~K. Excluding
probable Class II pre-main-sequence stars, emission-line stars show a
correlation between temperature and fractional infrared
luminosity. This correlation among mid-IR excess sources might imply
an evolutionary progression from high temperature and moderate
fractional infrared luminosity to low temperature and small fractional
infrared luminosity consistent with inner disk clearing.

We estimate the incidence of main-sequence stars found within the
Tycho-2 Spectral Catalog with mid-IR excesses is 1.0$\pm0.3$$\%$. This
statistic varies among spectral types B8--K in both the GLIMPSE and
MSX surveys with B stars exhibiting a excess fraction of
2.4$\pm1.0$$\%$ in GLIMPSE and 2.2$\pm2.2$$\%$ in MSX. The mid-IR
excess fraction among A stars is smaller in the GLIMPSE sample,
1.0$\pm0.5$$\%$, but statistically equivalent to the MSX sample,
2.5$\pm1.5$$\%$.  We did not find any main-sequence F stars, in either
survey, that exhibited a mid-IR excess ($<$0.4\%).  In the MSX sample,
1.3$\pm0.7$$\%$ of main-sequence G stars exhibited a mid-IR excess,
while only 0.7$\pm0.7$$\%$ of K stars exhibited a mid-IR excess.  The
GLIMPSE sample did not contain any main-sequence G or K stars that
exhibited a mid-IR excess ($<$1.0\% and $<$4.1\% respectively). The
statistics of field FGK.  main-sequence stars are consistent with
those of Bryden \etal\ (2006) determined using longer wavelengths. The
mean excess fraction measured here is much lower than the 15-20\% of
main-sequence stars that exhibit far-IR excesses at $\lambda \geq 60$
$\mu$m (Lagrange \etal \ 2000). The rarity of main-sequence stars that
exhibit mid-, but not near-IR, excesses suggests that this stage of
star formation is short-lived, and further investigation of these
sources will help identify their place in the evolutionary process of
star and planet formation.

\acknowledgments

We would like to thank Lynne Hillenbrand for her helpful discussion.
We would like to thank Dylan Semler for his assistance observing at
WIRO. We acknowledge the comments from a critical referee which
inspired significant improvements to this manuscript. Support for this
work, part of the \textit{SST} Space Telescope Legacy Science Program,
was provided by NASA through Contract Numbers (institutions) 1224653
(UW), 1225025 (BU), 1224681 (UMd), 1224988 (SSI), 1242593 (UCB),
1253153 (UMn), 1253604 (UWy), 1256801 (UWW) by the Jet Propulsion
Laboratory, California Institute of Technology under NASA contract
1407.  B.U. acknowledges support from a NASA Graduate Student
Researchers Program fellowship, grant NNX06AI28H.  This research has
made use of the SIMBAD database, operated at CDS, Strasbourg,
France. This publication makes use of data products from the Two
Micron All Sky Survey, which is a joint project of the University of
Massachusetts and the Infrared Processing and Analysis
Center/California Institute of Technology, funded by the National
Aeronautics and Space Administration and the National Science
Foundation.

\clearpage

\begin{deluxetable}{cccc}
\rotate
\tabletypesize{\tiny}
\setlength{\tabcolsep}{0.01in}
\tablewidth{8.0in}
\tablecaption{Revised Spectral/Luminosity Classes}
\tablehead{
\colhead{ID} &
\colhead{Tycho Spectra} &
\colhead{WIRO Spectra} &
\colhead{Classification Basis} }
\startdata
G010.4468+00.0629 & B3/5 IB & B3/5 V & No CII 4267 but had HeI 4388 \\
G015.0981+00.3409 & B7/8 (III) & B5Ve & H$\beta$ absorption with emission core, weak HeI 4713 \\
G026.7861--00.7723 & G0 & G0I & Narrow lines \\
G031.3962--00.4802 & F2/3 & F2II & Ba II 4554 present \\ 
G036.8722--00.4112 & B8 & B8V & Broad deep Balmer lines. MgII 4481 $>$ HeI 4471 \\
G047.3677+00.6199 & B9 & B8Ve & H$\beta$ absorption with emission core, HeI 4471 $\geq$ MgII 4481, HeI 4471 $>$ HeI 4388 \\
G047.4523+00.5132 & A0 & A0V & MgII 4481 and Balmer series only lines present in spectral range \\
G051.6491--00.1182 & A0 & A0V & MgII 4481 and Balmer series only lines present in spectral range \\
G060.5455--00.0757 & B0II & O9/9.5 Ib & HeII 4542 and HeII 4200 present, HeII 4686 $>$ HeI 4716 \\
G063.5770--00.3387 & B9 & B9V & MgII 4481 $>$ HeI 4471, broad Balmer lines \\
G063.6150--00.8305 & B5 & B5Ve & H$\beta$ absorption with emission core, HeI 4481 $>$ MgII 4471 \\
G064.3921+00.0872 & B8 & B8+G & Mg 4481 $\simeq$ He 4471, Ca 4227 present, Fe I lines present, strong Balmer lines \\
\enddata
\tablecomments{ (1) GLIMPSE ID; (2) Tycho-2 Spectral Catalog Classification; (3) Revised Classification from WIRO Observations;
(4) Classification Basis} 
\end{deluxetable}

\begin{deluxetable}{ccccccccccccccccccc}
\rotate
\tabletypesize{\tiny}
\setlength{\tabcolsep}{0.01in}
\tablewidth{8.0in}
\tablecaption{Photometric Parameters for Main-Sequence Stars  B8 or later with Mid-IR Excesses From GLIMPSE}
\tablehead{
\colhead{Table \# } &
\colhead{ID} &
\colhead{Spec.} &
\colhead{V} &
\colhead{B} &
\colhead{J} &
\colhead{[J]$\sigma$} &
\colhead{H} &
\colhead{[H]$\sigma$} &
\colhead{K} &
\colhead{[K]$\sigma$} &
\colhead{[3.6]} &
\colhead{$\sigma$[3.6]} &
\colhead{[4.5]} &
\colhead{$\sigma$[4.5]} &
\colhead{[5.8]} &
\colhead{$\sigma$[5.8]} &
\colhead{[8.0]} &
\colhead{$\sigma$[8.0]} \\
\colhead{\ } &
\colhead{} &
\colhead{} &
\colhead{[mag]} &
\colhead{[mag]} &
\colhead{[mag]} &
\colhead{[mag]} &
\colhead{[mag]} &
\colhead{[mag]} &
\colhead{[mag]} &
\colhead{[mag]} &
\colhead{[mag]} &
\colhead{[mag]} &
\colhead{[mag]} &
\colhead{[mag]} &
\colhead{[mag]} &
\colhead{[mag]} &
\colhead{[mag]} &
\colhead{[mag]} \\
\colhead{(1)} &
\colhead{(2)} &
\colhead{(3)} &
\colhead{(4)} &
\colhead{(5)} &
\colhead{(6)} &
\colhead{(7)} &
\colhead{(8)} &
\colhead{(9)} &
\colhead{(10)} &
\colhead{(11)} &
\colhead{(12)} &
\colhead{(13)} &
\colhead{(14)} &
\colhead{(15)} &
\colhead{(16)} &
\colhead{(17)} &
\colhead{(18)} &
\colhead{(19)}  }
\startdata
1 & G011.2691+00.4208 & B9 V      &  8.29 &  8.38 &  7.88 &  0.02 &  7.90 &  0.05 &  7.82 &  0.03 &  7.64 &  0.04 &  7.74 &  0.04 &  7.68 &  0.04 &  7.48 &  0.05 \\
2 & G014.4239-00.7657 & B8        & 10.94 & 10.92 & 10.32 & 0.03 & 10.29 & 0.03 & 10.24 &  0.03 & 10.23 &  0.07 & 10.11 &  0.05 & 10.16 &  0.09 &  9.77 &  0.10 \\
3 & G027.0268+00.7224 & A0 V      &  9.47 &  9.81 &  8.01 &  0.02 &  7.59 &  0.04 &  7.41 &  0.02 &  7.31 &  0.04 &  7.27 &  0.04 &  7.28 &  0.05 &  7.07 &  0.03 \\
4 & G036.8722-00.4112 & B8V        &  8.69 &  8.95 &  7.94 &  0.02 &  7.88 &  0.03 &  7.82 &  0.02 &  7.72 &  0.06 &  7.67 &  0.06 &  7.65 &  0.04 &  7.55 &  0.03 \\
5 & G042.9876+00.4201 & B8V        &  9.00 &  9.30 &  8.24 &  0.02 &  8.19 &  0.02 &  8.14 &  0.02 &  7.86 &  0.05 &  7.80 &  0.06 &  7.75 &  0.04 &  7.59 &  0.04 \\
6 & G047.3677+00.6199 & B8Ve        &  7.93 &  7.98 &  7.71 &  0.03 &  7.69 &  0.05 &  7.68 &  0.02 &  7.45 &  0.04 &  7.25 &  0.03 &  7.11 &  0.04 &  6.75 &  0.02 \\
7 & G047.4523+00.5132 & A0V        &  9.74 &  9.94 &  9.00 &  0.03 &  8.93 &  0.03 &  8.85 &  0.02 &  8.76 &  0.05 &  8.84 &  0.06 &  8.84 &  0.04 &  8.56 &  0.03 \\
8 & G051.6491-00.1182 & A0V        &  8.11 &  8.17 &  7.80 &  0.03 &  7.84 &  0.05 &  7.82 &  0.03 &  7.81 &  0.04 &  7.78 &  0.06 &  7.82 &  0.04 &  7.63 &  0.03 \\
\enddata
\tablecomments{(1) Table \#; (2) GLIMPSE ID \#; (3) Spectral Type from
Tycho-2 Spectral Catalog; (4) V magnitude;
(5) B magnitude; (6) 2MASS J  magnitude;
(7) 1$\sigma$ uncertainty for J; (8) 2MASS H magnitude; (9) $1\sigma$ uncertainty for
H; (10) 2MASS K magnitude; (11) $1\sigma$ uncertainty for K;
(12) [3.6] magnitude; (13) $1\sigma$ uncertainty for [3.6];
(14) [4.5] magnitude; (15) $1\sigma$ uncertainty for [4.5]; 
(16) [5.8] magnitude; (17) $1\sigma$ uncertainty for [5.8];
(18) [8.0] magnitude; (19) $1\sigma$ uncertainty for [8.0]}
\end{deluxetable}

\begin{deluxetable}{ccccccccccccccccccc}
\rotate
\tabletypesize{\tiny}
\setlength{\tabcolsep}{0.01in}
\tablewidth{8.0in}
\tablecaption{Photometric Parameters for Main-Sequence Stars Earlier than B8 and Post Main-Sequence Stars with Mid-IR Excesses in GLIMPSE}
\tablehead{
\colhead{Table \# } &
\colhead{ID} &
\colhead{Spec.} &
\colhead{V} &
\colhead{B} &
\colhead{J} &
\colhead{[J]$\sigma$} &
\colhead{H} &
\colhead{[H]$\sigma$} &
\colhead{K} &
\colhead{[K]$\sigma$} &
\colhead{[3.6]} &
\colhead{$\sigma$[3.6]} &
\colhead{[4.5]} &
\colhead{$\sigma$[4.5]} &
\colhead{[5.8]} &
\colhead{$\sigma$[5.8]} &
\colhead{[8.0]} &
\colhead{$\sigma$[8.0]} \\
\colhead{\ } &
\colhead{} &
\colhead{} &
\colhead{[mag]} &
\colhead{[mag]} &
\colhead{[mag]} &
\colhead{[mag]} &
\colhead{[mag]} &
\colhead{[mag]} &
\colhead{[mag]} &
\colhead{[mag]} &
\colhead{[mag]} &
\colhead{[mag]} &
\colhead{[mag]} &
\colhead{[mag]} &
\colhead{[mag]} &
\colhead{[mag]} &
\colhead{[mag]} &
\colhead{[mag]} \\
\colhead{(1)} &
\colhead{(2)} &
\colhead{(3)} &
\colhead{(4)} &
\colhead{(5)} &
\colhead{(6)} &
\colhead{(7)} &
\colhead{(8)} &
\colhead{(9)} &
\colhead{(10)} &
\colhead{(11)} &
\colhead{(12)} &
\colhead{(13)} &
\colhead{(14)} &
\colhead{(15)} &
\colhead{(16)} &
\colhead{(17)} &
\colhead{(18)} &
\colhead{(19)}  }
\startdata
1 & G010.0898+00.5289 & B1/2 (I)N &  8.49 &  8.87 &  7.02 &  0.02 &  6.78 &  0.05 &  6.48 &  0.02 & \nodata \tablenotemark{a} & \nodata & \nodata \tablenotemark{a} & \nodata &  5.18 &  0.04 &  4.83 &  0.02 \\
2 & G010.4468+00.0629 & B3/5 V   &  8.40 &  8.67 &  7.45 &  0.02 &  7.28 &  0.04 &  7.09 &  0.02 & \nodata \tablenotemark{a} & \nodata & \nodata \tablenotemark{a} & \nodata &  6.26 &  0.04 &  5.98 &  0.02 \\
3 & G010.6669-00.0339 & B5 II     &  9.15 &  9.44 &  7.89 &  0.02 &  7.76 &  0.05 &  7.58 &  0.02 &  7.26 &  0.04 &  7.11 &  0.06 &  6.86 &  0.04 &  6.64 &  0.03 \\
4 & G011.1473+00.2092 & B9 IB/II  &  9.59 &  9.76 &  8.92 &  0.02 &  8.88 &  0.04 &  8.81 &  0.02 &  8.66 &  0.03 &  8.77 &  0.05 &  8.78 &  0.04 &  8.55 &  0.03 \\
5 & G011.7709-00.9755 & B5 III/V  &  9.19 &  9.35 &  8.58 &  0.02 &  \nodata & \nodata &  8.49 &  0.03 &  8.49 &  0.06 &  8.57 &  0.06 &  8.41 &  0.05 &  8.29 &  0.03 \\
6 & G011.7928+00.6478 & B7/8 IB   & 10.14 & 10.30 &  9.34 &  0.02 &  9.26 &  0.02 &  9.21 &  0.02 &  8.98 &  0.05 &  9.01 &  0.07 &  9.10 &  0.05 &  8.84 &  0.04 \\
7 & G012.1238+00.4899 & B3 (II)E  &  9.01 &  9.29 &  7.68 &  0.02 &  7.52 &  0.05 &  7.28 &  0.03 & \nodata \tablenotemark{a} & \nodata &  6.74 &  0.06 &  6.51 &  0.04 &  6.26 &  0.03 \\
8 & G012.1292+00.0934 & B0?VPE    & 10.12 & 10.32 &  8.49 &  0.02 &  8.16 &  0.04 &  7.91 &  0.03 &  7.52 &  0.06 &  7.38 &  0.05 &  7.29 &  0.04 &  7.03 &  0.03 \\
\enddata
\tablecomments{(1) Table \#; (2) GLIMPSE ID \#; (3) Spectral Type from
Tycho-2 Spectral Catalog; (4) V magnitude;
(5) B magnitude; (6) 2MASS J  magnitude;
(7) 1$\sigma$ uncertainty for J; (8) 2MASS H magnitude; (9) $1\sigma$ uncertainty for
H; (10) 2MASS K magnitude; (11) $1\sigma$ uncertainty for K;
(12) [3.6] magnitude; (13) $1\sigma$ uncertainty for [3.6];
(14) [4.5] magnitude; (15) $1\sigma$ uncertainty for [4.5]; 
(16) [5.8] magnitude; (17) $1\sigma$ uncertainty for [5.8];
(18) [8.0] magnitude; (19) $1\sigma$ uncertainty for [8.0]}
\tablenotetext{a}{Saturated band not non-detection}
\end{deluxetable}

\begin{deluxetable}{ccccccccccccccccccc}
\rotate
\tabletypesize{\tiny}
\setlength{\tabcolsep}{0.01in}
\tablewidth{8.0in}
\tablecaption{Photometric Parameters for Main-Sequence Stars B8 or later with Mid-IR Excesses From MSX}
\tablehead{
\colhead{Table \# } &
\colhead{ID} &
\colhead{Spec.} &
\colhead{V} &
\colhead{B} &
\colhead{J} &
\colhead{[J]$\sigma$} &
\colhead{H} &
\colhead{[H]$\sigma$} &
\colhead{K} &
\colhead{[K]$\sigma$} &
\colhead{MSX A} &
\colhead{$\sigma$ A} &
\colhead{MSX C} &
\colhead{$\sigma$ C} &
\colhead{MSX D} &
\colhead{$\sigma$ D} &
\colhead{MSX E} &
\colhead{$\sigma$ E} \\
\colhead{\ } &
\colhead{} &
\colhead{} &
\colhead{[mag]} &
\colhead{[mag]} &
\colhead{[mag]} &
\colhead{[mag]} &
\colhead{[mag]} &
\colhead{[mag]} &
\colhead{[mag]} &
\colhead{[mag]} &
\colhead{[mag]} &
\colhead{[\%]} &
\colhead{[mag]} &
\colhead{[\%]} &
\colhead{[mag]} &
\colhead{[\%]} &
\colhead{[mag]} &
\colhead{[\%]} \\
\colhead{(1)} &
\colhead{(2)} &
\colhead{(3)} &
\colhead{(4)} &
\colhead{(5)} &
\colhead{(6)} &
\colhead{(7)} &
\colhead{(8)} &
\colhead{(9)} &
\colhead{(10)} &
\colhead{(11)} &
\colhead{(12)} &
\colhead{(13)} &
\colhead{(14)} &
\colhead{(15)} &
\colhead{(16)} &
\colhead{(17)} &
\colhead{(18)} &
\colhead{(19)}  }
\startdata
1 & G007.2369+01.4894 &              A1 V  &  6.88 &  6.97 &  6.20 &  0.02 &  5.53 &  0.04 &  4.78 &  0.02 &  1.55 &  4.10 &  0.66 &  5.00 &  0.50 &  6.10 & -0.59 &  6.00 \\
2 & G008.3752-03.6697 &              F3 V  &  9.15 &  9.79 &  7.32 &  0.02 &  6.93 &  0.03 &  6.40 &  0.02 &  3.31 &  4.10 &  2.40 &  5.60 &  2.26 &  6.50 &  1.93 &  8.50 \\
3 & G009.4973+05.2441 &              A7 V  &  7.60 &  7.86 &  6.76 &  0.02 &  6.60 &  0.03 &  6.54 &  0.02 &  5.96 &  6.30 &  \nodata & \nodata &  \nodata & \nodata &  \nodata & \nodata \\
4 & G025.6122+01.3020 &              B9 V  &  6.36 &  6.39 &  6.21 &  0.02 &  6.18 &  0.04 &  6.16 &  0.02 &  5.74 &  4.90 &  \nodata & \nodata &  \nodata & \nodata &  \nodata & \nodata \\
5 & G040.9995-00.0423 &              F5 V  &  7.22 &  7.56 &  6.47 &  0.02 &  6.36 &  0.02 &  6.30 &  0.02 &  5.82 &  4.90 &  \nodata & \nodata &  \nodata & \nodata &  \nodata & \nodata \\
6 & G042.0859-71.0408 &              A9 V  &  8.44 &  8.72 &  7.43 &  0.02 &  7.30 &  0.02 &  7.25 &  0.03 &  6.51 &  7.70 &  \nodata & \nodata &  \nodata & \nodata &  \nodata & \nodata \\
7 & G054.5163+02.4470 &             K1 IV  &  7.32 &  8.49 &  5.10 &  0.02 &  4.53 &  0.04 &  6.51 &  0.05 &  4.33 &  4.20 &  3.21 & 46.20 &  \nodata & \nodata &  \nodata & \nodata \\
8 & G056.7848+00.8671 &              G8 V  &  9.34 & 10.76 &  6.76 &  0.03 &  6.12 &  0.02 &  5.96 &  0.02 &  5.09 &  4.40 &  \nodata & \nodata &  \nodata & \nodata &  \nodata & \nodata \\
\enddata
\tablecomments{ (1) Table \#; (2) MSX ID \#; (3) Spectral Type from
Tycho-2 Spectral Catalog; (4) V magnitude;
(5) B magnitude; (6) 2MASS J  magnitude;
(7) 1$\sigma$ uncertainty for J; (8) 2MASS H magnitude; (9) $1\sigma$ uncertainty for
H; (10) 2MASS K magnitude; (11) $1\sigma$ uncertainty for K;
(12) MSX A Band magnitude using zero-points from Cohen \etal\ (2000); (13) $1\sigma$ \% Flux Density uncertainty for MSX A;
(14) MSX C Band magnitude; (15) $1\sigma$ \% Flux Density uncertainty for MSX C; 
(16) MSX D Band magnitude; (17) $1\sigma$ \% Flux Density uncertainty for MSX D;
(18) MSX E Band magnitude; (19) $1\sigma$ \% Flux Density uncertainty for MSX E}
\end{deluxetable}

\begin{deluxetable}{ccccccccccccccccccc}
\rotate
\tabletypesize{\tiny}
\setlength{\tabcolsep}{0.01in}
\tablewidth{8.0in}
\tablecaption{Photometric Parameters for Main-Sequence Stars Earlier than B8 with Mid-IR Excesses From MSX}
\tablehead{
\colhead{Table \# } &
\colhead{ID} &
\colhead{Spec.} &
\colhead{V} &
\colhead{B} &
\colhead{J} &
\colhead{[J]$\sigma$} &
\colhead{H} &
\colhead{[H]$\sigma$} &
\colhead{K} &
\colhead{[K]$\sigma$} &
\colhead{MSX A} &
\colhead{$\sigma$ A} &
\colhead{MSX C} &
\colhead{$\sigma$ C} &
\colhead{MSX D} &
\colhead{$\sigma$ D} &
\colhead{MSX E} &
\colhead{$\sigma$ E} \\
\colhead{\ } &
\colhead{} &
\colhead{} &
\colhead{[mag]} &
\colhead{[mag]} &
\colhead{[mag]} &
\colhead{[mag]} &
\colhead{[mag]} &
\colhead{[mag]} &
\colhead{[mag]} &
\colhead{[mag]} &
\colhead{[mag]} &
\colhead{[\%]} &
\colhead{[mag]} &
\colhead{[\%]} &
\colhead{[mag]} &
\colhead{[\%]} &
\colhead{[mag]} &
\colhead{[\%]} \\
\colhead{(1)} &
\colhead{(2)} &
\colhead{(3)} &
\colhead{(4)} &
\colhead{(5)} &
\colhead{(6)} &
\colhead{(7)} &
\colhead{(8)} &
\colhead{(9)} &
\colhead{(10)} &
\colhead{(11)} &
\colhead{(12)} &
\colhead{(13)} &
\colhead{(14)} &
\colhead{(15)} &
\colhead{(16)} &
\colhead{(17)} &
\colhead{(18)} &
\colhead{(19)}  }
\startdata
1 & G058.7278-00.4506 &          B2 V:NNE  & 10.07 & 10.39 &  8.90 &  0.02 &  8.78 &  0.02 &  8.49 &  0.02 &  6.93 &  7.50 &  3.43 & 53.90 &  \nodata & \nodata &  \nodata & \nodata \\
2 & G059.7235-02.0661 &            B3) IV  &  4.85 &  4.71 &  5.39 &  0.24 &  5.05 &  0.02 &  4.95 &  0.02 &  4.66 &  4.30 &  \nodata & \nodata &  \nodata & \nodata &  \nodata & \nodata \\
3 & G065.5291-04.0730 &           B7 V(E)  &  5.88 &  5.78 &  5.97 &  0.02 &  6.05 &  0.02 &  6.00 &  0.02 &  5.40 &  4.60 &  \nodata & \nodata &  \nodata & \nodata &  \nodata & \nodata \\
4 & G071.0310-01.6810 &         B1 V:PNNE  &  7.18 &  7.24 &  6.62 &  0.02 &  6.53 &  0.02 &  6.36 &  0.02 &  5.45 &  4.70 &  \nodata & \nodata &  \nodata & \nodata &  \nodata & \nodata \\
5 & G073.0727+03.7595 &              B3 V  &  5.11 &  4.98 &  5.41 &  0.03 &  5.48 &  0.02 &  5.49 &  0.03 &  4.60 &  4.30 &  \nodata & \nodata &  \nodata & \nodata &  \nodata & \nodata \\
6 & G073.9148+02.0441 &              B3 V  &  4.89 &  4.76 &  5.07 &  0.04 &  5.16 &  0.02 &  5.04 &  0.02 &  4.01 &  4.20 &  3.70 &  9.50 &  3.67 & 10.10 &  \nodata & \nodata \\
7 & G074.2184+01.1861 &            B0 IV)  &  8.43 &  8.73 &  7.18 &  0.03 &  7.05 &  0.02 &  6.76 &  0.02 &  5.73 &  4.90 &  \nodata & \nodata &  \nodata & \nodata &  \nodata & \nodata \\
8 & G074.8713+00.9892 &             B5 IV  &  6.43 &  6.34 &  6.51 &  0.02 &  6.55 &  0.01 &  6.57 &  0.02 &  6.10 &  5.40 &  \nodata & \nodata &  \nodata & \nodata &  \nodata & \nodata \\
\enddata
\tablecomments{ (1) Table \#; (2) MSX ID \#; (3) Spectral Type from
Tycho-2 Spectral Catalog; (4) V magnitude;
(5) B magnitude; (6) 2MASS J  magnitude;
(7) 1$\sigma$ uncertainty for J; (8) 2MASS H magnitude; (9) $1\sigma$ uncertainty for
H; (10) 2MASS K magnitude; (11) $1\sigma$ uncertainty for K;
(12) MSX A Band magnitude; (13) $1\sigma$ \% Flux Density uncertainty for MSX A;
(14) MSX C Band magnitude; (15) $1\sigma$ \% Flux Density uncertainty for MSX C; 
(16) MSX D Band magnitude; (17) $1\sigma$ \% Flux Density uncertainty for MSX D;
(18) MSX E Band magnitude; (19) $1\sigma$ \% Flux Density uncertainty for MSX E}
\end{deluxetable}

\begin{deluxetable}{ccccccccccc}
\tabletypesize{\tiny}
\setlength{\tabcolsep}{0.01in}
\tablewidth{6.0in}
\tablecaption{24 Micron Photometry}
\tablehead{
\colhead{Table \# } &
\colhead{ID} &
\colhead{Spec.} &
\colhead{[24]} &
\colhead{[24]} &
\colhead{$\sigma$ [24]} \\
\colhead{\ } &
\colhead{} &
\colhead{} &
\colhead{[mJy]} &
\colhead{[mag]} &
\colhead{[mJy]} \\
\colhead{(1)} &
\colhead{(2)} &
\colhead{(3)} &
\colhead{(4)} &
\colhead{(5)} &
\colhead{(6)} }
\startdata
1 & G011.2691+00.4208 & B9 V      &  18.4 & 6.47 & 0.3 & \\
2 & G014.4239--00.7657 & B8        & 71.9 & 4.99 & 13.5 \\
3 & G027.0268+00.7224 & A0 V      &  17.7 & 6.51 & 1.1 \\
4 & G036.8722--00.4112 & B8V        &  21.7 & 6.29 & 0.5 \\
5 & G042.9876+00.4201 & B8V        &   4.7 & 7.95 & 0.2 \\
6 & G047.3677+00.6199 & B8Ve        &  44.6 & 5.51 & 0.8 \\
7 & G047.4523+00.5132 & A0V        &  8.5 & 7.31 & 0.7 \\
8 & G051.6491--00.1182 & A0V        & 17.8 & 6.51 & 2.3 \\
\enddata
\tablecomments{(1) Table \#; (2) GLIMPSE ID \#; (3) Spectral Type from
Tycho-2 Spectral Catalog; (4) [24] flux measurement in mJy;
(5) [24] measurement in magnitudes using 7.14 Jy zero-point ; (6) 1$\sigma$ uncertainty for [24] in magnitudes}
\tablenotetext{a}{3$\sigma$ upper limit}
\tablenotetext{b}{Data has not yet been released for this region of MIPSGAL}

\end{deluxetable}

\begin{deluxetable}{cccccccccccccc}
\tabletypesize{\tiny}
\setlength{\tabcolsep}{0.01in}
\tablewidth{7.0in}
\tablecaption{Derived Parameters for Mid-IR
Excess Sources Later than B8 in GLIMPSE and MSX}
\tablehead{
\colhead{Table \# } &
\colhead{ID} &
\colhead{Spec.} &
\colhead{HD \# } &
\colhead{$E(K-[8.0])$} &
\colhead{$\frac{E(K-[8.0])}{\sigma_{(K-[8.0])}}$} &
\colhead{Distance} &
\colhead{$A_{V}$} &
\colhead{Temp} &
\colhead{$\frac{L_{IR}}{L_{*}}$} &
\colhead{T$_{eff}$} &
\colhead{Log g} &
\colhead{R} &
\colhead{Comments} \\
\colhead{\ } &
\colhead{} &
\colhead{} &
\colhead{} &
\colhead{[mag]} &
\colhead{} & 
\colhead{[pc]} &
\colhead{[mag]} &
\colhead{[K]} &
\colhead{} &
\colhead{[K]} &
\colhead{} &
\colhead{arcsec} &
\colhead{} \\
\colhead{1} &
\colhead{2} &
\colhead{3} &
\colhead{4} &
\colhead{5} &
\colhead{6} &
\colhead{7} &
\colhead{8} &
\colhead{9} &
\colhead{10} &
\colhead{11} &
\colhead{12} &
\colhead{13} &
\colhead{14} }
\startdata
1 & G007.2369+01.4894 & A1V & 163296 & 2.96 & 64.89 & 63$^{+ 10}_{- 10}$ & 2.09$^{+ 0.01}_{- 1.05}$ & 493$^{+ 4}_{- 5}$ & 0.14$^{+ 0.002}_{- 0.002}$ & 8200 & +4.29 & 0.8 & M,e,p,o,x,c,g \\
2 & G008.3752--03.6697 & F3V & 167905 & 2.82 & 61.82 & 53$^{+ 10}_{- 10}$ & 1.76$^{+ 0.01}_{- 0.01}$ & 566$^{+ 7}_{- 8}$ & 0.20$^{+ 0.004}_{- 0.005}$ & 6440 & +4.34 & 0.6 & M,o,c,g \\\
3 & G011.2691+00.4208 & B9V & 165854 & 0.34 & 4.52 & 278$^{+ 45}_{- 45}$ & 0.52$^{+ 0.06}_{- 0.06}$  & 473$^{+ 13}_{- 14}$ & 0.0010$^{+ 0.0001}_{- 0.0001}$ & 9520 & +4.14 & 0.6 & G,e \\
4 & G014.4239--00.7657 & B8 & 168246 & 0.48 &  4.25 & 1008$^{+ 155}_{- 155}$ & 0.93$^{+ 0.05}_{- 0.06}$ & 191$^{+ 5}_{- 4}$ & 0.013$^{+ 0.001}_{- 0.001}$ & 11900 & +4.04 & 0.6 & G\\
5 & G027.0268+00.7224 & A0V  & 172030 & 0.22 & 3.68 & 208$^{+ 35}_{- 35}$ & 2.22$^{+ 0.13}_{- 0.13}$ & 486$^{+ 17}_{- 19}$ & 0.0014$^{+ 0.0001}_{- 0.0002}$ & 9520 & +4.14 & 0.4 & G,e \\
6 & G036.8722--00.4112 & B8V & \nodata & 0.26 &  4.09 & 323$^{+ 50}_{- 50}$ & 1.24$^{+ 0.07}_{- 0.07}$ & 388$^{+ 11}_{- 10}$ & 0.0013$^{+ 0.0001}_{- 0.0001}$ & 11900 & +4.04 & 0.3 & G\\
7 & G047.3677+00.6199\tablenotemark{a} & B8Ve & 180398 & 0.95 & 15.98 & 318$^{+ 50}_{- 50}$ & 0.55$^{+ 0.07}_{- 0.03}$ & 787$^{+ 20}_{- 16}$ & 0.0034$^{+ 0.0001}_{- 0.0002}$ & 11900 & +4.14 & 0.4 & G,e \\
8 & G047.4523+00.5132 & A0V & 231033 & 0.26 &  4.08 & 438$^{+ 70}_{- 70}$ & 0.98$^{+ 0.06}_{- 0.05}$ & 355$^{+ 8}_{- 8}$ & 0.0012$^{+ 0.0001}_{- 0.0001}$ & 9520 & +4.14 & 0.5 & G\\
\enddata

\tablecomments{ (1) Table Number; (2) GLIMPSE Catalog ID; (3) Spectral
type; (4) HD Number (5) $E(K-[8.0])$; (6) significance of excess; (7)
Spectrophotometric distance; (8) Visual extinction; (9) Best fit
blackbody model temperature for excess; (10) Best fit infrared
luminosity; (11) Effective temperature for the adopted Kurucz (1993)
model; (12) Log g for the adopted Kurucz model; (13) Radial distance
between optical and IR source positions in arcsecond; (14) Other comments: G: GLIMPSE source M: MSX
source e: emission line star v: variable star n: star in nebula m:
multiple stars [from SIMBAD database] o: Oudmaijer \etal \ (1992) star
p: previous mid-IR spectral work x: Rosat X-ray source c: Clarke \etal
\ (2005) star g: Group I designation} 
\tablenotetext{a}{The model fit is marginal and model flux is below [24].}  
\tablenotetext{b}{There is no data available for $\lambda >$ 8 $\mu$m} 
\tablenotetext{c}{This MSX star is shown to have a false A band or [8.0] excess, see text.}
\tablenotetext{d}{This star is a known Be star with the excess arising
from a circumstellar shell.}
\end{deluxetable}

\clearpage

\clearpage

\clearpage

\clearpage

\clearpage

\clearpage

\clearpage

\clearpage

\clearpage

\clearpage

\clearpage

\clearpage

\clearpage

\end{document}